\documentclass[hyper,12pt]{JHEP3}
\usepackage{epsfig}
\usepackage{amsmath}
\usepackage{amssymb}
\usepackage{epsf}
\usepackage{graphicx}
\usepackage{cite}
\usepackage{graphics}
\usepackage{epstopdf}

\newcommand{\be}{\begin{eqnarray}}
\newcommand{\ee}{\end{eqnarray}}

\newcommand{\nn}{\nonumber} 
\newcommand{\beq}{\begin{equation}}
\newcommand{\eeq}{\end{equation}} 
\newcommand{\bi}{\begin{itemize}}
\newcommand{\ei}{\end{itemize}} 
\newcommand{\beqa}{\begin{eqnarray}} 
\newcommand{\eeqa}{\end{eqnarray}}

\def\bea{\begin{eqnarray}}
\def\eea{\end{eqnarray}}
\def\bean{\begin{eqnarray*}}
\def\eean{\end{eqnarray*}}

\newcommand{\ie}{{\it i.e.}}

\newcommand{\morder}[1]{{\cal O}\left(#1 \right)}
\newcommand{\eq}[1]{(\ref{#1})}

\newcommand{\ave}[1]{\langle{#1}\rangle}

\newcommand{\half}{\frac{1}{2}}  \newcommand{\halft}{{\textstyle \frac{1}{2}}}
\newcommand{\lsim}{\lesssim} \newcommand{\gsim}{\gtrsim}

\def\COMMENT#1{}

 \def\esim{\,\mathrel{\rlap{\lower0.2em\hbox{$-$}}\raise0.15em\hbox{\footnotesize $\hskip0.04em\sim$}}\,}
 \def\gsim{\mathrel{\rlap{\lower0.2em\hbox{$\sim$}}\raise0.2em\hbox{$>$}}}
 \def\ksim{\mathrel{\rlap{\lower0.2em\hbox{$\sim$}}\raise0.2em\hbox{$<$}}}

\title{\center{Energy losses in relativistic plasmas: QCD versus QED}}

\renewcommand{\thefootnote}{\fnsymbol{footnote}}
\author{S. \textsc{Peign\'e} and A.V. \textsc{Smilga}\footnote{On leave of absence from ITEP, Moscow, Russia.}
\\ SUBATECH, UMR 6457, Universit\'e de Nantes \\
Ecole des Mines de Nantes, IN2P3/CNRS. \\ 4 rue Alfred Kastler, 44307 Nantes
cedex 3, France \\ E-mail: \email{peigne@subatech.in2p3.fr,smilga@subatech.in2p3.fr}}

\abstract{\small{We present a mini-review of the problem of evaluating the energy loss of a ultrarelativistic charged particle 
crossing a thermally equilibrated high temperature 
$e^+e^-$ or quark--gluon plasma. The average energy loss $\Delta E$ depends on the particle energy $E$ and mass $M$, the plasma 
temperature $T$, the QED (QCD) coupling constant $\alpha$ ($\alpha_s$), and the distance $L$ the particle travels in the medium. 
Two main mechanisms contribute to the energy loss: elastic collisions and bremsstrahlung. 
For each contribution, we use simple physical arguments to obtain the 
functional dependence $\Delta E(E,M,T,\alpha_{(s)},L)$ in different regions of the parameters. The suppression of bremsstrahlung
due to the Landau-Pomeranchuk-Migdal effect is relevant in some regions. In addition, radiation by heavy particles
is often suppressed for kinematical reasons. Still, when the travel distance $L$ is not too small, and 
for large enough energies ($E \gg M^2/(\alpha T)$ in the Abelian case and $E \gg M/ \sqrt{\alpha_s}$ in the non-Abelian case),  
radiative losses dominate over collisional ones. 
We rederive the known results and make some new observations. We emphasize, in particular, 
that for light particles ($m^2 \ll \alpha T^2$), the difference in the behavior of $\Delta E(E,m,T,\alpha_{(s)},L)$ in QED and QCD 
is mostly due to the different way the problem is usually posed in these two cases. In QED, it is natural to study the energy 
losses of an electron coming from infinity. In QCD, the quantity
of physical interest is the medium-induced energy loss of a parton produced {\it within} the medium.
Somewhat unexpectedly, considering the case of an electron produced within a QED plasma,
we find the same behavior as  
for the medium-induced radiative energy loss in  QCD (in particular $\Delta E_{\rm rad} \propto L^2$ at small $L$), 
despite drastically different behaviors of the photon and gluon radiation spectra,
the latter being due to the fact that the bremsstrahlung cones for soft gluons are broader than for soft photons. We also show that the 
average radiative loss of an ``asymptotic light parton'' crossing a QCD plasma is  similar to that of an asymptotic electron crossing a 
QED plasma. For the radiative loss of heavy particles ($M^2 \gg \alpha T^2$), the difference between QED and QCD is more evident, 
even when the same physical situation is considered.}}

\begin{document}

\renewcommand{\thefootnote}{\arabic{footnote}}
\setcounter{footnote}0
\setcounter{page}{1}

\section{Introduction}
\label{sec1}

Evaluating the energy loss of a charged particle passing through usual matter 
is a standard problem in nuclear physics 
that is very well studied both theoretically and experimentally \cite{Jackson}.
It is known, for example, that for heavy particles (protons) of not too high energy, the main contribution to the energy loss is 
due to collisions with individual atomic electrons, while for light particles (electrons) 
of similar energies it is due to bremsstrahlung. 

\begin{figure}[t]
\begin{center}
\includegraphics[width=3in]{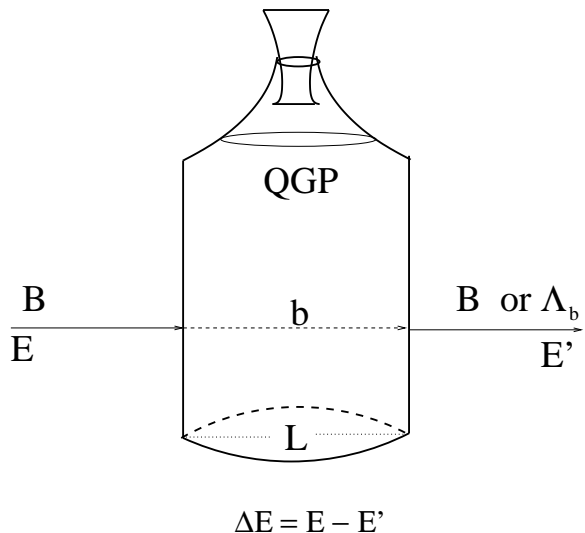}
\end{center}
\caption{A heavy quark passes a thermos bottle filled with QGP.}
\label{bottle}
\end{figure}

The energy loss problem can also be posed for a particle passing a hot ultrarelativistic plasma.\footnote{In all our study we 
will consider thermally equilibrated and non-expanding (static) plasmas.} 
In particular, one may ask what happens 
if a particle carrying {\it color} charge passes a hot QCD medium. In the limit where the medium temperature $T$ is very high, 
$T \gg \Lambda_{_{\rm QCD}}$, this medium is a {quark-gluon plasma} (QGP), \ie  ,
a system of quarks and gluons with a small effective Coulomb-like interaction, 
$\alpha_s(T) \ll 1$.\footnote{We will use the term QGP in this restricted sense 
which is a natural generalization of the conventional definition of a plasma \cite{LPit}. When E.~Shuryak first proposed 
this name, he thought about this analogy \cite{Shur}. Unfortunately, in realistic heavy-ion collisions,
the temperature reached is not high enough to have $\alpha_s(T) \ll 1$, and whether the system can be 
reliably described perturbatively is questionable. 
In this paper we will only consider a situation where the effective coupling is small and perturbation theory applies.} 
Of course, colored particles do not exist 
as asymptotic states, but one can imagine several thought experiments where the energy lost by an energetic parton crossing a QGP could 
in principle be measured. Consider a fast heavy meson (say, a $B$-meson), consisting of a heavy quark and 
a light antiquark, coming from infinity (\ie, created in the remote past) and entering a tiny thermos bottle filled with 
QGP on the laboratory table, as depicted in Fig.~\ref{bottle}. 
In the hot environment, the heavy quark sheds away its light partner and travels through the plasma losing energy. 
When it leaves the bottle, it picks up a light 
antiquark or two light quarks to form a heavy colorless hadron. The difference in energy between the 
incoming meson and the outgoing heavy hadron roughly coincides with the heavy quark energy loss.

This situation is probably the cleanest one from the theoretical viewpoint.  
If the heavy quark mass is very large, $M \gg T$, the density of such quarks in the plasma is exponentially suppressed, 
and thus the passing heavy quark is tagged. 
But it is not possible, of course, to do such an experiment in reality. Hot QCD matter is produced in heavy-ion collisions 
for a few $10^{-23}\, $s, and studying its interaction with a $B$-meson beam is impossible. 
It is possible, however, to access the energy loss of heavy or light particles produced {\it within} the plasma 
(via a hard partonic subprocess). A quark-antiquark pair created with large relative transverse momentum in a hard partonic
process gives rise to two distinct
hadron jets. In heavy-ion collisions, the energy loss of the quark between its production and 
its escape from the QGP softens the $p_T$-spectrum of the leading hadrons in the associated jet, when compared to 
proton-proton collisions. This effect, called {\it jet-quenching} and first anticipated by Bjorken 
\cite{Bjorken}, has now been well-established by the RHIC experiments 
\cite{phenix-light,star-light,phenix-heavy,star-heavy}.\footnote{See Ref.~\cite{quenching-review} for a recent review on 
jet-quenching.} A similar effect in cold nuclear matter 
and at lower energies, namely the attenuation of hadron energy distributions 
in deep inelastic scattering off nuclei, has been observed by the HERMES \cite{HERMES} and CLAS \cite{CLAS} collaborations. 

Considering a parton initially produced in a (perturbative) QGP, there are two distinct cases.  
Either the produced parton is light, or it is a heavy quark. For light partons of sufficiently high energy, $E \gg T$, 
the dominant energy loss mechanism (at not too small $L$) is gluon bremsstrahlung. 
This problem was earlier considered in 
Refs.~\cite{Baier:1994bd,Baier:1996kr,Baier:1996sk,Baier:1998kq,Zakharov:1996fv,Zakharov:1997uu,Zakharov:1998sv,Zakharov:2000iz,Gyulassy:1999zd,Gyulassy:2000fs,Gyulassy:2000er}. 
To evaluate correctly radiative losses, it is important to take into account the generalization of 
the Landau-Pomeranchuk-Migdal (LPM) effect \cite{LP,Migdal,Feinberg} to QCD. 
In brief, bremsstrahlung is a process where a radiation field accompanying a charged particle is shed away. 
But a newborn ``undressed'' particle must first 
``dress''  with its proper field coat  before it can radiate again. 
If the time needed for such dressing ({\it formation time}) is large,  the radiation intensity 
and thereby  the radiative energy loss are suppressed. For heavy quarks, bremsstrahlung is further suppressed compared to the 
light parton case \cite{Dokshitzer:2001zm} and the relative contribution of collisional losses increases. 
When the mass of the particle  is not too large, $M \ll \sqrt{\alpha ET}$ in QED and $M \ll \sqrt{\alpha_s}\, E$ in QCD, and 
the travel distance $L$ is not too small, radiative losses still dominate over collisional ones.

Our goal is to rederive and explain the results in a relatively simple way. 
We will not attempt to perform precise calculations (as far as radiative losses are concerned, 
they are very difficult, maybe impossible to do in a model-independent way) and will only give the physical reasoning 
elucidating the parametric dependence of $\Delta E(E,M,T,\alpha_{(s)}, L)$ in different regions of the parameters. 
This physical emphasize and an accurate analysis of the radiation spectra for light and heavy particles are some 
distinguishing features of our review compared to the review presented in Ref.~\cite{BSZ}.

As emphasized above, studying the energy loss of a parton produced in a QGP is of phenomenological interest to 
heavy-ion collisions. It is, however, also instructive to discuss the problem of energy loss in QED. In this case,
it is more natural to consider an asymptotic (on-shell) particle entering and then leaving a domain containing a 
ultrarelativistic $e^+e^-$ plasma, but the production of a QED particle 
{\it within} a QED plasma can in principle also be considered.\footnote{Think for instance of the energy loss of a charged lepton produced in a heavy-ion collision (though it would be small in this case, due to the smallness of the QGP size compared to the 
lepton mean free path).} Thus, we found it useful to calculate the energy loss of a ultrarelativistic charged 
particle in all different situations we can think of, even though some of them are academic. 

We will first discuss in section \ref{sec2} 
the collisional contribution to the energy loss, considering the cases of QED, QCD, and of heavy and light
particles. For collisional losses, the way the particle is produced (in the remote past outside the plasma, or initially inside the 
plasma) is not important. In sections \ref{sec3} and \ref{sec4} we discuss the radiative energy loss of 
an asymptotic particle crossing a high temperature plasma. Section \ref{sec3} is devoted to QED, where this physical 
situation is more natural. We study in section \ref{sec4} the similar problem in QCD, where the ``asymptotic'' partons
should be understood as constituents of colorless hadrons when entering the plasma. 
In section \ref{sec5} we consider the case of a particle produced in a plasma, both in QED and QCD. 
We show there that the quadratic dependence of the induced radiative loss on the plasma size $L$ at small enough $L$ is not a feature 
specific to QCD. It also holds in QED with the same parametric dependence (up to logarithms). The differential gluon and photon 
radiation spectra are qualitatively different, however. Since it might be useful to jet-quenching phenomenology, 
the induced gluon energy spectra corresponding to light and heavy quark radiation are discussed in some detail. 
Finally, we briefly summarize and give some general remarks in section \ref{sec6}.

\section{Collisional energy loss of a fast charged particle}
\label{sec2}

The first calculation of the collisional loss of a fast charge crossing a hot
plasma is due to Bjorken \cite{Bjorken}. This was done in the context of QCD, and was the basis of Bjorken's proposal 
to use jet-quenching as a signature of the QGP. Bjorken's result for the collisional
loss (per unit distance) of an energetic light parton (light quark ${\rm q}$ or gluon ${\rm g}$) reads
\beq
\label{bjorkenresult}
\left. \frac {dE_{\rm coll}}{dx} \right|_{\rm q, g} = C_R \pi \, \alpha_s^2  T^2 \left(1+\frac{n_f}{6} \right) 
\ln{\frac{ET}{\mu^2}}  \, ,
\eeq
where $n_f$ is the number of thermally equilibrated quark flavours, 
$C_R =C_F = 4/3$ (quark) or $C_R = N_c =3$ (gluon), and $\mu$ is an effective infrared cut-off.
To the logarithmic accuracy, to which \eq{bjorkenresult} is derived, $\mu$ can be taken as the Debye 
screening mass in the QGP, $\mu \sim  gT$. 

More detailed studies of the collisional losses in QED and QCD plasmas were performed for instance  
 in Refs.~\cite{TG,Mrow,BTqedqcd,Thoma:2000dx,Peshier:2006hi,PPqed,PPqcd}. 
We review below this problem, emphasizing the difference between the two cases of a {\it tagged} (heavy) or {\it untagged} 
(light) particle.

\subsection{Hot QED plasma}

\subsubsection{Ultrarelativistic muon}

\begin{figure}[t]
\begin{center}
\includegraphics[width=10cm]{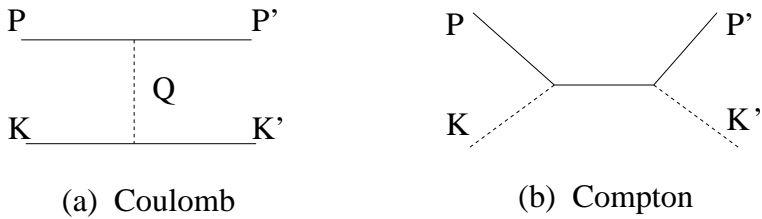}
\end{center}
\caption{The typical graphs for collisions of a muon with plasma particles.}
\label{fig-coll}
\end{figure}

We consider the case of a ultrarelativistic muon of mass $M$ and 
four-momentum $P=(E, \vec{p})$ passing a hot $e^+e^-$ plasma of 
temperature $T$, with $E \gg M \gg T$. The second inequality ensures that there are no muons in the heat bath. 
The muon can lose its energy in either Coulomb 
collisions with electrons and positrons (Fig.~\ref{fig-coll}a), or Compton collisions with photons (Fig.~\ref{fig-coll}b).

Consider first the losses due to Coulomb scattering. The differential 
Coulomb\footnote{Following the usage adopted in the literature on this subject, we use the word ``Coulomb'' in a generalized sense, including also {\it screened Coulomb}, \ie, Yukawa.} cross section is of the form 
\be
\label{Couldiff}
\frac{d\sigma_{\rm Coulomb}}{dt} \ \sim \ \frac {\alpha^2}{(t- \mu^2)^2} \ ,
\ee
where $t \equiv Q^2 = (K-K')^2$ is the Mandelstam invariant momentum transfer, and $\mu \sim eT$
is the Debye screening mass in the QED plasma. The total Coulomb
 scattering cross section reads
\be
\sigma_{\rm Coulomb} \sim \frac{\alpha^2}{\mu^2} \sim \frac{\alpha}{T^2} \ .
\label{Coultot}
\ee

The momenta $K$ and $K'$ in Fig.~\ref{fig-coll} refer to thermal particles and are of order $T$.
Note that they are slightly off mass shell, $K^2 \sim K'^2 \sim \mu^2 $, due to medium effects, which
does not affect, however, the estimates below. Since the integral for the total collisional loss is saturated
by the region $|t| \gg \mu^2$ (as we will see in a moment), we can write $t =  -2 K Q = -2 (|\vec{K}| Q_0 - 
\vec{K} \cdot \vec{Q})$. We have in average 
$Q_0 \sim -  t/(2 |\vec{K}|) \sim - t/T$. The mean energy loss in a single scattering is thus
\be
\label{QEDmeanloss}
\langle \Delta E \rangle_{\rm 1\  scat.} \ \sim \frac{1}{\sigma_{\rm Coulomb}} \int dt\, \frac{d\sigma_{\rm Coulomb}}{dt}
 \, \frac{-t}{T} \sim \alpha T \ln{\frac{|t|_{\rm max}}{\mu^2}} \, .
\ee
The logarithm arises from the broad logarithmic interval $\mu^2 \ll |t| \ll |t|_{\rm max}$, implying 
(to logarithmic accuracy) that the energy transfer in a single collision is small compared to $E$. 

Introducing the Coulomb mean free path,  
\be
\label{mfpath}
\lambda_{\rm Coulomb} \equiv \frac{1}{n\sigma_{\rm Coulomb}} \sim \frac{1}{\alpha T}  \, , 
\ee
where $n \sim T^3$ is the particle density in the plasma, we estimate 
the rate of energy loss per unit distance as 
\be
\label{coulomblossestimate}
\frac{dE_{\rm Coulomb}}{dx} \sim \frac{\langle \Delta E \rangle_{\rm 1 \ scat.}}{\lambda_{\rm Coulomb}}
\sim \alpha^2 T^2 \ln{\frac{|t|_{\rm max}}{\mu^2}} \, .
\ee

The maximal transfer $|t|_{\rm max}$ is given by 
\be
|t|_{\rm max} = \frac{(s-M^2)^2}{s} \sim \frac{E^2T^2}{M^2 + \morder{ET}} \, ,
\ee
where we used $s = (P+K)^2 = M^2 + 2 P\cdot K = M^2 + \morder{ET}$. Thus, we see that the 
logarithmic factor in \eq{coulomblossestimate} takes a different form in the mass domains
$M^2 \ll ET$ and $M^2 \gg ET$, 
\bea
\label{coulomblossestimate2}
\frac{dE_{\rm Coulomb}}{dx} \sim \alpha^2 T^2 
\left \{ 
\begin{array}{cc}
\ln{\frac{ET}{\mu^2}} \ \ \ &  (M^2 \ll ET) \\ \\
2\ln{\frac{ET}{\mu M}} \ \ \ & (M^2 \gg ET) 
\end{array}
\right.
\ \ \, .
\eea

\vskip 5mm

Two remarks are in order here. 
\begin{itemize}

\item[(i)] Strictly speaking, using ~\eq{Couldiff} for the differential cross section is not correct. First, it is written
for static and scalar plasma particles, while the particles are moving and have spin. Second, the graph in 
Fig.~\ref{fig-coll}a was evaluated with the 
model expression $1/(t-\mu^2)$ for the photon propagator, while the actual expression is more involved. Third,
neither $S$-matrix nor 
cross section is well defined in medium, because there are no asymptotic states, and the relevant
quantity is not $\sigma^{\rm tot}$, but the damping rate $\zeta$ of the 
ultrarelativistic collective excitation with muon quantum numbers, 
related to the muon mean free path by $\lambda = \zeta^{-1}$. 
This damping rate was evaluated in Ref.~\cite{damping}, with the result\footnote{The calculation was done for QCD, 
but the Abelian result is directly obtained from the result for $\zeta_q$ 
by setting $c_F =1$ in Ref.~\cite{damping}.}
\be
\label{lamlog}
\lambda = \zeta^{-1}  =  \left[ \half {\alpha T}  \ln{(C/\alpha)}) \right]^{-1}  \, . 
\ee
Thus, the mean free path involves an extra logarithm of the coupling constant as compared to
our estimate \eq{mfpath}. This is because \eq{lamlog} depends on magnetic interactions in addition to
Coulomb scattering (see also Ref.~\cite{BI}).  
Thus, strictly speaking, it is \eq{lamlog} 
which should be taken as the definition of $\lambda$ to be used in sections \ref{sec3} to \ref{sec5}. 
However, keeping track of the logarithms of the coupling constant is a difficult problem,
 which we will not address.
The only logarithms we will keep are those depending on the particle energy $E$. Hence, in all our study
we will use $\lambda \sim 1/(\alpha T)$, which coincides with the Coulomb mean free path \eq{mfpath}.

\item[(ii)] A more standard way to define the mean free path is not as in \eq{mfpath}, but rather by $\lambda^{\rm tr} = 1/(n\sigma^{\rm tr})$, where 
\be
\label{transport}
\sigma^{\rm tr} = \ \int d\sigma (1-\cos \theta)
\ee
is the transport cross section involving an additional suppression factor for small angle scattering. The 
transport mean free path $\lambda^{\rm tr}$ conveniently describes standard transport phenomena associated with collisions 
(viscosity, electric conductivity, $\ldots$).\footnote{For a thorough discussion of transport phenomena in a QGP, see 
\cite{Yaffe}.} For the problem of collisional energy loss, the use of the
definition \eq{transport} is equally warranted, when deriving \eq{coulomblossestimate}
(the estimates for $\langle \Delta E \rangle_{\rm 1 \ scat}$ and $\lambda$ would
be different, but their ratio would not change).
The scale \eq{mfpath} appears to be rather elusive also for radiative losses. Almost all results there depend not on
$\lambda$ as such, but on the combination  $\hat{q} = \mu^2/\lambda$, which is a transport coefficient.
 The ``isolated'' scale $\lambda$ only enters arguments of certain logarithms.
 We will come back to the discussion of this point, when concluding in section \ref{sec6}. We only
stress now that, irrespectively of whether it is observable or not, the notion
of mean free path as defined in \eq{mfpath} or \eq{lamlog} proves to be very convenient and instructive, and
we will stick to this definition throughout the paper.

\end{itemize}

Let us now turn to collisional losses due to Compton scattering (Fig.~\ref{fig-coll}b), and focus on the 
region $M^2 \ll ET$. The differential Compton scattering cross section reads, for $M^2 \ll |u| \ll s \sim ET$, 
\be
\label{Compdiff}
\frac{d\sigma_{\rm Compton}}{dt} \ \sim \ \frac {\alpha^2}{s u} \ .
\ee
This gives the total Compton cross section
\be
\label{sigmaCompt}
\sigma_{\rm Compton} \ \sim \ \frac{\alpha^2}{s} \ln \frac s{M^2} \ \sim \ \frac{\alpha^2}{ET } \ln{\frac{ET}{M^2}} \ ,
\ee
arising to logarithmic accuracy from the domain $M^2 \ll |u| \ll s \simeq |t|$ (recall that $s+t+u = 2M^2$). 
Similarly to \eq{QEDmeanloss}, we find for the energy loss in a single Compton scattering,
\be
\label{Comptonmeanloss}
\langle \Delta E \rangle_{\rm 1\ Compton \  scat.} \ \sim \frac{1}{\sigma_{\rm Compton}} \int dt\, \frac{d\sigma_{\rm Compton}}{dt}
 \, \frac{-t}{T} \sim  \frac{s}{T}  \sim E \, .
\ee
Thus the characteristic energy transfer is of order $E$. (The experimental projects to produce energetic photons 
by scattering energetic electrons off laser beams are based on this property of Compton scattering \cite{photonbeams}).
Introducing the Compton mean free path
\be
\lambda_{\rm Compton} \equiv \frac{1}{n\sigma_{\rm Compton}} \sim \frac{E}{\alpha^2 T^2 \ln{\frac{ET}{M^2}}} 
\gg \lambda_{\rm Coulomb} \, ,
\label{Comptonmpath}
\ee
we obtain
\be
\frac{dE_{\rm Compton}}{dx} \sim \frac{E}{\lambda_{\rm Compton}} \sim \alpha^2 T^2 \ln \frac{ET}{M^2} \, .
\label{Comptonlog}
\ee
Note that the logarithm in \eq{Comptonlog} arises from the same logarithmic integral (over $u$) as for the total Compton 
cross section \eq{sigmaCompt}, and that it is present only in the mass domain $M^2 \ll ET$. 

We see that the losses due to Coulomb and Compton scattering are of the same order, although the
two processes differ drastically. Compton scattering is rare, $\lambda_{\rm Compton} \gg \lambda_{\rm Coulomb}$, 
but, as mentioned above, it is very efficient in transferring energy. 
Summing the Coulomb and Compton contributions to the collisional loss in the domain $M^2 \ll ET$, we 
get\footnote{See Ref.~\cite{PPqed} for the exact calculation, where the constant beyond logarithmic accuracy is also evaluated.}
\be
\label{muoncoll}
\left. \frac {dE_{\rm coll }}{dx} \right|_{\mu^{-}} = \frac{\pi}{3} \alpha^2 T^2 \left[ \ln{\frac{ET}{\mu^2}}
+ \frac{1}{2} \ln{\frac{ET}{M^2}} + \morder{1} \right]  \ \ \ \ \ \ \ (M^2 \ll ET) \, .
\ee

For Compton scattering, when $M^2 \ll ET$ we have $|u| \ll s$ to leading logarithmic accuracy, implying that 
$|t|\simeq |t|_{\rm max} \simeq s$. Thus, in this mass region Compton scattering is characterized by a final state
consisting of a {\it soft} muon and a {\it hard} photon (of energy $\simeq E$) ejected from the plasma. 
This is to be confronted with the characteristic Coulomb scattering process, where the energy transfer is small. 
Beyond the leading logarithm, configurations where the final muon and the scattered thermal particle share similar fractions 
of the initial energy $E$ also contribute to $dE/dx$, for both Coulomb and Compton contributions \cite{PPqed}. 

Finally, let us note that for $M^2 \gg ET$, the Compton logarithm in \eq{muoncoll} should be dropped, 
and the Coulomb logarithm is modified, see \eq{coulomblossestimate2}, giving
\be
\label{muoncolllargeM}
\left. \frac {dE_{\rm coll }}{dx} \right|_{\mu^{-}} = \frac{2\pi}{3} \alpha^2 T^2 \left[ \ln{\frac{ET}{\mu M}}
+ \morder{1} \right]  \ \ \ \ \ \ \ (M^2 \gg ET) \, .
\ee

\subsubsection{Ultrarelativistic electron}

It is tempting to estimate the collisional loss of an energetic electron crossing an $e^+e^-$ plasma 
by replacing  in \eq{muoncoll} the muon mass $M \gg \mu$ by the electron thermal mass 
in the medium $m_{\rm th} \sim e T \sim \mu$. However, as noted above, the leading Compton contribution 
corresponds to the situation where the incoming particle loses almost all its energy.  Thus, it seems impossible 
to distinguish the final soft electron from a thermal electron. In other words, when an energetic electron becomes
soft after interacting with the plasma, it effectively disappears. 
In addition, the incoming electron can be annihilated with a thermal positron. 
A similar situation arises when discussing positron energy loss in usual matter.

One way to better define an {\it observable} energy loss in this case is to require the final electron
to be hard enough,
say, with energy $E' > E/2$. This constraint  makes it very unlikely that the final energetic electron is a thermal electron,
due to the exponential suppression of the Fermi-Dirac thermal weight, 
$n_F(E') \simeq \exp{(-E'/T)} \ll 1$ for $E'> E/2 \gg T$. Demanding the presence of an energetic electron in the 
final state allows one to discard the annihilation channel, as well as the leading (logarithmic) Compton 
contribution. Only the Coulomb contribution, which to logarithmic accuracy corresponds to small
energy transfers, should be kept. 

With this setup, we obtain from \eq{muoncoll}
\beq
\label{electroncoll}
\left. \frac {dE_{\rm coll }}{dx} \right|_{e^{-},\, E'>E/2} = \frac{\pi}{3} \alpha^2 T^2 \left[ \ln{\frac{ET}{\mu^2}}
 + \morder{1} \right]  \, .
\eeq

\subsection{Quark gluon plasma}

\subsubsection{Tagged heavy quark}

Here we consider the case of a ultrarelativistic heavy quark ($E \gg M \gg T$) crossing a QGP. 
For clarity we focus on the limit $M^2 \ll ET$. (As in QED, for $M^2 \gg ET$ the Compton logarithm 
should be dropped in what follows, and the Coulomb logarithm modified, see \eq{coulomblossestimate2}.) 
Purely collisional energy loss does not depend much on the production mechanism of the heavy quark. 
The latter can be produced in a hard process within the medium (like in heavy-ion collisions) 
or preexist within a heavy meson coming from infinity.
The crucial difference between these two situations will reveal itself when studying
the radiative energy loss of a quark induced by its rescattering in the plasma. 

The calculation of the heavy quark collisional loss in a QGP is similar to the case of 
a muon crossing an $e^+e^-$ plasma. The main change consists in the running
of the coupling $\alpha_s$. This is essential to take into account, not only to improve the accuracy
of predictions, but also to 
obtain the correct energy dependence of $dE/dx$. 

The contribution to $dE/dx$ from Coulomb scattering of the fast heavy quark off
thermal quarks and gluons is easily inferred from the QED case. 
The Coulomb differential cross section is $\propto \alpha_s^2$, and the scale
at which to evaluate $\alpha_s$ is given by the invariant momentum transfer $t$ 
itself. Due to the running of $\alpha_s$, the logarithmic integral appearing in the 
fixed coupling (QED) expression \eq{QEDmeanloss} is thus modified to \cite{Peshier:2006hi}
\beq
\label{rulet1}
\alpha^2 \int_{\mu^2}^{ET} \frac{d|t|}{|t|} \rightarrow \int_{\mu^2}^{ET} \frac{d|t|}{|t|} \alpha_s^2(t) \, . 
\eeq
Using $\alpha_s(t) \sim 1/\ln{(|t|/\Lambda^2)}$, the r.h.s. of the latter equation can be 
exactly integrated, and we can rewrite it as
\beq
\label{rulet2}
\alpha^2 \ln\frac{ET}{\mu^2} \rightarrow \alpha_s(\mu^2) \alpha_s(ET) \ln\frac{ET}{\mu^2} \ .
\eeq

A similar discussion applies to the contribution from Compton scattering off 
thermal gluons.\footnote{In QCD, the terms {\it Coulomb} and {\it Compton} refer not to different
processes, as they do in QED, but to different kinematical regions associated with the 
same process. Thus, the amplitude ${\cal M}( {\rm Qg} \to {\rm Qg})$ is dominated by the 
Coulomb diagram with soft gluon exchange when $|t|$ is small. The same amplitude is dominated by 
the Compton-like diagram corresponding to $u$-channel exchange when $|u|$ is small.} 
To logarithmic accuracy, Compton scattering is dominated by $u$-channel exchange. 
The relevant scale  determining the coupling in the differential cross section for this contribution is $\sim \morder{u}$. 
The total Compton scattering cross section in QCD  is
obtained from the QED expression \eq{sigmaCompt} by replacing
\beq
\label{ruleu1}
\alpha^2 \int_{M^2}^{ET} \frac{d|u|}{|u|} \rightarrow \int_{M^2}^{ET} \frac{d|u|}{|u|} \alpha_s^2(u) \ .
\eeq
In other words,
\beq
\label{ruleu2}
\alpha^2 \ln\frac{ET}{M^2} \rightarrow \alpha_s(M^2) \alpha_s(ET) \ln\frac{ET}{M^2} \ .
\eeq

Using \eq{rulet2} and \eq{ruleu2} in \eq{muoncoll}, we obtain for the fast heavy quark collisional loss 
in the limit $M^2 \ll ET$ (after performing the thermal average over 
the target quarks and gluons and introducing color factors \cite{PPqcd}), 
\bea
\left. \frac {dE_{\rm coll}}{dx} \right|_{Q} &=& \frac{4\pi}{3} T^2 \left[ \left(1+\frac{n_f}{6} \right) \alpha_s(\mu^2) \alpha_s(ET) 
\ln{\frac{ET}{\mu^2}} \right. \nn \\ 
&& \hskip 1cm + \left. \frac{2}{9} \alpha_s(M^2) \alpha_s(ET) \ln{\frac{ET}{M^2}} + \morder{\alpha_s^2} \right]  \, .
\label{heavyQcoll}
\eea 
The term beyond logarithmic accuracy $\sim \morder{\alpha_s^2}$ was determined in Ref.~\cite{PPqcd}.
Similarly to the QED case, the Compton leading logarithm  in \eq{heavyQcoll} corresponds to final 
state configurations with a soft (but {\it tagged}) heavy quark jet and a  hard jet 
initiated by a gluon of energy $\simeq E$ knocked out of the plasma.\footnote{We note that such configurations 
are presently not counted in the RHIC  experimental setup, due 
to a lower energy cut-off used in the selection of heavy quark tagged events. 
Under those experimental conditions, only the {Coulomb} leading logarithm should be kept in \eq{heavyQcoll}.}

\subsubsection{(Untagged) light parton}

Similarly to the case of QED, it would be misleading to pretend obtaining the 
energy loss of an energetic light parton by replacing in \eq{heavyQcoll} the mass 
$M$ by the parton thermal mass $\sim g T$. First, even though this is not done in practice, it is
theoretically possible to observe the events with a soft {tagged} heavy quark. But for a light (and hence untagged) 
parton, this is impossible. Second,  light quarks may annihilate with the light antiquarks in the heat bath,
and this adds to the intrinsic uncertainty of what energy loss of a light parton is.

As far as Compton scattering is concerned, the situation is even worse than in QED, where the detection 
of an energetic photon in the final state would at least signal that a Compton scattering occurred. 
In QCD, it is very difficult to distinguish the hadron jet initiated by a hard final gluon produced in Compton scattering from 
the hadron jet initiated by a hard final quark having undergone soft Coulomb exchanges.  

For a light (or, more generally, untagged) parton, the {\it observable} energy loss must be defined, 
at the partonic level, with respect to the leading (\ie, most energetic) parton. 
When $E' < E/2$, or equivalently $|u| < s/2$, 
the corresponding energy loss is thus $\Delta E = E- |\vec{K}'| = E'- |\vec{K}| \simeq E' \sim |u|/(2 |\vec{K}|)$.  
When $E' > E/2$, we have $\Delta E = E- E' \sim |t|/(2 |\vec{K}|)$, as in the case of a 
tagged parton. The Compton contribution to $dE/dx$ is thus of the form
\beq
\label{lossleading} 
\frac {dE_{\rm Compton}}{dx} \sim n \int_{M^2}^s d|u| \, \frac{\alpha_s^2}{s |u|} \, \left[ \frac{|u|}{T} \Theta(s/2 - |u|)
+ \frac{|t|}{T} \Theta(|u| - s/2) \right] \, .
\eeq
We observe that, in contrast to the tagged case, the Compton contribution has no logarithmic enhancement. 
The reason is clear: the logarithm $\sim \ln{(ET/M^2)}$ in \eq{heavyQcoll} came from the small $u$ region. But 
with the ``untagged'' physical definition \eq{lossleading}, small $u$ corresponds to small energy transfer, and does not give
any important (logarithmic) contribution to the energy loss. Within such a definition, we also 
note that the annihilation channel contribution does not yield any logarithm either. 

Thus, as far as leading logarithms are concerned, we are left with only the $t$-channel Coulomb logarithm 
arising from the broad interval $\mu^2 \ll |t| \ll ET$. This Coulomb logarithm can be read off 
from \eq{heavyQcoll}. Generalization to the gluon case is obvious and we obtain 
\beq
\label{lightQcoll}
\left. \frac {dE_{\rm coll}}{dx} \right|_{\rm q, g} = C_R \pi  T^2 \left[ \left(1+\frac{n_f}{6} \right) 
\alpha_s(\mu^2) \alpha_s(ET) \ln{\frac{ET}{\mu^2}} + \morder{\alpha_s^2} \right]  \, ,
\eeq
with $C_R = 4/3$ ($3$) being the quark (gluon) color charge. 
Replacing $\alpha_s(\mu^2) \alpha_s(ET) \to \alpha_s^2$ in \eq{lightQcoll}, we recover the fixed $\alpha_s$ 
Bjorken's result \eq{bjorkenresult}. We stress that Bjorken's result is valid in the logarithmic
approximation and specific to the {\it untagged} experimental setup.

\section{Radiative loss of a fast asymptotic charged particle -- QED}
\label{sec3}

We will now discuss the radiative energy loss due to bremsstrahlung. In this section, we 
consider the case of an {\it asymptotic} charged particle (\ie, produced in the
remote past) crossing a plasma layer of finite size. 
This is a natural experimental setup in QED, where it is possible to prepare an on-shell energetic electron 
entering the plasma with its already formed proper field ``coat''. We thus consider the QED case first.
In section \ref{sec4} we will study the somewhat academic but instructive case of an ``asymptotic quark'' crossing a QGP.

\subsection{Fast electron}

Here we evaluate the radiative energy loss 
of an on-shell energetic electron going through a ultrarelativistic $e^+e^-$ plasma. 
We will assume $E \gg T$ and $\mu \sim eT \gg m$, where $m \equiv m_e$ is the electron mass. 
The electron is scattered by the plasma particles, changes the direction of its motion and emits 
bremsstrahlung photons.

\vskip 5mm
\centerline{{\bf \boldmath{$L \ll \lambda$}: Bethe-Heitler regime}}
\vskip 3mm

\begin{figure}[t]
\begin{center}
\includegraphics[width=10cm]{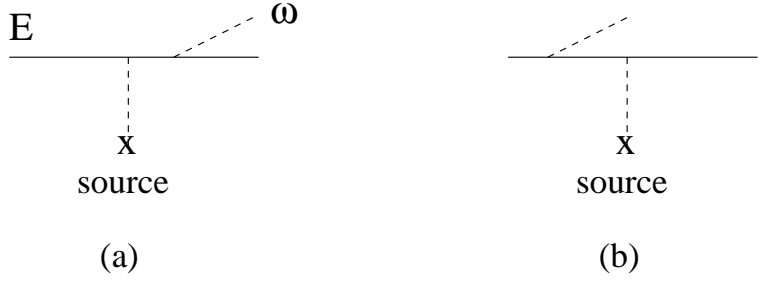}
\end{center} 
\caption{The two diagrams for photon radiation induced by electron Coulomb scattering.}
\label{QEDemission}
\end{figure}

Let us first consider the case of a very thin plasma layer of size $L \ll \lambda$, 
where  the electron mean free path  $\lambda$ is the characteristic distance between subsequent elastic 
scatterings.\footnote{In our paper, $L$  denotes the
distance travelled by the particle in the plasma, to be distinguished from the plasma size $L_p$. 
For thermal equilibration 
to occur, the size of the medium should be much larger than the {\it transport} mean free path of plasma particles 
(which is of the same order
as the transport mean free path of an energetic electron),
$L_p \gg \lambda^{\rm tr} \sim 1/(\alpha^2 T) \gg \lambda$.  Thus, the situation $L \ll \lambda$, implying 
$L \ll L_p$, is quite unrealistic for an asymptotic electron, for which we expect $L \sim L_p$, 
except if the electron is crossing the plasma near its edge. 
(It is somewhat more natural to consider the limit $L \ll \lambda$ in the case of 
a particle produced within a plasma, as we will do in section \ref{sec5}.)

However, in order to better understand what happens in the more physical situation  with larger $L$, we find it instructive 
to consider first the case $L \ll \lambda$. 
} 
As discussed in section \ref{sec2}, $\lambda$ is given by \eq{lamlog} or rather by \eq{mfpath} since we neglect logarithms of the coupling constant 
in our study. 
The probability that the incoming electron undergoes a Coulomb scattering is 
$\sim L/\lambda \ll 1$, and the probability to have several scatterings is further suppressed.
The average energy loss after crossing the length $L$ is thus 
\beq
\label{DEBH}
\Delta E(L \ll \lambda) \sim \ \frac{L}{\lambda} \Delta E^{\,\rm rad}_{1 \, \rm{scat.}} \ ,
\eeq
where $\Delta E^{\,\rm rad}_{1 \, \rm{scat.}}$ is the radiative energy loss induced by a single Coulomb 
scattering.\footnote{The radiative energy losses induced by a single scattering will be referred to as 
Bethe-Heitler (BH) losses in our paper.} It is obtained from the 
photon radiation spectrum derived by calculating the two diagrams of Fig.~\ref{QEDemission}, 
\be
\Delta E^{\,\rm rad}_{1 \, \rm{scat.}} = \int \frac{dI_{\rm rad}}{d\omega} \, \omega \, d\omega \, . 
\ee 

The result for the radiation spectrum 
does not depend much on the nature of scatterers. It depends mainly on the characteristic scattering
cross section and thus on the characteristic momentum transfer $q_{\perp}$. As discussed previously, we may 
consider only Coulomb scatterings. Thus, we can replace the plasma particles by static sources
and use the form \eq{Couldiff} for the differential cross section. The integral giving the total Coulomb cross section 
\eq{Coultot} is saturated by the values $|t| \simeq q_\perp^2 \sim \mu^2$ and, in most estimates (with few exceptions to be spelled out
later), we can assume $q_\perp$ to be on the order of the Debye mass $\mu$. 

In the soft photon approximation $\omega = |\vec{k}| \ll E$, which is sufficient for our purposes,\footnote{In fact, 
the characteristic frequency of emitted photons contributing to $\Delta E_{\rm rad}$ is of order $E$
(see e.g. the spectrum \eq{interpolate} below). However, this modifies the result obtained in the soft photon 
approximation only by numerical factors, which is not our concern here.} the photon radiation intensity reads
\be
dI_{\rm rad} = \sum_{i=1,2} e^2 \left| \frac{P \cdot \varepsilon_i}{P\cdot k} - 
\frac{P' \cdot \varepsilon_i}{P'\cdot k}\right|^2 
\, \frac{d^3\vec{k}}{(2\pi)^3 2\omega} \, , 
\label{photonproba} 
\ee
where $\varepsilon_i$ are two transverse photon polarization vectors. 
When the photon radiation ``angle'' $\vec{\theta} \equiv \vec{k}_{\perp}/\omega$ and the electron scattering ``angle'' 
$\vec{\theta}_s \equiv \vec{q}_{\perp}/E$ are small, $\theta, \theta_s \ll 1$, the sum over photon polarizations gives
\beqa
&& \hskip 1cm dI_{\rm rad} = \frac{\alpha}{\pi^2} \frac{d\omega}{\omega} \, d^2\vec{\theta} \, \vec{J}_e^{\,\, 2}  \, ,
\label{BHspectrum} \\  
&& \hskip 1cm  \vec{J}_e = \frac{\vec{\theta}'}{\theta'^2+\theta_m^2} - \frac{\vec{\theta}}{\theta^2+\theta_m^2} \, , \label{Je} \\
\vec{J}_e^{\,\, 2}  &=& \frac{\theta_s^2}{(\theta^2+\theta_m^2)(\theta'^2+\theta_m^2)} 
- \left( \frac{\theta_m}{\theta^2+\theta_m^2} - \frac{\theta_m}{\theta'^2+\theta_m^2} \right)^2 \ ,
\label{J2}
\eeqa
where $\theta_m \equiv m/E$ and $\vec{\theta}' \equiv \vec{\theta}- \vec{\theta}_s$.

We will be interested in the small and large mass limits, 
corresponding respectively to $\theta_m \ll \theta_s$ and $\theta_m \gg \theta_s$. Thus, we quote:
\beqa
&& \theta_m \ll \theta_s \ \Rightarrow \ \int \frac{d^2\vec{\theta}}{2\pi} \, \vec{J}_e^{\,\, 2} 
\simeq \ln{\frac{\theta_s^2}{\theta_m^2}} \ \Rightarrow \ \omega \, \frac{dI_{\rm rad}}{d\omega} \simeq 
\frac{2 \alpha}{\pi} \, \ln{\frac{\theta_s^2}{\theta_m^2}}  \, ,
\label{largemu} 
\\
&& \theta_m \gg \theta_s \ \Rightarrow \ \int \frac{d^2\vec{\theta}}{2\pi} \, \vec{J}_e^{\,\, 2}  
\simeq \frac{1}{3} \, \frac{\theta_s^2}{\theta_m^2} \ \Rightarrow \ \omega \, \frac{dI_{\rm rad}}{d\omega} \simeq 
\frac{2 \alpha}{3 \pi} \, \frac{\theta_s^2}{\theta_m^2} \, . 
\label{largem}
\eeqa
Eq.~\eq{largemu} is valid to logarithmic accuracy. 
The logarithm arises from the first term of \eq{J2} and from the angular regions $\theta_m \ll \theta \ll \theta_s$
and $\theta_m \ll \theta' = |\vec{\theta}-\vec{\theta}_s| \ll \theta_s$. 
The asymptotic expressions \eq{largemu} and \eq{largem} are conveniently incorporated in the following interpolating formula for 
the energy spectrum,
\beq
\label{interpolate} 
\omega \, \frac{dI_{\rm rad}}{d\omega} =  \frac{2 \alpha}{\pi} \int \frac{d^2\vec{\theta}}{2\pi} \, \vec{J}_e^{\,\, 2} 
\simeq \frac{2 \alpha}{\pi} \, \ln{\left( 1 + \frac{\theta_s^2}{3 \theta_m^2} \right)} \sim 
\alpha \ln{\left( 1 + \frac{q_\perp^2}{3 m^2} \right)} \, .
\eeq
The exact expression displays the same qualitative behavior, but 
is more complicated \cite{photonradtextbooks}. 

When $q_\perp \sim \mu \gg m$ (implying $\theta_s \gg \theta_m$), we obtain
\be
\Delta E^{\, \rm{rad}}_{1 \,{\rm scat.}} \sim \int^E \frac{dI_{\rm rad}}{d\omega} \, \omega \, d\omega \sim  \alpha \, E \, 
\ln{\frac{\mu^2}{m^2}}  \, ,
\label{deltaE1} 
\ee
and for the Bethe-Heitler (BH) energy loss per unit distance,
\be
\frac{dE_{\rm BH}}{dx}(L \ll \lambda)  \ \sim \ \frac{\Delta E^{\,\rm rad}_{1\, {\rm scat.}}}{\lambda} \ 
\sim \alpha^2 ET \ln{\frac {\mu^2}{m^2}} \ \, .
\label{BHlossQED}
\ee

Let us now discuss a possible modification of this result due to Compton scattering. Essentially, there is none.
(This is in contrast to the collisional loss. We have seen in section 2 that the contributions due to Compton and Coulomb scattering
are of the same order there.) 
Indeed, for Compton scattering, $\Delta E^{\,\rm{rad}}_{1\, {\rm scat.}}$ is of the same order as for Coulomb
scattering, 
$\Delta E^{\, \rm{rad}}_{1\, {\rm scat.}} \sim \alpha E$, but the corresponding mean free path \eq{Comptonmpath}
is much larger and 
\beq
\label{Comptonlossrate}
\frac{dE^{\, \rm rad}_{\rm Compton}}{dx}(L \ll \lambda)\ \sim \ \frac{\alpha E}{\lambda_{\rm Compton}} 
\ \sim\ \alpha^3 T^2 
\eeq
is much smaller than the BH contribution \eq{BHlossQED} induced by Coulomb scattering. (The Compton
contribution \eq{Comptonlossrate} is even smaller than the collisional loss \eq{electroncoll}, and will 
not be mentioned any more in this paper.) 

We see that, when $L \ll \lambda$, the BH radiative loss \eq{BHlossQED} of an energetic electron crossing a 
ultrarelativistic plasma is much larger than its collisional loss \eq{electroncoll},
\beq
\label{ratioplasma}
L \ll \lambda \ \Rightarrow \ \frac {dE_{\rm BH}}{dE_{\rm coll}} \ \sim \ \frac ET \gg 1 \ .
\eeq

It is instructive to compare this with the energy losses of a ultrarelativistic particle 
in usual matter \cite{Jackson}, consisting of nonrelativistic electrons and {static} nuclei. 
The basic difference between usual matter and an 
$e^+e^-$ plasma is that, in the former, screening effects do not play an important role. For sure, the electric
fields of individual electrons and nuclei are screened at atomic distances $\sim 1/(\alpha m)$, but these are 
comparatively large distances (see the footnote below).
Screening in usual matter can affect the arguments of certain logarithms, but is otherwise unimportant for order of magnitude
estimates. In the following digression on usual matter, we neglect logarithms. 

Consider first collisional losses. Weighing the $e^- e^- \to e^- e^- $ Coulomb differential cross 
section $d\sigma/dt \sim \alpha^2/t^2$ by the energy transfer $\Delta E (t)$, we get
\beq
\label{collosmattdiff}
\left( \frac {dE}{dx} \right)_{\rm coll}^{\rm usual\ matter} \ \sim 
n Z  \int 
\frac{\alpha^2}{t^2} \, \Delta E(t) \, dt \, ,
\eeq
where $n$ is the number of atoms per unit volume and $Z$ is the number of electrons in an atom. 
Using $t = -2K\cdot Q \simeq -2(m Q_0 -\vec{K}\cdot \vec{Q})$, we have 
$Q_0 = \Delta E(t) \simeq |t|/(2m)$. We obtain, up to some logarithm,
\beq
\label{collosmatt}
\left( \frac {dE}{dx} \right)_{\rm coll}^{\rm usual\ matter} \ \sim \ \frac{n\, Z \alpha^2}{m} \, .
\eeq
Let us note that the order of magnitude of collisional energy loss 
in a hot plasma is recovered from \eq{collosmatt} by replacing $Z \to 1$, $n \to T^3$ and $m \to T$.

Now we discuss radiative losses. 
Let us assume (as we also did for a hot plasma) that the matter layer is thin enough, such that
$\Delta E(L) \ll E$. In addition, let us first assume the scattered energetic particle 
to be an electron and the medium hydrogen ($Z=1$). 
Then the radiative losses due to scattering by electrons and nuclei (protons)  
are of the same order. When $q_{\perp} \ll m$,  the characteristic energy loss in a single scattering is 
$\sim \alpha E$ times the suppression factor $\sim q_{\perp}^2/m^2 \simeq |t|/m^2$, see \eq{largem}. 
Hence, instead of \eq{collosmattdiff} and \eq{collosmatt} we obtain for the BH radiative loss
\beq
\label{radlosmatt}
\left( \frac {dE}{dx} \right)^{\rm hydrogen}_{\rm BH} \ \sim 
n  \int \frac {\alpha^2}{t^2} \, (\alpha E) \,\frac {|t|}{m^2}\, d|t| \ \sim \frac {n\, \alpha^3 E}{m^2} \, .
\eeq
Neglecting logarithms, the estimate (\ref{BHlossQED}) for radiative losses in a ultrarelativistic (thin) plasma 
is obtained from \eq{radlosmatt} by replacing $n \to T^3$ and $m \to \mu \sim e T$, instead of $m \to T$ as we did 
in the case of collisional losses.\footnote{This can be understood by noting that 
in the plasma case, because of screening at the scale $\mu$, only the 
region $|t| \gsim \mu^2 \gg m^2$ contributes. (In usual matter, screening occurs at some scale $|t|_{\rm min} \ll m^2$.) 
Thus, the factor $|t|/m^2$  in \eq{radlosmatt} disappears and the $t$-integral is saturated by $|t| \sim \mu^2$.}

For usual matter, the characteristic ratio of radiative and collisional losses is 
\beq
\label{ratiomat}
\frac {dE_{\rm BH}}{dE_{\rm coll}} \ \sim \ \frac {\alpha E}m \, , 
\eeq
with an additional suppression factor $\alpha$ compared to \eq{ratioplasma}, due to the different screening properties of 
plasma and usual matter. That is why radiative losses in usual matter dominate only at comparatively high energies $E \gg m/\alpha$. 
For electrons in hydrogen, the critical energy where the radiative and collisional 
losses are equal is $E_c \sim 350\,{\rm MeV}$ \cite{PDP}.
 
The physics of collisional and radiative losses differ in one important respect. 
The energetic particle loses a tiny fraction of its energy
in a collision with an individual electron, much like a cannon ball loses a tiny fraction 
of its energy in a collision with an individual air molecule. 
The drag force $dp/dt = dE/dx$ is an adequate  physical quantity to describe this. On the other hand, about half of the original 
particle energy is lost during a single radiation act. The estimate \eq{radlosmatt} refers to an {\it average} drag force, while 
the fluctuations of this quantity are very large. That is why radiative losses are usually not described in terms
of the drag force \eq{radlosmatt}, but in terms of the radiation length $X_0$ 
-- the average distance at which about half of the energy (more exactly, the fraction $1 - 1/e$) is lost.
It is especially sensible bearing in mind that $-dE_{\rm BH}/dx \propto E$ and the energy thus decreases exponentially.

We will see later that, though the radiative energy losses of light partons passing a QCD plasma strongly fluctuate by the same reason
as electron radiative losses, the notion of radiation length is not convenient there, because the linear BH law \eq{radlosmatt} is not
realized and the energy dependence of $dE/dx$ is more complicated. We can also mention right now that the radiation spectrum
of heavy enough quarks (but not of heavy Abelian charged particles!) turns out to be soft, so that the physics is more similar
to the physics of collisional loss and the drag force fluctuations are suppressed.

When $Z >1$ and the incoming particle is heavy, $M \gg m$, two new effects come into play. First, radiation mainly occurs when
the incoming particle is scattered on a heavy nucleus, due to the enhancement factor 
$Z^2$ (square of the nucleus charge) compared to the hydrogen case. The collisional loss is still
due to scattering off electrons and is only enhanced by the number of electrons $Z$. Second, the radiation intensity  
is suppressed at small momentum transfers by the factor $\sim q_{\perp}^2/M^2$. We obtain:
\beq
\label{ratioM}
\frac {dE_{\rm BH}}{dE_{\rm coll}} \ \sim \ \frac {Z \alpha E m} {M^2} \ .
\eeq

The suppression factor $q_{\perp}^2/M^2 \sim \mu^2/M^2$ is effective also for radiative losses of a massive particle passing
through a ultrarelativistic QED plasma (see \eq{largem} and \eq{BHmassive}). The meaning of the condition $\mu \sim eT \gg m$ 
imposed at the beginning of the section has now become
clear. When $\mu \gg M$, a particle of mass $M$ passing through a plasma can be considered as light. 
When $M \gg \mu$, it can be considered as heavy, and its radiative loss in a QED plasma 
is suppressed (at least for $L \ll \lambda$, see \eq{BHmassive} below) by the factor\footnote{This suppression, 
which is due, as can be inferred from \eq{largem}, to the 
suppression of radiation in the cone $\theta < \theta_M \equiv M/E$, is sometimes called {\it dead cone} effect \cite{Dokshitzer:2001zm}.
} $\sim \mu^2/M^2$. 

\vskip 5mm
\centerline{{\bf \boldmath{$L \gg L^*$}: LPM regime}}
\vskip 3mm

After this digression we come back to the case of a hot plasma, and consider a medium of very large size. 
The estimate \eq{DEBH} is correct for small $L \ll \lambda$, the factor $L/\lambda$ bearing the meaning of 
the electron scattering {\it probability}. One can be tempted to extend it also to the
region $L \gg \lambda$, with the factor $L/\lambda$ being interpreted as 
the {\it number} of electron scatterings, but this is not correct
 because the basic assumption on which it is based, 
namely, that a photon is emitted in each scattering, does not hold. 

Indeed, the radiation process  takes time. The particle is not ready to radiate again until it grows 
its accompanying
coat of radiation field. Strictly speaking, it takes an infinite time for the coat to be fully developed such that 
the particle can be treated
as an asymptotic state. But if we are interested only in emission of photons in a particular wave-vector range, we are not 
obliged to wait
eternity and should only be sure that the corresponding Fourier harmonics of the radiation field are already present in the 
coat. The length at which a harmonic corresponding to radiation of the photon of frequency $\omega$ 
at angle $\theta$ is formed, the {\it formation length}, reads\footnote{For our purposes, it is better to talk about formation length rather than formation time. In the rest frame of the virtual electron, 
the formation time/length is of order $1/\sqrt{\kappa^2}$. In the laboratory frame, 
it is multiplied by the Lorentz factor $E/\sqrt{\kappa^2}$. In general, 
the angle $\theta$ appearing in \eq{Lform} is the angle between the photon and 
the electron from which it is radiated (the final electron in Fig.~\ref{QEDemission}a).}  
\beq
\label{Lform}
\ell_f(\omega, \theta) \sim \frac{E}{\kappa^2} \simeq \frac{1}{\omega \theta^2} \simeq \frac{\omega}{k_{\perp}^2}  \, ,
\eeq
where $\kappa^2 \simeq 2 E \omega (1 - \cos \theta) \simeq E \omega \theta^2$ is the virtuality of the internal 
electron in Fig.~\ref{QEDemission}a. 

The formation length $\sim 1/(\omega \theta^2)$ can be interpreted as the length at which a photon of energy $\omega$ emitted at 
angle $\theta$ acquires a phase of order 1 in the frame moving with the particle. The latter condition is, indeed,
\be
\label{disbalance}
\langle {\rm phase \ disbalance} \rangle  \ \sim \omega t - k_\| \ell \sim \ell (\omega - \sqrt{\omega^2 - \omega^2 \theta^2 } )
\sim \ell \omega \theta^2  \sim 1 \ ,
\ee
where we assumed that the electron moves with the speed of light, $t = \ell$ (one can do it as long as 
$k_{\perp} \simeq \omega \theta \gg m$).

The crucial observation is that, even if the electron is scattered {\it several} times before travelling the distance 
$\ell_f(\omega, \theta)$, it cannot emit (to leading order in $\alpha$) more than {\it one} photon
of energy $\omega$ at angle $\theta$. Indeed, in this case one cannot talk about independent photon 
emissions, and we are dealing with coherent emission of a single photon in a multiple scattering process. 
As a result, photon emission with $\ell_f(\omega, \theta) \gg \lambda$ is suppressed compared to the 
situation where it would be additive in the number of elastic scatterings. This effect is known as the
Landau-Pomeranchuk-Migdal (LPM) effect \cite{LP,Migdal,Feinberg}. 

At fixed $\omega$ and when $\mu \gg m$, the radiation spectrum \eq{interpolate} induced by a single scattering 
arises from the angular domains $\theta_m \ll \theta \ll  \theta_s$ and 
$\theta_m \ll \theta' \equiv |\vec{\theta}-\vec{\theta}_s| \ll \theta_s$, \ie, from formation lengths
\beq
\label{timerange1}
\frac{E^2}{\omega \mu^2} \ll \ell_f(\omega, \theta) \ll \frac{E^2}{\omega m^2} \, .
\eeq
In the integrated spectrum, we have $\omega \sim E$, and thus the BH radiative loss \eq{BHlossQED} 
arises from photon formation lengths
\beq
\label{timerange2}
\frac{E}{\mu^2} \ll \ell_f(E, \theta) \ll \frac{E}{m^2} \, .
\eeq
For $E \gg T$, we have $E/\mu^2 \gg \lambda$, and thus the radiated photons contributing to the BH loss 
\eq{BHlossQED} are formed far away from the medium layer of size $L \ll \lambda$. 

What happens when the medium size increases, $L \gg \lambda$? 
Since, for a thin medium of size $L \ll \lambda$, the typical formation length of radiated photons is larger than $E/\mu^2$, we would 
naively expect the energy loss to be approximately given by $\Delta E^{\,{\rm rad}}_{\rm 1 \, {\rm scat.}}$ (see \eq{deltaE1}) 
as long as $L < E/\mu^2$, the entire medium acting as a {\it single} effective scattering center. 
However, this is not so. The critical size $L^*$ beyond which
$\Delta E(L)$ strongly differs from $\Delta E^{\,{\rm rad}}_{\rm 1 \, {\rm scat.}}$ 
happens to be much smaller than $E/\mu^2$. It is approximately given by the 
geometric average of $\lambda$ and $E/\mu^2$, 
\beq
\label{LstarE}
\lambda \ \ \ll \ \  L^* \ \sim \ \sqrt{\frac {\lambda E}{\mu^2}} \sim \frac{1}{\alpha T} \sqrt{\frac{E}{T}}  
\ \ \ll \ \ \frac{E}{\mu^2} \, .
\eeq
To find how this scale arises, consider the limit $L \to \infty$. In this case, all radiated photons 
are formed in the medium, and the estimate of the typical photon formation length $\ell_f^{\rm med}$ 
is modified compared to the vacuum case. To see that, substitute for $\theta^2$ in the estimate \eq{Lform} 
the typical electron deviation angle $\theta_s^2$ at the length $\ell_f^{\rm med}$, 
\beq
\label{Lformmed} 
\ell_f^{\rm med} \sim \frac{1}{\omega \theta_s^2 (\ell_f^{\rm med})} \sim \frac{1}{\omega \, {N} \, \mu^2/E^2}  \ \, ,
\eeq
where ${N} = \ell_f^{\rm med}/\lambda$ is the number of electron scatterings on the distance 
$\ell_f^{\rm med}$. The last estimate in \eq{Lformmed} 
relies on the Brownian motion picture, namely on the assumption that the transverse momentum transfers 
in successive elastic scatterings are not correlated. We obtain 
\beq
\label{Lformmedofomega}
\ell_f^{\rm med}(\omega) \sim \sqrt{\frac{\lambda E^2}{\omega \mu^2}}\, .
\eeq
Substituting $\omega \sim E$, we thus get the estimate \eq{LstarE} for the characteristic
in-medium formation length $L^*$. 

Scattering centers located within a distance $\sim  L^*$ 
act as one { effective} scattering center, inducing {\it single} photon
radiation. For $L \gg L^*$, the medium contains $L/L^*$ effective scattering centers, and 
the estimate for the radiative energy loss is obtained by multiplying this number
by the energy lost in single photon radiation,\footnote{This energy 
is given by \eq{deltaE1}, but {\it without} the mass-dependent logarithmic factor. 
This is because our estimate $\ell_f^{\rm med} \sim  L^*$ displays no logarithmic spread, 
contrary to \eq{timerange2}. This can also be expressed (using \eq{Lform}) by noting that the 
electron travelling in the medium is off mass shell, with a characteristic virtuality of order 
$\kappa^2_{\rm med} \sim E/L^* \sim \sqrt{E \mu^2/\lambda} \sim \alpha \sqrt{ET^3}$. This is in
contrast with the logarithmic spread $m^2 \ll \kappa^2 \ll \mu^2$ in vacuum.}
\beq
\label{DElargeL}
\Delta E (L \gg L^*) \ \sim \ \alpha E \frac{L}{L^*}  \ \sim \ 
\alpha \, L \, \sqrt{\frac{\mu^2}{\lambda} E}\ \sim \ \alpha^2 \, L \, \sqrt{ET^3} \ .
\eeq

This estimate has been obtained by assuming that the typical momentum transfer
squared after ${N}$ electron scatterings is $q_{\perp}^2({N}) \sim {N} \mu^2$ 
($\mu$ being the typical transfer in one scattering). 
In fact, this is correct only when the $q_{\perp}$-distribution in single elastic scattering 
decreases rapidly enough at large $q_{\perp}$, such that the average $\langle q_{\perp}^2 \rangle$
is well-defined. However, for Coulomb scattering, the integral
\beq
\langle q_{\perp}^2 \rangle \ =\ \int dq_{\perp}^2 \, q_{\perp}^2 \, \frac{\mu^2}{(q_{\perp}^2+\mu^2)^2} 
\eeq
diverges logarithmically when the ultraviolet cut-off 
$\left. q_{\perp}^2 \right|_{\rm max} \sim |t|_{\rm max} \sim ET \to \infty$. 
It is possible to show (see Ref.~\cite{Baier:1996sk}, or Appendix A for an alternative derivation) 
that the typical transfer after $N$ 
Coulomb scatterings scales as $({N} \ln{{N}}) \, \mu^2$ instead of ${N} \mu^2$. 
The presence of this logarithm somewhat modifies the estimate \eq{LstarE} of $L^*$, but otherwise does not 
affect the above heuristic derivation. We thus find:
\beqa
L^* \ \sim \ \sqrt{\frac {\lambda E}{\mu^2 \ln (L^*/\lambda)}}
\ \sim \ \sqrt{\frac{\lambda E}{\mu^2 \ln{(E /(\lambda \mu^2))}}} \ \ \ \, , && 
\label{Lstarwithlog} \\
\Delta E (L \gg L^*) \ \sim \ \alpha E \frac{L}{L^*}  \ \sim \ 
\alpha^2 \, L \, \sqrt{E T^3  \, \ln{(E/T)}} \ \ \, . && 
\label{DElargeLCoulomb}
\eeqa

\begin{figure}[t]
\begin{center}
\includegraphics[width=8cm]{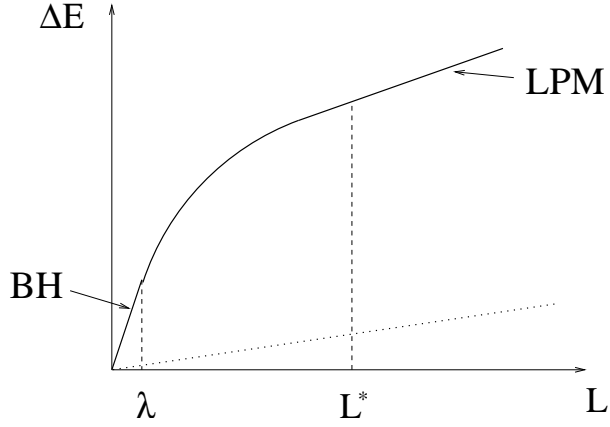}
\end{center}
\caption{Schematic plot of the radiative energy loss of an asymptotic light ($m \ll \mu$) and fast 
($E \gg T$) charge crossing a hot $e^+e^-$ plasma, as a function of the distance $L$ travelled in the plasma. The dotted line represents 
the collisional energy loss $\Delta E_{\rm coll}(L) \sim \alpha^2 T^2 L$.} 
\label{DEvsLQED}
\end{figure}

The parametric dependence of the latter result agrees with that found in Ref.~\cite{Baier:1996vi}. 
At both small $L \ll \lambda$ (see \eq{DEBH}) and large $L \gg L^*$ (see \eq{DElargeLCoulomb}), 
the radiative energy loss is proportional to $L$. But the proportionality coefficients in those 
two regions are different. At large $L$, the slope is 
smaller as a result of LPM suppression. The behavior of $\Delta E(L)$ is 
represented schematically  in Fig.~\ref{DEvsLQED}.

It is worth noting that for $L \gg L^*$, the in-medium {\it energy spectrum} of the radiated photons can 
be easily obtained from \eq{Lformmedofomega},
\be
\label{LPMspectrum}
\omega \frac{dI_{\rm rad}}{d\omega}(L) \ \sim \ \alpha \, \frac{L}{\ell_f^{\rm med}(\omega)} \ \sim \ 
\alpha \sqrt{\frac{\omega\,\omega_c}{E^2}} \ \ \ \ \ \ \ \ \ \ \ 
\left( \omega > \frac {E^2}{\omega_c} \right) \, ,
\ee
where we introduced the energy scale
\be
\label{omegac}
\omega_c \sim \frac{L^2 \mu^2}{\lambda} \, .
\ee
Including the correct logarithmic factor as discussed above, the spectrum \eq{LPMspectrum} reads
\beq
\label{LPMspectrum2}
\omega \frac{dI_{\rm rad}}{d\omega}(L) \ 
\sim \ \alpha^2 \, L \,  \sqrt{ \frac{\omega}{E^2} \, T^3 \, \ln{\left(\frac{E^2}{\omega T}\right)}} \ \ \, .
\eeq
Integrating the latter spectrum up to $\omega \sim E$, we recover \eq{DElargeLCoulomb}.
The LPM suppression is more pronounced at low $\omega$, where the photon formation length \eq{Lformmedofomega} 
is larger. 

In the case of usual matter, the LPM effect  
was observed experimentally for energetic electrons crossing thin metal foils, first at SLAC \cite{SLAC-LPM}, 
and more recently at the CERN SPS \cite{SPS-LPM}. An accurate description of the data based on rigorous theoretical 
calculations is available \cite{Migdal,ZakharovLPMqed}. 

\vskip 5mm
\centerline{{\bf Intermediate region \boldmath{$\lambda \ll L \ll L^*$}}}
\vskip 3mm

The region $\lambda \ll L \ll L^*$ is transitional between the BH and LPM regimes. 
The energy loss in this region is the same as that induced by a single effective scattering of typical momentum transfer
$q_{\perp}^2({N}) \sim ({N} \ln{N}) \, \mu^2$, with ${N} = L/\lambda$. We obtain from \eq{deltaE1}
the energy loss
\beq
\label{intermediate}
\Delta E(\lambda \ll L \ll L^*) \ \sim \ \alpha E \, \ln{\left( \frac{L}{\lambda} \frac{\mu^2}{m^2} \right)} \, ,
\eeq
which in analogy to \eq{timerange2} arises from photon formation lengths 
\beq
\label{timerange3}
\frac{E}{\mu^2 L/\lambda} = \left. \ell_f\right|_{\rm min}\ll \ell_f \ll \frac{E}{m^2} \, .
\eeq
The result \eq{intermediate} holds as long as the radiated photon ``sees'' only one effective scattering
center, \ie, as long as $\left. \ell_f\right|_{\rm min} \gg L$. This condition is precisely equivalent
to $L \ll L^*$. 

The regimes $L \ll \lambda$ and $\lambda \ll L \ll L^*$ smoothly match when $L \sim \lambda$ (see \eq{BHlossQED} and \eq{intermediate}). 
Comparing now \eq{intermediate} and \eq{DElargeLCoulomb}, we see that the former estimate  of $\Delta E(L = L^*)$  involves an extra 
logarithmic factor and dominates at this point.  It is associated with
photons radiated outside the medium.
Adding to \eq{intermediate} the LPM linear term \eq{DElargeLCoulomb}, we obtain 
\beq
\label{linear}
\Delta E(L \gsim L^*) \ \sim \ \alpha E \, \ln{\left( \frac{L^*}{\lambda} \frac{\mu^2}{m^2} \right)} + \alpha E \, \frac{L}{L^*} \, .
\eeq
The linear regime sets in when the second term starts to dominate. This happens at the scale 
\beq
L \sim  L^* \ln{\left( \frac{L^*}{\lambda} \frac{\mu^2}{m^2} \right)} \, ,
\eeq
which is somewhat larger than $L^*$.

\subsection{Energetic muon}
\label{asymQEDheavy}

What happens if the ultrarelativistic particle going through the plasma is heavy, $M \gg \mu \sim eT$? 
As already mentioned, the intensity of 
the BH radiation for a massive QED particle is suppressed by the factor $\sim \mu^2/M^2$, see \eq{largem}.
For $L \ll \lambda$, we thus have
\beq
\label{BHmassive}
\Delta E(L \ll \lambda) \ \sim \ \alpha E \, \frac{L}{\lambda} \frac{\mu^2}{M^2} \ \sim \ \frac{\alpha^3 T^3 E}{M^2} \, L \, .
\eeq
The suppression factor $\sim \mu^2/M^2$ in \eq{largem} 
arises from an integral over the photon radiation angle of the type 
\beq
\label{angularint}
\theta_s^2 \int_{\theta_M^2} d \theta^2/\theta^4 \ \sim \ \theta_s^2/\theta_M^2 \, ,
\eeq
which is saturated by the angles $\theta^2 \sim \theta_M^2 \equiv M^2/E^2$. 

Note now that, for a massive particle, the expression \eq{Lform} for the vacuum formation length is modified,
\be
\label{lfheavy}
\ell_f(\omega, \theta) \ \to \ \ell_f(\omega, \theta, M)  \sim \frac{1}{\omega (\theta^2  + \theta_M^2)} \ .
\ee
The characteristic formation length of emitted photons is thus of order
\beq
\label{formQEDheavy}
\ell_f^{\rm heavy} \ \sim \ \frac{1}{\omega \theta_M^2} \ \sim \ E/M^2 \, .
\eeq
When $M^2 \ll \alpha ET$, we have 
$\ell_f^{\rm heavy} \gg \lambda \gg L$, \ie, the photons are mostly formed outside the medium.

Consider now the behavior of the radiative energy loss for larger lengths. 
There are two distinct cases, depending on whether $M^2$ is smaller or larger than $\alpha \sqrt{ET^3}$.
The appearance of this scale is easy to understand from our previous study of the light particle case. 
We showed in particular that the result \eq{DElargeL} -- or more accurately \eq{DElargeLCoulomb} -- in the domain 
$L \gg L^*$ corresponds to an electron in-medium virtuality of order $\kappa^2_{\rm med} \sim \alpha \sqrt{ET^3}$. 
We thus expect strong modifications of $\Delta E(L)$, as compared 
to the light particle case, when $M^2 \gg \alpha \sqrt{ET^3}$, and milder modifications 
when $M^2 \ll \alpha \sqrt{ET^3}$. 

\vskip 3mm 
\centerline{{\bf A)} \ \boldmath{$M^2 \ll \alpha \sqrt{ET^3}$} } 
\vskip 3mm

The condition $M^2 \ll \alpha \sqrt{ET^3}$ is equivalent to saying that the formation length \eq{formQEDheavy} 
exceeds the scale $L^*$ given in \eq{LstarE} or \eq{Lstarwithlog},
\beq
\label{M4}
\frac{E}{M^2} \gg L^* \sim \sqrt{\frac{\lambda E}{\mu^2}} \ \Leftrightarrow \ M^2 \ll \sqrt{\frac{\mu^2 E}{\lambda}}  
\sim \alpha \sqrt{ET^3} \, .
\eeq

When $L$ somewhat exceeds $\lambda$, the medium still acts as a single 
effective scattering center, transferring the momentum 
$q_{\perp}^2 \sim (L/\lambda) \mu^2$ to the massive particle. The energy loss will be given by $\alpha E$ times the 
suppression factor $q_{\perp}^2/M^2 \sim (L/\lambda)\cdot(\mu^2/M^2)$ as long as $q_{\perp}^2 / M^2 \ll 1$. The result
{\it coincides} with \eq{BHmassive}, which is thereby valid up to the scale $L^{**}$ defined by  
\beq
\label{newscale}
L^{**} \equiv \lambda \frac{M^2}{\mu^2} \gg \lambda \, .
\eeq
We can thus write:
\beq
\label{BHmassive2}
\Delta E(L \ll L^{**})  \ \sim  \ \alpha E \, \frac{L}{\lambda} \frac{\mu^2}{M^2} \ \sim \ \frac{\alpha^3 T^3 E}{M^2} \, L \, .
\eeq

As soon as the suppression factor disappears, the problem is reduced to the already discussed  case
of light particles. The vacuum formation length \eq{formQEDheavy} is modified in the medium to $L^*$. 
For large lengths $L \gg L^{*}$, the dependence of $\Delta E$ on $L$ is linear with the same slope as for light particles,
\beq
\label{DElargeLM}
\Delta E(L \gg L^{*}) \ \sim \ \alpha E \, L/L^*  \,  .
\eeq 

As for a light particle, we can infer that in the
intermediate region $L^{**} \ll L \ll L^{*}$, we have (see \eq{intermediate}):
\beq
\label{intermediateM}
\Delta E(L^{**}\ll L \ll L^*) \ \sim \ \alpha E \, \ln{\left( \frac{L}{\lambda} \frac{\mu^2}{M^2} \right)} 
\ \sim \ \alpha E \, \ln{\left( \frac{L}{L^{**}} \right)} \, .
\eeq

\begin{figure}[t]
\begin{center}
\includegraphics[width=10cm]{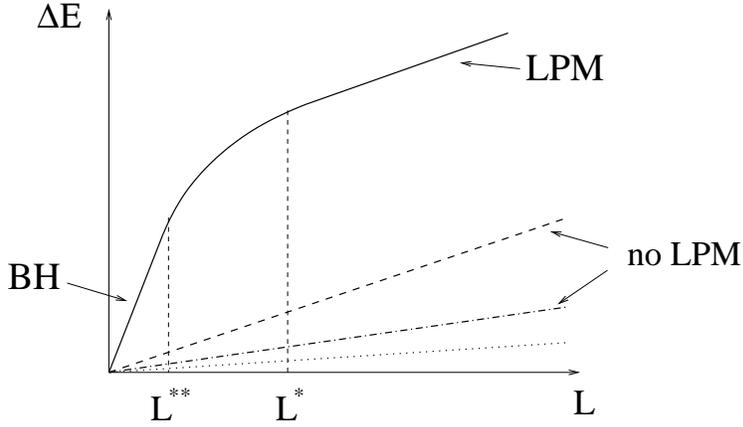}
\end{center}
\caption{Radiative energy loss of a heavy charged  QED particle. 
{\it Solid line:} moderately heavy particle, $\mu^2 \sim \alpha T^2 \ll M^2 \ll \alpha \sqrt{ET^3}$.
{\it Dashed line:} $M^2 \sim \alpha \sqrt{ET^3}$.
{\it Dash-dotted line:} 
 $M^2  \gg  \alpha \sqrt{ET^3}$. {\it Dotted line:} Collisional loss.}
\label{DEvsLM}
\end{figure}

The overall behavior is displayed by the solid line in Fig.~\ref{DEvsLM}. It is similar to the case of 
a light particle (Fig.~\ref{DEvsLQED}), with the scale $L^{**}$ playing the role of $\lambda$. Note also the relation 
\beq
L^{*} = \sqrt{L^{**} \, \ell_f^{\rm heavy}} \, ,
\eeq
which is analogous to \eq{LstarE}. 

\vskip 3mm 
\centerline{{\bf B)} \ \boldmath{$M^2 \gg \alpha \sqrt{ET^3}$} } 
\vskip 3mm

This case corresponds to the formation length \eq{formQEDheavy} being smaller than $L^*$,
\beq 
\label{rangeM}
\frac{E}{M^2} \ll L^*  \ \Leftrightarrow \ M^2 \gg \alpha \sqrt{ET^3} \, .
\eeq

If the vacuum formation length \eq{formQEDheavy} is less than the would-be in-medium
formation length $L^*$, the LPM effect never plays a role. The linear law (\ref{BHmassive}) is valid for {\it all} lengths.
Indeed, it naturally extends up to the scale $L = E/M^2$, where the suppression factor 
$(L/\lambda) \cdot (\mu^2/M^2) \sim L/L^{**}$ is still there.
When $L$ further increases, the suppression factor stays frozen, but the number of emitted photons grows, 
\be
\label{slope}
\Delta E(L \gg E/M^2) \ \sim \ \frac{L}{E/M^2} \left( \frac{E/M^2}{L^{**}} \right) \alpha E 
\ \sim \  \alpha E \, \frac{L}{\lambda} \frac{\mu^2}{M^2} \, . 
\ee    
This coincides with \eq{BHmassive}. The behavior of $\Delta E(L)$ in this case is represented 
by the dash-dotted curve in Fig.~\ref{DEvsLM}. 

We summarize the results of this section by briefly discussing the different slopes which appear in Figs.~\ref{DEvsLQED} and 
\ref{DEvsLM}. Let us consider a fixed energy $E$ and progressively increase the particle mass. When the particle is 
light (Fig.~\ref{DEvsLQED}) the slope for $L \ll \lambda$ is larger than the slope for $L \gg L^*$ by a factor 
$\sqrt{E/T}$. When the mass increases (Fig.~\ref{DEvsLM}, solid line), 
the linear regime at small $L$ extends now to $L^{**}$ (instead of 
$\lambda$) and has a slope reduced by a factor $\mu^2/M^2$. This slope will still be larger than the slope
at $L \gg L^*$ as long as $M^2 \ll \alpha \sqrt{ET^3}$. When $M^2  \sim \alpha \sqrt{ET^3}$ 
(Fig.~\ref{DEvsLM}, dashed line), $L^{**}$ coincides with 
$L^*$, the two slopes also coincide and $\Delta E(L)$ is given by \eq{BHmassive} for all $L$. When $M$ further increases 
(Fig.~\ref{DEvsLM}, dash-dotted line), $\Delta E(L)$ is still given by \eq{BHmassive}, in particular its slope decreases 
as $\sim 1/M^2$.

Comparing this slope with the slope $\sim \alpha^2 T^2$ for collisional energy loss 
(see section \ref{sec2}), we see that the radiative losses of a massive QED particle 
parametrically dominate over collisional ones 
(dotted line in Fig.~\ref{DEvsLM}) when the mass is not too large, or equivalently at large enough energies:
\beq 
\label{collvsradE}
M^2 \ll \alpha E T \ \Leftrightarrow \  E \ \gg \  \frac {M^2}{\alpha T} \ .
\eeq
The latter condition for the dominance of radiative losses in a QED plasma is actually equivalent to the 
vacuum formation length $\ell_f^{\rm heavy} \sim E/M^2$ being larger than the mean free path $\lambda$. 
It is also interesting to note that \eq{collvsradE} 
can be put in correspondence with the case of a massive particle crossing 
usual matter, see ~\eq{ratioM} with the replacements $Z \to 1$ and $m \to T$.

\section{Radiative loss of an ``asymptotic parton''}
\label{sec4}

The experiment where an incoming energetic parton (constituent of an asymptotic hadron) enters a preexisting QGP, 
suffers radiative energy loss, and then escapes the QGP is, unfortunately, not feasible. We nevertheless view the problem of 
evaluating the radiative loss of such an ``asymptotic parton'' crossing a QGP as an instructive and useful
theoretical exercise. We will spend some time on it before going to the more realistic case of a parton produced 
in the medium in section \ref{sec5}. 

The problem of the energy loss of a parton coming from infinity is somewhat better posed (at least, at the level
of a thought experiment) for a tagged heavy quark 
than for a light parton. When the quark is heavy (\ie, when the quark mass
satisfies $M \gg \Lambda_{\rm QCD}$) one can think, 
as was discussed in the Introduction, of a heavy meson (or heavy baryon) scattering off
a thermos bottle containing a QGP. The heavy quark energy loss in the QGP will be roughly the same as the 
energy difference between the initial heavy meson and the final fast hadron with the same flavour. 

If the projectile is a light meson (say, a pion), it contains at least two light valence 
quarks, and the energy loss through the QGP of a single quark is not 
observable, because of the absence of tagging. One can still imagine a thought experiment where the net energy loss  
of the light projectile constituents passing through some medium 
could be observed. Consider the situation where the two valence quarks 
of the incoming pion materialize as two separate jets after crossing
the medium. For the jets to be well separated, they should have a large relative transverse momentum.
This can happen if the pion enters the medium in a compact configuration (\ie, with a large relative 
transverse momentum between the quarks) and loses coherence due to in-medium rescatterings. 
 Although the sum of the total energies of the final jets should coincide with the initial 
pion energy, the energy loss of the light quarks after pion dissociation should affect the 
energy distribution of the {\it leading hadrons} within the jets. The pion diffractive dissociation process in cold nuclear matter, 
$\pi + A  \to$ 2 jets + $A$, has been studied experimentally \cite{piondissoEXP}, as a tool to access the pion wave function 
\cite{Ashery:2005wa}. Light quark  energy loss in nuclear matter should in principle affect this process.
One can contemplate a similar experiment with a QGP. 
Thus, the energy loss of an ``asymptotic light parton'' crossing a QGP
is {\it in principle} observable. 

In the following, we will successively study the cases of a light and heavy parton,
{\it light} and {\it heavy} referring to whether the parton is lighter or heavier than the 
Debye mass $\mu \sim gT$ in the QGP. We will assume $\mu \gg \Lambda_{\rm QCD}$, so that within this definition 
a light parton, $m \ll \mu$, can also be heavy in the usual sense, $m \gg \Lambda_{\rm QCD}$. 
We follow the same lines as in QED (section \ref{sec3}) and study the behavior 
of the radiative loss in different regions of the distance $L$ travelled by the colored particle.

\subsection{Light parton}
\label{sec41}

We will consider the case of a light quark rather than of a gluon, since it 
is technically more convenient (we will also discuss heavy quarks and it will be instructive to see 
what is changed 
when the mass is increased). In the following formulae, the parameter $m$ stands for the quark mass 
when $\Lambda_{\rm QCD} \ll m \ll \mu$, but should be replaced by $\Lambda_{\rm QCD}$ if the quark
is really light in the usual sense, $m \lsim \Lambda_{\rm QCD}$. 
All the results for a gluon are qualitatively the same as for a light quark, with the replacement
$m \to \Lambda_{\rm QCD}$ and with modified color factors -- but those are not our concern in this paper. 

\vskip 5mm
\centerline{{\bf \boldmath{$L \ll \lambda$}: Bethe-Heitler regime}}
\vskip 3mm

\begin{figure}[t]
   \begin{center}
 \includegraphics[width=4cm]{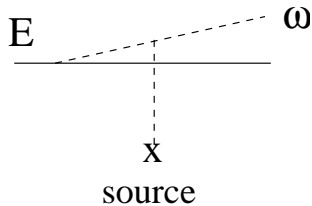}
    \end{center}
\caption[*]{The diagram with a three-gluon vertex contributing to the third term of \eq{gluonamp}.}
\label{3graph}
\end{figure}

For $L \ll \lambda$,  it is sufficient to determine the radiative loss induced by one scattering, as in 
\eq{DEBH}. The gluon radiation amplitude ${\cal M}_{\rm rad}$ induced by a single elastic scattering includes, 
besides the graphs of Fig.~\ref{QEDemission} 
(with proper color factors), also the graph with a three-gluon vertex shown in Fig.~\ref{3graph}. This graph is 
evaluated for instance in Ref.~\cite{Gunion-Bertsch} in the case of a massless quark. The generalization to a massive quark
is trivial and the sum of the three diagrams gives for $\omega \ll E$ ( we will work consistently in this approximation, as
we did in QED)
\be
\label{gluonamp}
{\cal M}_{\rm rad} \propto 
\left[ \frac{\vec{\theta}}{\theta^2+\theta_m^2} t^a t^b - \frac{\vec{\theta}'}{\theta'^2+\theta_m^2}t^b t^a 
- \frac{\vec{\theta}''}{\theta''^2+\theta_m^2} [t^a,t^b] \right] \cdot \vec{\varepsilon} \ \ \ \, .
\ee
In the latter expression, the parameters $\vec{\theta} = \vec{k}_\perp/\omega$, 
$\vec{\theta}' = \vec{\theta}- \vec{\theta}_s = \vec{\theta}- \vec{q}_\perp/E$ and $\theta_m = m/E$ 
are the same as in \eq{Je}. We have also denoted $\vec{\theta}'' = \vec{\theta}- \vec{\theta}_g$, 
with $\vec{\theta}_g = \vec{q}_{\perp}/\omega$. The parameter $\theta_g$ can be interpreted
as the scattering angle of the virtual gluon of energy $\simeq \omega$ in Fig.~\ref{3graph}.  
The color factors can be conveniently grouped in terms of the commutator $[t^a,t^b]$ and anticommutator 
$\{ t^a,t^b \}$ of color matrices. Eq.~\eq{gluonamp} can be rewritten as 
\be
\label{gluonamp2}
{\cal M}_{\rm rad} \propto \left[ 
[t^a,t^b] \left( \vec{J}_q - \halft \vec{J}_e \right) + \{ t^a , t^b \} \halft \vec{J}_e \right] \cdot \vec{\varepsilon}
\ \ \ \, ,
\ee
where $\vec{J}_e$ (already given in \eq{Je}) and $\vec{J}_q$ read:
\beqa
\label{je} 
\vec{J}_e &=& \frac{\vec{\theta}'}{\theta'^2+\theta_m^2} - \frac{\vec{\theta}}{\theta^2+\theta_m^2}
\ \ \ , \\ 
\label{jq} 
\vec{J}_q &=& \frac{\vec{\theta}''}{\theta''^2+\theta_m^2} - \frac{\vec{\theta}}{\theta^2+\theta_m^2}
\ \ \, .
\eeqa

The soft  gluon radiation intensity can be obtained by squaring \eq{gluonamp2}, summing/averaging over 
color indices, and normalizing by the elastic scattering cross section. For $N_c$ quark colors, we obtain:
\beqa
dI_{\rm rad} &=& \frac{\alpha_s}{\pi^2} \frac{d\omega}{\omega} \, d^2\vec{\theta} \, 
\left\{ N_c\,( \vec{J}_q - \halft \vec{J}_e )^2 + \frac{N_c^2-2}{N_c} ( \halft \vec{J}_e )^2 \right\} \nn \\
 &=& \frac{\alpha_s}{\pi^2} \frac{d\omega}{\omega} \, d^2\vec{\theta} \, 
\left\{ N_c \, \vec{J}_q^{\,\, 2} - \frac{1}{2N_c} \vec{J}_e^{\,\, 2} 
+ \frac{N_c}{2} \left[ ( \vec{J}_q - \vec{J}_e )^2 - \vec{J}_q^{\,\, 2} \right] 
\right\} \, .
\label{gluonproba}
\eeqa

In the latter expression, the terms have been organized to facilitate the integral over angles. 
Considering a light quark with $m \ll \mu$
and using \eq{largemu}, we see that the first term of \eq{gluonproba} contributes to the energy spectrum as
\be
\label{spectrumlightQCD}
\omega \left. \frac{dI_{\rm rad}}{d\omega} \right|_{\rm broad} \sim \alpha_s \ln{\frac{\theta_g^2}{\theta_m^2}}  \sim \alpha_s 
\ln{\left( \frac{\mu^2}{m^2} \frac{E^2}{\omega^2}\right)} \, ,
\ee
the logarithm arising from the {\it broad} angular domain $\theta_m \ll \theta \ll \theta_g$. 
The second term of \eq{gluonproba} is similar to the QED case, see \eq{BHspectrum}. Its contribution to the 
spectrum is 
\be
\label{spectrumlightQED}
\omega \left. \frac{dI_{\rm rad}}{d\omega} \right|_{\rm narrow} \sim \alpha_s \ln{\frac{\theta_s^2}{\theta_m^2}}  \sim \alpha_s 
\ln{\frac{\mu^2}{m^2}} \, ,
\ee
arising from the {\it narrow} angular domain $\theta_m \ll \theta \ll \theta_s$. 
The third term of \eq{gluonproba} does not bring any logarithm in the energy spectrum and will be neglected in
the following. By integrating \eq{spectrumlightQCD} and \eq{spectrumlightQED} over $\omega$, we see that the terms 
$\propto \vec{J}_q^{\,\, 2}$ and $\propto \vec{J}_e^{\,\, 2}$ contribute similarly to the radiative loss,\footnote{This is because 
in the integrated spectrum we have $\omega \sim E$, and the broad and narrow angular domains then coincide.}
\be
\label{QCDlikecont}
\Delta E^{\,\rm rad}_{1 \, \rm{scat.}} \sim \alpha_s E \ln{\frac{\mu^2}{m^2}} \, .
\ee
Thus, in the light quark case $m \ll \mu$, we have, similarly to the QED case (see \eq{DEBH} and \eq{deltaE1}),
\beq
\label{BHQCDlight}
\Delta E(L \ll \lambda) \ \sim \ \frac{L}{\lambda} \Delta E^{\,\rm rad}_{1 \, \rm{scat.}} \ \sim \ 
\alpha_s E \, \frac{L}{\lambda} \ln{\frac{\mu^2}{m^2}} \, .
\eeq

To recapitulate, at fixed $\omega \ll E$, we have $\theta_s \ll \theta_g$, and the differential energy spectrum
receives two logarithmic contributions: a QED-like contribution, from the narrow region $\theta_m \ll \theta \ll \theta_s$,  
and a contribution specific to QCD, from the broad region $\theta_m \ll \theta \ll \theta_g$. 
The second contribution (note that it dominates at large $N_c$) reads
\beq
\omega \left. \frac{dI_{\rm rad}}{d\omega} \right|_{\rm broad}
\sim \alpha_s \int d^2\vec{\theta} \, \vec{J}_q^{\,\, 2}\, .
\label{gluonprobaapprox}
\eeq
Using $\vec{k}_{\perp} \simeq \omega \vec{\theta}$ and neglecting the quark mass, 
the spectrum \eq{gluonprobaapprox} can be expressed as 
\beq
\left. \omega \frac{dI_{\rm rad}}{d\omega \, d^2 \vec{k}_{\perp}} \right|_{k_\perp \gg \, \omega \mu /E}
\sim \alpha_s 
\frac{q_{\perp}^2}{k_{\perp}^2 (\vec{k}_{\perp} - \vec{q}_{\perp})^2} \, ,
\label{gunionbertsch}
\eeq
which is the well-known Gunion-Bertsch spectrum \cite{Gunion-Bertsch}.

We emphasize that the spectrum \eq{gluonprobaapprox} is obtained from the QED spectrum \eq{BHspectrum} by replacing the 
electron scattering angle $\theta_s \equiv q_{\perp}/E \sim \mu/E$ by the virtual gluon scattering angle 
$\theta_g \equiv q_{\perp}/\omega \sim \mu/\omega$ (compare \eq{je} and \eq{jq}). 
This fact has interesting consequences. 

\begin{itemize}
\item[(i)] At fixed $\omega \ll E$, the broad angular domain $\theta_m \ll \theta \ll \theta_g$ 
contributing to \eq{gluonprobaapprox} translates to the interval in gluon formation lengths 
\beq
\label{timerange1QCD}
\frac{\omega}{\mu^2} \ll \ell_f(\omega) \ll \frac{E^2}{\omega m^2} \, ,
\eeq
to be compared to \eq{timerange1} in the QED case. Thus, in QCD the gluon starts to be formed at 
much smaller lengths, 
\beq
\label{gluonlength}
\left. \ell_f(\omega) \right|_{\rm min} \sim  \frac{\omega}{\mu^2}  \, ,
\eeq
than the photon in QED. Since $\left. \ell_f(\omega) \right|_{\rm min} \propto \omega$ in QCD, 
we expect the LPM suppression of energy loss  
to {\it increase} with increasing $\omega$  (in QED, the opposite was true, see the remark after 
\eq{LPMspectrum2}). 
We will discuss this in more detail below. 
\item[(ii)] As long as $\omega \sim E$ in the {\it integrated} spectrum (in which case the broad QCD domain 
$\theta \ll \theta_g$ coincides with the narrow QED domain $\theta \ll \theta_s$), the typical formation lengths in QED and QCD
do not differ (see \eq{timerange2}),
\beq
\label{timerange2QCD}
\frac{E}{\mu^2} \ll \ell_f(\omega \sim E) \ll \frac{E}{m^2} \, .
\eeq
We thus anticipate that the average radiative loss of a light quark has a parametric dependence similar to that
of an electron, for all travel distances $L$. 
\item[(iii)] On the other hand, if the characteristic frequencies $\omega$ in the integrated spectrum were much smaller than $E$, 
the broad angular domain would indeed be broader than the 
narrow one, and we would obtain different parametric behaviors for the radiative loss in QCD and QED. 
We will see in the second part of this section that it is exactly what happens for heavy quarks.
\end{itemize}

\vskip 3mm
\centerline{{\bf \boldmath{$L \gg L^*$}: LPM regime}}
\vskip 3mm

The derivation of the light quark energy loss when $L \gg L^*$ follows the same lines as in 
QED, with some differences which we discuss below. First, similarly to the photon case, for $L \gg L^*$ a single gluon is emitted 
in a multiple scattering process composed of ${N} \sim \ell_f^{\rm med}(\omega)/\lambda$ individual 
scatterings. This amounts to exchanging ${N}$ gluons in the $t$-channel. This ${N}$-gluon state can be either
color octet or color singlet (higher representations do not contribute to the quark scattering
amplitude). In the color singlet case, the physics is the same as in QED and the radiated gluon is emitted within a narrow cone\footnote{In contrast 
to what happens in the BH regime, this cone does not produce any logarithmic factor, see the footnote related to 
\eq{DElargeL}.} of angle 
$\theta_s^2({N}) \sim {N} \mu^2/E^2$. For the color octet, the physics is the same as for the process with one gluon
exchange discussed above where two cones are present, but with the narrow and broad cones angles being now of
order $\theta_s^2({N}) \sim {N} \mu^2/E^2$ and $\theta_g^2({N}) \sim {N} \mu^2/\omega^2$, 
respectively. 
In the large $N_c$ limit, the probability to have a singlet 
$t$-channel ${N}$-gluon state is suppressed by $1/N_c^2$ compared to the probability to have an octet. 
This suppression is of the same order as the suppression of the Abelian contribution yielding the narrow radiation cone in single gluon
exchange (the second term in \eq{gluonproba}). In other words, the dynamics of gluon emission in a multiple scattering process 
is roughly the same as for single scattering, the only difference being that the characteristic momentum transfer $\mu^2$ is multiplied 
by the factor ${N} \sim \ell_f^{\rm med}(\omega)/\lambda$. 
 
The in-medium gluon formation length $\ell_f^{\rm med}(\omega)$ is obtained from \eq{Lform} 
by replacing the gluon radiation angle $\theta^2$ by $\theta_g^2({N}) \sim {N} \mu^2/\omega^2$, 
\beq
\label{LformmedQCD} 
\ell_f^{\rm med}(\omega) \sim \frac{\omega}{{N} \mu^2} \ \Rightarrow \ \ell_f^{\rm med}(\omega) \sim 
\sqrt{\frac{\omega \lambda}{\mu^2}} \, .
\eeq
The gluon formation length increases with $\omega$, an opposite behavior compared to QED, see
\eq{Lformmedofomega}. The gluons with $\omega \sim E$ are formed within the length $L^*$ given in 
\eq{LstarE}, notwithstanding, leading to the same parametric dependence of the radiative loss as in QED,
\beq
\label{DElargeLQCD}
\Delta E (L \gg L^*) \ \sim \ \alpha_s E \frac{L}{L^*}  \ \sim \ 
\alpha_s \, L \, \sqrt{\frac{\mu^2}{\lambda} E}\ \sim \ \alpha_s^2\, L \, \sqrt{ET^3} \, .
\eeq

There are, however, differences between the QED case of an electron and the QCD case of a light quark.
\bi
\item[(i)] Similarly to QED, due to the long tail in the Coulomb scattering potential, the typical 
transverse momentum exchange $q_{\rm typ}^2({N})$ after ${N}$ scatterings is not exactly ${N} \mu^2$. 
As shown in Appendix A, it is $\sim ({N} \ln{N})\,\mu^2$ in QED, but in QCD this simple dependence is replaced by 
\eq{qtypnQCD} due to the running of $\alpha_s$. For very large ${N}$, $q_{\rm typ}^2({N})$ is given by the expression 
\eq{qtypngg} that does not involve the factor $\ln {N}$. This means that the factor 
$\sim \sqrt{\ln E}$ in the energy loss of energetic light partons in the large $L$ region
disappears for asymptotically large energies $\ln (E/T) \gg \ln (\mu/\Lambda_{\rm QCD})$. We have instead 
of \eq{Lstarwithlog} and \eq{DElargeLCoulomb}, 
\beqa
\hskip 0.5cm L^* \ &\sim& \ \sqrt{\frac{\lambda E}{\mu^2 \ln{\frac{\mu}{\Lambda_{\rm QCD}}}}}  \, , 
\label{LstarQCDwithlog}
\\
\Delta E (L \gg L^*) \ &\sim& \ \alpha_s E \frac{L}{L^*}  \ \sim \ 
\alpha_s^2 \, L \, \sqrt{E T^3  \, \ln{\frac{\mu}{\Lambda_{\rm QCD}}}}  \ \, . 
\label{DElargeLCoulombQCD0}
\eeqa
When the energy is large but not asymptotically large, the dependence is more complicated,
\beq 
\label{DElargeLCoulombQCD}
\Delta E (L \gg L^*)  \ \sim \ 
\alpha_s^2 \, L \, \sqrt{E T^3  \,
\frac{\alpha_s{\left(\mu^2\sqrt{\frac{E}{\lambda\mu^2}}\right)}}{\alpha_s{(\mu^2)}}\, \ln{\frac{E}{T}} } \ \, . 
\eeq
\item[(ii)] Due to the difference between the gluon \eq{LformmedQCD} and photon \eq{Lformmedofomega} formation lengths,
the LPM gluon energy spectrum reads, instead of \eq{LPMspectrum},
\be
\label{LPMspectrumQCD}
\omega \frac{dI_{\rm rad}}{d\omega}(L) \ \sim \ \alpha_s \frac{L}{\ell_f^{\rm med}(\omega)} \ \sim \ 
\alpha_s \sqrt{\frac{\omega_c}{\omega}}  \hskip 2cm (\omega > \lambda \mu^2) \,  ,
\ee
where $\omega_c$ is still given by \eq{omegac}, but depends now on the parameters $\mu$ and $\lambda$ 
appropriate to a non-Abelian plasma. 
As expected, the LPM suppression in QCD increases with increasing $\omega$, contrary to QED. Note that 
below $\lambda \mu^2$, the formation length is less than $\lambda$ and 
we are in the incoherent BH regime where the spectrum is given by \eq{spectrumlightQCD} times the number of 
rescatterings $L/\lambda$. 

Inserting the logarithmic factor discussed above, we obtain
\beq
\label{LPMspectrumQCD2}
\omega \frac{dI_{\rm rad}}{d\omega}(L) \ \sim \  \alpha_s^2 \, L \, \sqrt{\frac{T^3}{\omega} \ln{\frac{\mu}{\Lambda_{\rm QCD}}} } \ \, ,
\eeq 
where we assumed $\ln (\omega/T) \gg \ln (\mu/\Lambda_{\rm QCD})$ (otherwise the expression is 
more complicated, similarly to \eq{DElargeLCoulombQCD}). 
\ei

\vskip 5mm
\centerline{{\bf Intermediate region \boldmath{$\lambda \ll L \ll L^*$}}}
\vskip 3mm

In this region, the reasoning used in the case of an electron can be directly transposed to 
the QCD case of a light quark, yielding the result
\beq
\label{intermediateQCD}
\Delta E(\lambda \ll L \ll L^*) \ \sim \ \alpha_s E \, \ln{\left( \frac{L}{\lambda} \frac{\mu^2}{m^2} \right)} \, .
\eeq

As in the QED case, the result \eq{intermediateQCD} corresponds to the medium acting as a 
single effective scattering center. The medium size dependence appears only through the total
momentum transfer $q_\perp^2 \sim ({N} \ln{N})\,\mu^2$, where ${N} = L/\lambda$. 

Eq.~\eq{intermediateQCD} represents the so-called {\it factorization term} mentioned previously in 
Refs.~\cite{Baier:1994bd,Baier:1996vi,Baier:1996kr,Baier:1998kq}. Although the main goal of 
Ref.~\cite{Baier:1998kq} is to address the radiative loss of a quark produced in the plasma, it is mentioned there 
that in the case of an ``asymptotic quark'' entering the medium, the factorization term can be dropped 
when calculating the {\it induced} energy loss. As a consequence, the result of Ref.~\cite{Baier:1998kq} 
for $\Delta E$ in the region $\lambda \ll L \ll L^*$ is $\Delta E \propto \alpha_s \omega_c$, 
instead of \eq{intermediateQCD}. But in fact, for an asymptotic quark there is no distinction between 
the {induced} and {total} radiative energy loss,\footnote{This is because an ``asymptotic, on-shell quark'' 
does not radiate in the absence of the medium. This is in contrast with a quark produced 
in a hard subprocess, which radiates even in vacuum. See section \ref{sec5} for more discussion on this point.} and 
the factorization term \eq{intermediateQCD} 
should be kept. Despite the fact that the factorization term has a smooth (logarithmic) dependence on
$L$, it actually dominates over the term calculated in \cite{Baier:1998kq}. Indeed, for 
$L \ll L^*$, we have $\alpha_s \omega_c \ll \alpha_s E$. 

\begin{figure}[t]
\begin{center}
\includegraphics[width=8cm]{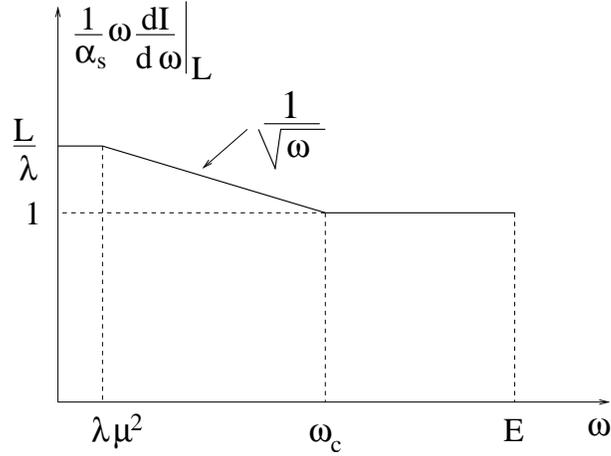}
\end{center}
\caption[*]{Gluon radiation spectrum by an asymptotic light quark crossing a hot QCD medium of size $L$, with $\lambda \ll L \ll L^*$ (double logarithmic plot).}
\label{fig-spec-asym}
\end{figure}

To illustrate this point, we represent in Fig.~\ref{fig-spec-asym} the gluon radiation 
spectrum in the region $\lambda \ll L \ll L^*$. For $L \ll L^*$ we have $\omega_c \sim L^2 \mu^2 /\lambda \ll E$, 
and the spectrum is given by \eq{LPMspectrumQCD} or \eq{LPMspectrumQCD2} in the interval $\lambda \mu^2 \ll \omega \ll \omega_c$. 
As mentioned above, for $\omega \ll \lambda \mu^2$, the spectrum is given by \eq{spectrumlightQCD} times $L/\lambda$. 
For $\omega \gg \omega_c$, the formation time of the gluon is larger than $L$, and 
the spectrum is the same as for a single effective scattering, \ie, it is obtained,
again, from \eq{spectrumlightQCD} by replacing 
$\mu^2 \to \mu^2 L/\lambda$. This flat part of the spectrum gives the dominant contribution to the light quark energy loss, 
see \eq{intermediateQCD}. 

Thus, the law $\Delta E(\lambda \ll L \ll L^*) \propto L^2$ \cite{Baier:1996kr,Zakharov:1997uu,Baier:1996sk,Baier:1998kq} 
for the induced energy loss of a light quark produced in the medium (we will review this case in Section \ref{sec5}), 
is not valid for an asymptotic quark. 
Combining \eq{BHQCDlight}, \eq{intermediateQCD} and \eq{DElargeLQCD}, we see that the average radiative loss 
of an asymptotic light quark is similar to that of an asymptotic electron crossing a QED plasma. It is represented in 
Fig.~\ref{DEvsLQED}.

\subsection{Heavy quark}
\label{secQCDheavyquarkasym}

We have found that the radiative energy loss of an asymptotic light quark crossing a QGP has the same parametric dependence 
(apart from a logarithmic factor when $L \gg L^*$) as for 
an electron crossing a hot $e^+e^-$ plasma, despite drastically different radiation 
spectra in these two cases. We will see now that the heavy quark radiative loss is 
{\it different} from the heavy muon loss in QED, and has a richer parametric dependence. 
The non-Abelian dynamics manifests itself more clearly for heavy than for light quark radiative energy loss. 

Consider first the BH regime, $L \ll \lambda$.

For heavy quarks, $M \gg \mu$, 
the gluon radiation intensity is suppressed, simi\-larly to heavy leptons in QED.
However, in the non-Abelian case, the suppression ({dead cone} effect) is not so strong since 
soft gluons are emitted in a cone broader than in QED, see \eq{gluonprobaapprox} and the related discussion. 
The estimate \eq{interpolate} for the QED radiation spectrum should be replaced by   
\be
\label{QCDspectrum}
\left. \omega \frac{dI_{\rm rad}}{d\omega} \right|_{1 \, \rm{scat.}} \sim \alpha_s \int d^2 \vec{\theta} \, \vec{J}_q^{\,\, 2} \sim 
\alpha_s \ln{\left( 1 + \frac{\theta_g^2}{\theta_M^2} \right)} 
 \sim \alpha_s \ln{\left( 1 + \frac{\mu^2}{M^2}\frac{E^2}{\omega^2}\right)} \, .
\ee
Integrating the latter spectrum over $\omega$ and multiplying by the scattering probability $L/\lambda$ we obtain 
\beq
\label{BHmassQCD}
\Delta E(L \ll \lambda) \ \sim \ \alpha_s E \, \frac{L}{\lambda} \frac{\mu}{M} \ \sim \ \frac{g^5 T^2 E}{M} \, L \, .
\eeq
Thus, for $L \ll \lambda$ the radiative loss of an asymptotic heavy quark is suppressed by a factor 
$\sim \mu/M$, instead of $\sim \mu^2/M^2$ in the QED case, see \eq{BHmassive}. 
A very important distinction of QCD compared to QED is that the spectrum
\eq{QCDspectrum} of a heavy quark is soft. Indeed, the characteristic energy of emitted gluons is
$\omega_{\rm char} \sim \mu E/M \ll E$, and the spectrum falls rapidly (as $1/\omega^2$) 
beyond this scale. (Instead, the photon radiation spectrum of a heavy charged particle
remains flat: mass effects bring about a uniform suppression $\sim \mu^2/M^2$ for all $\omega$.)  
The result \eq{BHmassQCD} arises from small gluon energies, $\omega \sim \omega_{\rm char}$. 
The typical gluon angles contributing to \eq{BHmassQCD} are of 
order $\theta^2_{\rm char} \sim \theta_g^2 \sim \mu^2/\omega^2_{\rm char} \sim \theta_M^2$. 
Hence the characteristic gluon formation length is 
\beq
\label{formQCDheavy}
\ell_f^{\rm QCD\ heavy} \ \sim \ \frac{1}{\omega_{\rm char} \, \theta^2_{\rm char}} \  \sim \ \frac{\omega_{\rm char}}{\mu^2}
\ \sim \ \frac E {\mu M} \, .
\eeq

Similarly to the Abelian case, 
the behavior of the radiative loss at larger lengths depends on the ordering of different length scales: 
the mean free path $\lambda$, the characteristic
formation length, or the scale $L^{**}$ where the suppression $\sim \mu/M$ disappears. The ordering of these scales
depends on the heavy quark mass. In the Abelian case, we had two mass regions \eq{M4}, \eq{rangeM}. In the non-Abelian case,
there are three distinct regions.

\vskip 3mm 
\centerline{{\bf A)} \ \boldmath{$M^2 \ll \alpha_s \sqrt{ET^3}$} } 
\vskip 3mm

\begin{figure}[t]
\begin{center}
\includegraphics[width=9cm]{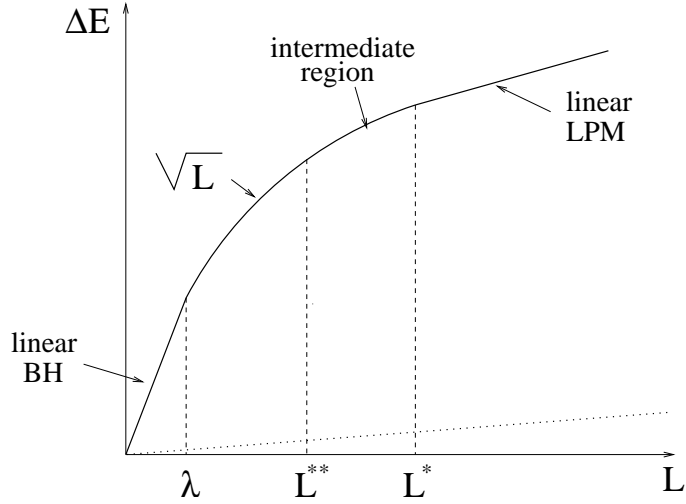}
\end{center}
\caption[*]{Radiative energy loss of an asymptotic heavy quark with $M^2 \ll \alpha_s \sqrt{ET^3}$.
({\it Dotted line:} collisional loss.)}
\label{DEvsLQCDM}
\end{figure}

In this case, the smallest length scale is $\lambda \sim 1/(\alpha_s T)$.
For $L \ll \lambda$, the law \eq{BHmassQCD} holds. When $L$ exceeds $\lambda$, the energy loss will be the same
as that induced by one effective scattering of momentum transfer $\mu_{\rm eff}^2 \sim (L/\lambda) \mu^2$, 
\be
\label{rootofL}
\Delta E(L) \sim \alpha_s E \, \frac {\mu \sqrt{L/\lambda}}{M} \ \sim\ \frac {\alpha_s^2 E \sqrt{LT^3}}{M} \ .
\ee
This law is valid in the region 
\be
\label{Ldoublestar}
\lambda \ll L \ll L^{**} = \lambda \frac{M^2}{\mu^2} \ \sim \ \frac {M^2}{\alpha_s^2 T^3} \ .
\ee
At the scale $L^{**}$, the suppression $\sim \mu_{\rm eff}/M$ disappears and the physics becomes the same
as for light quarks. When $L$ exceeds $L^{**}$, but is still less than
the in-medium formation length $L^*$ given by \eq{LstarQCDwithlog}, 
there is still one effective scattering, and we are in the intermediate region. 
When $L \gg L^*$, the number of effective scatterings grows as $L/L^*$, and the radiative loss
is given by the light quark estimate \eq{DElargeLCoulombQCD0}. A schematic plot of the energy dependence is drawn in 
Fig.~\ref{DEvsLQCDM}.  

\vskip 3mm 
\centerline{{\bf B)} \ \boldmath{$\alpha_s \sqrt{ET^3}  \ll M^2 \ll \alpha_s E^2$}} 
\vskip 3mm

When $M^2$ exceeds $\alpha_s \sqrt{ET^3}$, the scale $L^*$ becomes smaller than the scale
$L^{**}$ where the dead cone suppression disappears. Before proceeding further, note that the estimate 
$L^* \sim \sqrt{\lambda E/\mu^2}$ for the in-medium formation length does not hold anymore in this case, and $L^*$ should be 
replaced by another scale $\tilde{L}^*$. 

Indeed, recall that the in-medium formation length is defined by the condition that it coincides with 
 the vacuum formation length for an effective scattering of transfer 
$\mu_{\rm eff}^2 \sim (L/\lambda)\, \mu^2$. For a light quark this condition reads
\be
L \ \sim \ \frac{E}{\mu_{\rm eff}^2} \ \ \Rightarrow \ \ L^* \ \sim \ \sqrt{\frac{\lambda E}{\mu^2}}  \, .
\ee
For a massive quark the vacuum formation length is given by \eq{formQCDheavy} and we obtain
\be
\label{Lstartilde}
L \ \sim \ \frac{E}{\mu_{\rm eff} M} \ \ \Rightarrow \ \ \tilde{L}^* \sim \left( \frac{\lambda E^2}{\mu^2 M^2} \right)^{1/3} 
\sim \left( \frac{L^{*\, 4}}{L^{**}}  \right)^{1/3} 
\sim \frac{1}{T} \left( \frac{E}{\alpha_s M} \right)^{2/3} \, .
\ee 
Note that the condition $\alpha_s \sqrt{ET^3}  \ll M^2 \ll \alpha_s E^2$ defining the mass interval under consideration
is equivalent to $\lambda \ll \tilde{L}^* \ll L^*$. 

For $L \gg \tilde{L}^*$, the size of an effective scattering center 
responsible for the emission of one gluon is $\sim \tilde{L}^*$, and the dead cone suppression factor $\sim \mu_{\rm eff}/M$ 
stays frozen at the value
\be
\label{supfactor}
\left( \frac {\mu_{\rm eff}}M \right)_{\rm max} \sim \frac{\mu}{M} \sqrt{\frac{\tilde{L}^*}{\lambda}} 
\sim \left( \frac {\alpha_s \sqrt{ET^3} }{M^2} \right)^{2/3} \, .
\ee
The energy loss displays in this region a linear dependence on $L$ with the slope \cite{Marquet}
\be 
\label{DELggLstar}
\Delta E(L \gg \tilde{L}^*) \sim \alpha_s E  \, \frac{L}{\tilde{L}^*} \, \left( \frac{\mu_{\rm eff}}{M} \right)_{\rm max} 
\sim \alpha_s L \left( \frac{\mu^2 E}{\lambda M}\right)^{2/3}
\sim \alpha_s^{7/3} \, T^2 \, L \left( \frac EM \right)^{2/3} \, . \nn \\
\ee

The schematic plot of $\Delta E(L)$ looks as in Fig.~\ref{DEvsLQED}, but with two qualitative distinctions:
{(i)}  the scale ${L}^*$ is replaced by $\tilde{L}^*$; 
{(ii)} in the region between $\lambda$ and $\tilde{L}^*$, $\Delta E(L) \propto \sqrt{L}$ (as in \eq{rootofL})
instead of $\Delta E(L) \propto \ln{L}$ as in \eq{intermediate}.

\begin{figure}[t]
\begin{center}
\includegraphics[width=8cm]{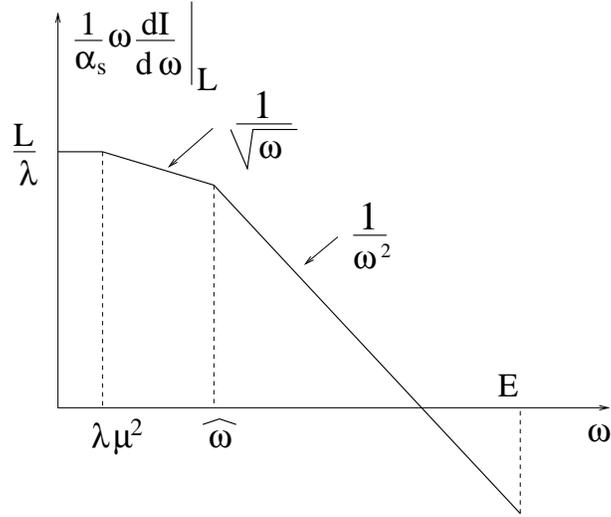}
\end{center}
\caption[*]{Gluon radiation spectrum of a heavy quark produced in a hot QGP for 
$\alpha_s \sqrt{ET^3} \ll M^2 \ll \alpha_s E^2$ and for $L \gg \tilde{L}^*$.}
\label{fig-spec-asym-heavy}
\end{figure}

Let us discuss now the energy spectrum of emitted gluons. Consider $L \gg  \tilde{L}^*$. We expect the spectrum to differ from the light quark case 
(see \eq{LPMspectrumQCD} and Fig.~\ref{fig-spec-asym}) as soon as the typical angle 
$\theta_{\rm typ}$ contributing to \eq{LPMspectrumQCD} becomes smaller than 
the parameter $\theta_M = M/E$. The typical angles associated with \eq{LPMspectrumQCD} read
\be
\label{typangle1sec4}
\theta_{\rm typ}^2 \ \sim \ \frac{1}{\omega \ell_f^{\rm med}(\omega)} \ \sim \ \sqrt{\frac{\mu^2/\lambda}{\omega^3}} \, ,
\ee
where we used ~\eq{LformmedQCD}. The condition $\theta_{\rm typ} < \theta_M$ is thus equivalent to 
\be
\label{omegahat}
\sqrt{\frac{\mu^2/\lambda}{\omega^3}}  < \frac{M^2}{E^2} \Leftrightarrow \omega  > \widehat{\omega} \ \equiv \ 
\left(\frac{\mu^2 E^4}{\lambda M^4} \right)^{1/3} \ \sim
\ T \left( \frac {\alpha_s E^2}{M^2} \right)^{2/3} \, .
\ee
Note that $L \gg \tilde{L}^*$ is equivalent to $\widehat{\omega} \ll \omega_c$. 
When $\omega  > \widehat{\omega}$ the spectrum is given by 
\be 
\label{diffspec1heavy} 
\omega \frac{dI}{d\omega}   \ \sim \ \alpha_s \frac{L}{\ell_f^{\rm med}} \cdot \frac{\theta_{\rm typ}^2}{\theta_M^2} 
\ \sim \ \alpha_s L \, \frac{\mu^2 E^2}{\lambda M^2} \cdot \frac{1}{\omega^2}
\ \sim \ \alpha_s \frac{\sqrt{\omega_c \, \widehat{\omega}^3}}{\omega^2}  \ \ \ 
(\omega > \widehat{\omega}) \ \ \, .
\ee
For $\omega < \widehat{\omega}$ the spectrum is given by \eq{LPMspectrumQCD}. The overall spectrum is depicted in Fig.~\ref{fig-spec-asym-heavy}.  
Note that the behavior $\propto \omega^{-2}$ that we find for $\omega > \widehat{\omega}$ differs from 
the behavior $\propto \omega^{-7/2}$ obtained in Ref.~\cite{Dokshitzer:2001zm}. Integrating the spectrum we recover the
average loss \eq{DELggLstar}, the latter being dominated by $\omega \sim  \widehat{\omega}$. Noting that 
\be
\label{omhat}
\hat{\omega} \ \sim \  E \, \left( \frac {\mu_{\rm eff}}M \right)_{\rm max} 
\ \sim \ \frac{\mu E}{M} \sqrt{\frac{\tilde{L}^*}{\lambda}} \, ,
\ee
the average loss \eq{DELggLstar} can be rewritten in a compact form,
\be 
\label{DELggLstar-trans}
\Delta E(L \gg \tilde{L}^*) \sim \alpha_s \, \hat{\omega} \, \frac{L}{\tilde{L}^*} \, ,
\ee
to be compared with the light quark estimate \eq{DElargeLQCD}.

\vskip 3mm 
\centerline{{\bf C)} \ \boldmath{$M^2 \gg \alpha_s E^2$}} 
\vskip 3mm

Finally, for very large masses, the LPM effect does not play any role 
and the BH linear law \eq{BHmassQCD} holds for {\it all} lengths. Indeed, when $M^2$ 
exceeds $\alpha_s E^2$, the formation length
$\tilde{L}^*$ becomes smaller than $\lambda$ and gluons are emitted in individual incoherent scatterings. 
In this case, radiative losses are suppressed compared to collisional ones, 
which can be checked by comparing the slope in \eq{BHmassQCD} with the slope $\sim \alpha_s^2 T^2$ for collisional loss. 
 Because of this, we stress here that 
the radiative loss \eq{BHmassQCD} might be difficult or impossible to observe (even in a thought experiment). 
See section \ref{sec6} for more discussion of this point. 

As was mentioned above, the spectrum of emitted gluons is soft in this case and given by \eq{QCDspectrum}
multiplied by the number of scatterings $L/\lambda$. The spectrum starts to fall down as $\sim \omega^{-2}$  at the scale 
$\mu E/M $ rather than at $\hat{\omega}$, as it did in the intermediate mass region.

The linear energy dependence $\Delta E(L) \propto E$ may suggest the description of heavy quark radiative 
losses in terms of radiation length, as it is usually done for ultrarelativistic electrons in
usual matter. But it is not convenient here for two reasons: (i) In contrast to ultrarelativistic
electrons, here collisional losses dominate over radiative ones; (ii) The spectrum of emitted gluons
is soft and energy loss fluctuations are much smaller than e.g. for electrons forming atmospheric showers.

\section{Radiative loss of a particle produced in a plasma}
\label{sec5}

Here we consider the case of a charged particle produced inside 
a plasma. This situation is much more natural for QCD, where an energetic parton can be produced 
in a hard partonic subprocess
inside the hot medium formed in heavy ion collisions. We however start by considering the 
less natural but simpler QED case of an electric charge produced in a QED plasma. We will then study the QCD case. 

\subsection{Hot QED plasma}
\label{sec51}

\subsubsection{Electron}

When an energetic charged particle is created in a hard process, it radiates brems\-strahlung
photons. This radiation occurs even when 
the particle is created in vacuum, and should be distinguished from the medium-induced radiative loss.

In this section, we will consider the situation of a fast and light charged particle produced in 
a hot QED plasma (think for instance of deep inelastic scattering off a QED plasma, or of direct lepton production in a QGP), 
and focus on its {\it medium-induced} radiative loss. This loss is 
associated with the components of the particle's radiation field coat which had the chance to be formed 
within the distance $L$ travelled by the newborn particle in the plasma, such that those components can be
released as emitted photons during subsequent rescatterings. In other words, only the photons 
whose formation length \eq{Lform} does not exceed $L$ contribute to the induced radiative loss, 
\beq
\label{prescription}
\ell_f(\omega, \theta) \sim \frac{1}{\omega \theta^2} \lsim L \, .
\eeq

In the case of an asymptotic particle studied in sections \ref{sec3} and \ref{sec4}, we found some contributions to 
the energy loss arising from $\ell_f \gg L$. In particular, in the 
BH regime $L \ll \lambda$, the electron energy loss 
\eq{BHlossQED} arises from photon formation lengths $\ell_f \gg E/\mu^2 \gg L$, see \eq{timerange2}. 
Due to the prescription \eq{prescription}, this contribution should now be disregarded. 
We should only count the photons whose formation length does not exceed $L$, which brings 
an additional suppression in {medium-induced} radiative losses. Thus, when a light particle is created inside the plasma,  
there is no BH regime whatsoever.\footnote{For heavy particles, this is not so, see the discussion below.} 
We will shortly see into what kind of behavior it is transformed.

On the other hand, the result \eq{DElargeL} (or, more accurately, \eq{DElargeLCoulomb}) 
for the radiative energy loss in a thick medium, $L \gg L^*$, 
should also be valid  when the particle is created in the plasma rather than in the remote past. Indeed, the result 
\eq{DElargeL} arises from photon formation lengths $\ell_f^{\rm med}(\omega) \sim  \ell_f^{\rm med}(E) \sim L^* \ll L$, thus 
satisfying the prescription \eq{prescription}. 
When $L \gg L^*$, the particle forgets the conditions of its birth. 

\begin{figure}[t]
\begin{center}
\includegraphics[width=10cm]{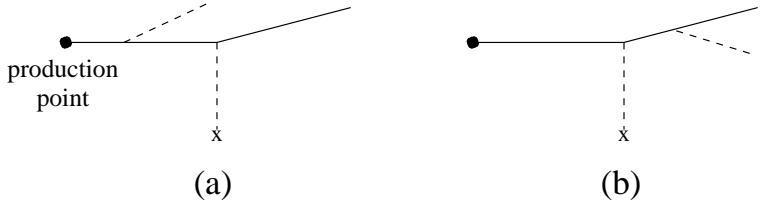}
\end{center}
\caption{Photon radiation of an electron produced and scattered in a QED plasma.}
\label{1scat}
\end{figure}

Consider first the region $L \ll \lambda$. 
In this case, the particle undergoes one scattering with probability $\sim L/\lambda$. The photon emission amplitude is given by the sum
of the two diagrams in Fig.~\ref{1scat}. For small photon frequencies $\omega \ll E$ and small scattering and emission
angles, it can be evaluated as ${\cal M} \propto e\,\vec{\varepsilon} \cdot \vec{J}(L)$ with 
\be
\label{JL}
\vec{J}(L) =  \frac{\vec{\theta'}}{\theta'^2} - \frac{\vec{\theta}}{\theta^2}
\left[ 1-  e^{- i\omega L \theta^2/2 } \right]\ .
\ee
Here $L$ is the distance travelled by the particle between its production and scattering. We assumed that the particle is massless. 
The term $\propto \vec{\theta}$ is the contribution of the graph in Fig.~\ref{1scat}a and the term 
$\propto \vec{\theta'}$ is the contribution of the graph in Fig.~\ref{1scat}b.

The result \eq{JL} for the amplitude is rigorously derived in Appendix B, but its structure looks rather natural
in the context of the above heuristic reasoning. When $L$ is large compared to the formation length
(\ref{prescription}), one can drop the rapidly oscillating factor $\sim \exp\{-i\omega L \theta^2/2\}$
and the current $\vec{J}(L)$ is reduced to the expression \eq{Je} for an asymptotic particle. On the other hand,
for a photon formation length larger than $L$, the contribution of the graph in Fig.~\ref{1scat}a is suppressed and 
only the second graph remains, corresponding to photon emission from the final electron line --  as in the absence of
rescattering. Since we are interested in the medium-induced radiation intensity, we should subtract the latter contribution
(at the cross section level), giving
\bea
\label{omdPom} 
\left. \omega \frac {dI}{d\omega} \right|_{\rm induced} \ 
&\sim&  \ \alpha \frac{L}{\lambda} \left \langle \int d^2 \vec{\theta} 
\left( \left|\vec{J}(L)\right|^2 - \left|\vec{J}(0)\right|^2 \right)  \right \rangle  \nonumber \\
&=& 2 \alpha \frac{L}{\lambda}  {\rm Re} \left \langle \int d^2 \vec{\theta} \, \frac{\vec{\theta}}{\theta^2} 
\left(  \frac{\vec{\theta}}{\theta^2} - \frac{\vec{\theta'}}{\theta'^2} \right) \left(1- e^{-i\omega L \theta^2/2} \right) 
\right \rangle  \, .
\ee
The averaring is done over the transverse momentum $\vec{q}_\perp$ exchanged in the scattering 
(we remind that $\vec{\theta}' \equiv \vec{\theta} 
- \vec{\theta}_s = \vec{\theta} - \vec{q}_\perp/E$). Let us average first over the azimuthal directions of $\vec{\theta}_s$. 
Using the identity
\be
\label{azimint}
\int \frac{d\phi}{2\pi} 
\left( \frac{\vec{\theta}}{\theta^2} - \frac{\vec{\theta} - \vec{\theta}_s }{(\vec{\theta} - \vec{\theta}_s)^2} \right) 
= \frac{\vec{\theta}}{\theta^2} \, \Theta(\theta^2_s - \theta^2)  
\ee
we obtain 
\be
\label{omdPom1} 
\left. \omega \frac {dI}{d\omega} \right|_{\rm induced} \ \sim \ \alpha \frac L\lambda \left \langle 
\int \frac {d \theta^2}{\theta^2}   \left(1- \cos(\omega L \theta^2/2) \right) \Theta(\theta^2_s - \theta^2)  \right \rangle \ .
\ee

We should now average over $\theta^2_s$ with the normalized probability $P(\theta_s^2)$, 
obtained by squaring the (momentum space) screened Coulomb (Yukawa) potential, see \eq{Couldiff},
\be
\label{Pthets}
P(\theta_s^2) \ =\ \frac {\mu^2/E^2}{(\theta^2_s + \mu^2/E^2)^2} \ .
\ee
The result is
\be
\label{omdPom2} 
\left. \omega \frac {dI}{d\omega} \right|_{\rm induced} \ 
\sim \ \alpha \frac {L \mu^2}{\lambda E^2} 
\int_0^\infty \frac {d \theta^2}{\theta^2(\theta^2 + \mu^2/E^2)} \left(1- \cos(\omega L \theta^2/2) \right)   \ .
\ee
When $\omega L \mu^2 /E^2 \ll 1$, which is true in the region $L \ll \lambda$ 
we are now considering, the integral
is saturated by  
the emission angles\footnote{The last inequality 
in \eq{inequalities} follows from $E \gg \lambda \mu^2 \sim T$. We also assume $EL \gg 1$, \ie, 
even though $L$ is  smaller than $\lambda$, it is still larger than the wavelength of the energetic particle.}
\be
\label{inequalities}
\theta^2 \gsim \frac 1{\omega L} > \frac 1{EL} \gg \frac 1{E\lambda}  \gg \frac {\mu^2}{E^2} \, .
\ee
The integral in \eq{omdPom2} is of order $\sim \omega L$. Thus, the spectrum is evaluated as 
\be
\label{omdPom3} 
\left. \omega \frac {dI}{d\omega} \right|_{\rm induced} \ 
\sim \ \alpha \frac{L^2 \mu^2}{\lambda} \cdot \frac{\omega}{E^2} \ \sim \  \alpha \frac {\omega \, \omega_c}{E^2} \, ,
\ee
 and the energy loss reads
\be
\label{DEmedLlllam}
\Delta E(L \ll \lambda) \ \sim \ \alpha \, \omega_c 
\sim \alpha^3 T^3 L^2\, . 
\ee

It is interesting to mention the analogy with the discussion of 
the energy loss of an {\it asymptotic} and {\it massive} particle in Section \ref{asymQEDheavy}. The effective cutoff 
$\theta^2 > 1/(EL)$ which arises here is similar to the ``dead cone'' cutoff $\theta^2 > M^2/E^2$ in the integral 
\eq{angularint}, with the parameter $\sqrt{E/L}$ playing the role of mass. 
One can actually obtain the estimate \eq{DEmedLlllam} by  replacing $M^2 \to E/L $ in \eq{BHmassive2}.

We see that $\Delta E(L)$ has a quadratic rather than linear dependence at small $L$. The reason for that
is quite transparent, as we qualitatively explained before. When a hard electron is just born, it has not grown its radiation
field coat yet and is not able to radiate. Roughly speaking, the {\it capacity} $dE/dx$ to radiate vanishes at $L=0$ 
and then grows linearly with $L$. The integration
\be 
\Delta E = \int_0^L dx \, \frac {dE}{dx}  
\ee 
yields another factor $\sim L$. 

\begin{figure}[t]
\begin{center}
\includegraphics[width=8cm]{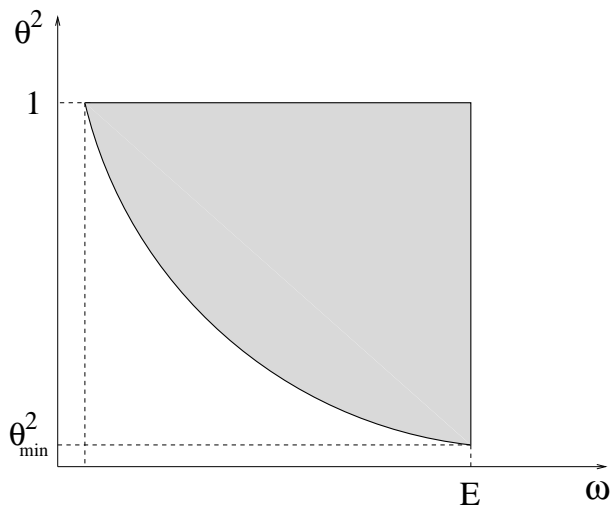}
\end{center}
\caption{Photon energies and emission angles satisfying $\ell_f(\omega, \theta) < L$.}
\label{domain}
\end{figure}
 
It is instructive to discuss a more heuristic derivation of the estimate \eq{DEmedLlllam} which does not 
use the exact expression \eq{JL} of the radiation amplitude, but simply consists in integrating the
spectrum \eq{J2} (derived for an asymptotic particle) over $\omega$ and $\theta$ with the 
constraint $1/(\omega \theta^2) < L$.
The corresponding integration domain  is depicted in Fig.~\ref{domain}. 
Since $\theta^2 \gg \mu^2/E^2$, the angular spectrum can be approximated by $\mu^2/(E^2 \theta^4)$.
The energy loss then reads
\beq
\label{doubleint}
\Delta E(L \ll \lambda) \ \sim \ \alpha  \frac{L}{\lambda} \int_0^E d\omega \int_{1/(\omega L)}^1 d\theta^2 \frac{\mu^2}{E^2\theta^4} 
\ \sim \ \alpha  \frac{L\mu^2}{\lambda E^2} \int_0^E d\omega \, \omega L  \ \sim \ \alpha \, \omega_c \  , 
\eeq
arising from the typical values $\theta^2 \sim \theta^2_{\rm min} \sim 1/(E L)$ and $\omega \sim E$. 
This coincides with (\ref{DEmedLlllam}). 

We want to emphasize, however, that although the latter argument is very simple, physically transparent,
and gives the correct result, it is heuristic and does not reproduce a certain
 subtle dynamical feature which is displayed in the 
accurate derivation based on \eq{JL}.\footnote{In other words, the diagrams in Fig. \ref{1scat} describing 
photon emission during a single scattering of a particle produced in a plasma dictate a slightly different 
dynamics, compared to the single scattering diagrams for an asymptotic 
particle in Fig. \ref{QEDemission} supplemented by the constraint
\eq{prescription}. On the other hand, the physical arguments that derive the multiple scattering
dynamics on the basis of the single scattering diagrams and the notion of formation length seem
to work in all cases.}  
Indeed, the argument leading to the heuristic estimate \eq{doubleint} implicitly assumes that the 
momentum transfer is fixed at the value $q_\perp^2 = \mu^2$, \ie, it refers to an 
hypothetical model where the Coulomb probability density \eq{Pthets} is replaced by 
$P(\theta_s^2) = \delta(\theta_s^2 - \mu^2/E^2)$ (corresponding to a scattering potential $V(r) \sim J_0(\mu r)$). 
On the other hand, substituting $\theta_s^2 \to \mu^2/E^2$, the expression \eq{omdPom1} which follows from a diagrammatic 
analysis would yield 
\be
 \label{omdPom4} 
 \left. \omega \frac {dI}{d\omega} \right|_{q_\perp^2 = \mu^2} \ 
  \sim  \ \alpha \frac {\omega^2 L^3 \mu^4}{\lambda E^4}  \, ,
\ee  
which leads to an energy loss suppressed compared to the expression \eq{doubleint}. 
Thus, the estimate \eq{doubleint} fails for the hypothetical model where $P(\theta_s^2) = \delta(\theta_s^2 - \mu^2/E^2)$. 

For the more realistic Yukawa potential, the results based on \eq{omdPom1} and \eq{doubleint} are the same, but the integrals 
are saturated in different kinematical regions. With the heuristic prescription (\ref{doubleint}), the characteristic radiation angle 
$\theta^2_{\rm rad} \sim 1/(\omega L)$ is much larger than the characteristic scattering angle 
$\theta_{\rm scat}^2 \sim \mu^2/E^2$. In the more accurate formula (\ref{omdPom2}), they are both large (see \eq{azimint}),
\be
\label{angles} 
\theta^2_{\rm rad} \sim \theta^2_{\rm scat} \sim \frac 1{\omega L} \ .
\ee     
A distinctive feature of the Yukawa scattering potential is that it allows very large transfers compared to 
the typical transfer $\sim \mu$. Thus, the constraints $\theta_s^2 \geq \theta^2$ (see \eq{azimint}) and 
$\theta^2 \sim 1/(\omega L) \gg \mu^2/E^2$ (see \eq{inequalities}) can be  realized simultaneously.

Let us now consider a larger medium, $L \gg \lambda$, and focus on the Yukawa scattering potential. 
The electron is now scattered ${N}$ times, with ${N} \sim L/\lambda$. If $L \ll L^*$, only one
photon is emitted.
The amplitude of photon emission in the multiple scattering process is evaluated 
accurately in Appendix B. It happens that the main contribution to the radiation spectrum 
arises from the region where {\it one} of the scattering
angles is large as in \eq{angles}, while all other scattering angles are relatively small $\sim \mu/E$. In other words, one
of the ${N}$ scattering momenta is $\sim \sqrt{E/L}$ (assuming $\omega \sim E$) and is much larger than the
characteristic momentum transfer $\mu_{\rm eff} \sim \mu \sqrt{N}$ due to all other 
scatterings.\footnote{The condition $\sqrt{E/L} \gg \mu \sqrt{L/\lambda}$ is equivalent to $L \ll L^*$.} 
Any of the ${N}$ scatterings can be distinguished in this way, which gives the factor ${N} \sim L/\lambda$ in the 
radiation spectrum. 
The latter is still given by the integral \eq{omdPom2}, with the factor $L/\lambda$ interpreted as the 
characteristic number of scatterings instead of the scattering probability. 
The estimate for the energy loss is still given by \eq{DEmedLlllam}, which thus holds in the extended range 
$L \ll L^*$,
\be
\label{DEmedLlllam-extended}
\Delta E(L \ll L^*) \ \sim \ \alpha \, \omega_c \sim \alpha^3 T^3 L^2\, . 
\ee

\begin{figure}[t]
\begin{center}
\includegraphics[width=7cm]{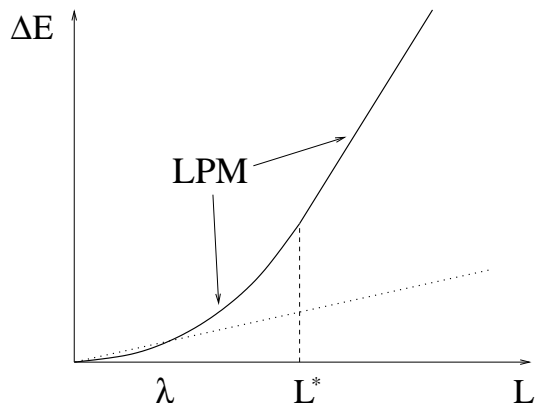}
\end{center}
\caption[*]{Radiative energy loss of an electron produced in a QED plasma. The quadratic dependence \eq{DEmedLlllam-extended} 
at $L \ll L^*$ is replaced  by the linear dependence \eq{DElargeLCoulomb} at $L \gg L^*$. ({\it Dotted line:} collisional loss.)}
\label{DEvsLinside}
\end{figure}

Finally, for $L \gg L^*$, the restriction \eq{prescription} is not effective, as we already mentioned. 
We obtain the same dependence  
\eq{DElargeLCoulomb} as for a particle coming from infinity. 
The results are represented in Fig.~\ref{DEvsLinside}. For large lengths, $\Delta E$ is linear in $L$, 
with $dE/dx \propto \sqrt{E}$ which is familiar from the previous sections. For $L$ smaller than $L^*$, 
$\Delta E(L) $ displays the quadratic dependence \eq{DEmedLlllam-extended}.

\subsubsection{Muon}

Consider now the radiative losses of a massive particle created in the plasma. As was also 
the case for a particle coming from infinity, the behavior of $\Delta E(L)$ is different in the regions 
\eq{M4} and \eq{rangeM}. 

\vskip 3mm
\centerline{{\bf A)} \boldmath{$M^2  \ll \alpha \sqrt{ET^3}$}}
\vskip 3mm

We will show that there is no difference with the case of light particles in this region, and the 
behavior of $\Delta E(L)$ is the same as in Fig.~\ref{DEvsLinside}. Recall that for a massive particle
the vacuum formation length is given by \eq{lfheavy}. 

When $L \ll L^*$, we showed previously that the induced radiative loss of a light QED particle arises
from $\omega \sim E$ and $\theta^2 \sim 1/(\omega L) \sim 1/(E L)$. Thus, as long as 
\be
\label{ELM2} 
1/(E L) \gg \theta_M^2 = M^2/E^2\ ,
\ee
the result \eq{DEmedLlllam} for a light particle will also apply to a heavy 
particle. Intuitively, this happens when $M$ is small compared to 
the effective ``mass'' $\sim \sqrt{E/L}$ of the light particle (see our discussion below \eq{DEmedLlllam}).
To see that the condition \eq{ELM2} is satisfied, note that $M^2 \ll \alpha \sqrt{ET^3}$ implies  
$L^* \ll E/M^2$. When 
$L \ll L^*$,  we have $M^2 \ll E/L^* \ll E/L$. Thus, the result \eq{DEmedLlllam-extended} 
is valid also for a moderately massive particle. 

The effects due to a nonzero mass are irrelevant in the range $M^2  \ll \alpha \sqrt{ET^3}$ also for large lengths, $L \gg L^*$. 
Indeed, when $L \gg L^*$, the electron radiative loss arises from 
$\omega \sim E$ and photon formation lengths  $1/(\omega \theta^2)$ of order $ L^*$, implying
$\theta^2 \sim  1/(E L^*) \gg \theta_M^2$, where we used again $L^* \ll E/M^2$. 
Thus, the light particle result \eq{DElargeLCoulomb} is valid in the region $M^2  \ll \alpha \sqrt{ET^3}$. 

An equivalent way to understand this is as follows. The length above which the mass can be neglected is determined 
by $\mu_{\rm eff}^2 \sim (L/\lambda) \, \mu^2 \sim M^2$, \ie, by the scale $L^{**}$, see \eq{Ldoublestar}. 
For $M^2  \ll \alpha \sqrt{ET^3}$, we have $L^{**} \ll L^*$, and mass effects can {\it a fortiori} be neglected 
when $L \gg L^*$.

\vskip 3mm
\centerline{{\bf B)} \boldmath{$M^2 \gg \alpha \sqrt{ET^3}$}}
\vskip 3mm

In this region, the characteristic vacuum formation length $\ell_f \sim E/M^2$ is 
smaller than the scale $L^*$ and shows up first. The quadratic law \eq{DEmedLlllam-extended} 
extends only up to the scale $L \sim E/M^2$, after which it is 
replaced by the law \eq{BHmassive}, the same as for asymptotic and heavy QED particles.

\subsection{Quark gluon plasma}

\subsubsection{Light parton}

An energetic, high $p_\perp$ light parton produced in a proton-proton collision 
can be ``observed'' via the jet of hadrons 
that it produces. Those hadrons are the products of the parton's bremsstrahlung induced 
by its sudden acceleration at the moment of its creation. In the process of building its asymptotic ($t \to  +\infty$)
field coat, the initially ``bare'' parton radiates quasi-collinear DGLAP gluons. 
The parton energy at the time of its production can in principle be determined 
by measuring the total jet energy. 

If the parton is produced in a finite size plasma, the parton energy loss due to its 
rescatterings in the 
hot medium can affect the structure of the hadron jet. In particular, such medium-induced 
energy loss leads to the suppression of large $p_\perp$ hadrons (jet-quenching) 
in ultrarelativisitic heavy-ion collisions \cite{quenching-review}, 
when compared to proton-proton collisions. Also, medium-induced gluon radiation 
will enhance the hadron multiplicity within the jet. 
Here we want to derive the medium-induced radiative energy loss of a light quark. 
As explained in section \ref{sec51}, the formation length of the medium-induced gluon radiation 
must be smaller than $L$, see \eq{prescription}. 

Consider first the region $L \ll \lambda$. 
The physics is the same as in the Abelian case with the only difference that the characteristic
Abelian radiation cone width $\sim \mu/E$ should be replaced by the non-Abelian one
$\sim \mu/\omega$ (see section \ref{sec41} and our comments below \eq{gunionbertsch}). 
Thus, the QED expression \eq{omdPom2} is transformed into
\be
\label{omdPom2QCD} 
\left. \omega \frac{dI}{d\omega} \right|_{\rm induced}^{\rm QCD} \ 
\sim \ \alpha_s \frac {L \mu^2}{\lambda \omega^2} 
\int_0^\infty \frac {d \theta^2}{\theta^2(\theta^2 + \mu^2/\omega^2)} \left(1- \cos(\omega L \theta^2/2) \right)  \, .
\ee
The spectrum \eq{omdPom2QCD} has two different forms depending on whether the gluon vacuum formation length
$\ell_f(\omega) \sim \omega/\mu^2$ is smaller or larger than $L$,
\bea
\label{domain1} 
\left. \omega \frac{dI}{d\omega} \right|_{\rm induced}^{\rm QCD} \ 
&\sim& \ \alpha_s \frac{L}{\lambda} \, \ln{\frac{L \mu^2}{\omega}}  \ \ \ \ (\omega < L \mu^2) \   , \\
\left. \omega \frac{dI}{d\omega} \right|_{\rm induced}^{\rm QCD} \ &\sim& \ \alpha_s \frac{\omega_c}{\omega}  \ \ \ \ \ \ \ \ 
(L\mu^2 < \omega  < E) \  .
\label{omdPom2QCDbis} 
\eea
The contributions of these regions to the induced energy loss are 
\bea
\label{DEmedLlllamQCD0}
\Delta E_{\rm QCD, 1}(L \ll \lambda) \ &\sim& \ \alpha_s \, \omega_c \, , \\
\Delta E_{\rm QCD, 2}(L \ll \lambda) \ &\sim& \ \alpha_s \, \omega_c \ln \frac{E}{L \mu^2}   \ .
\label{DEmedLlllamQCD}
\eea
The second contribution is logarithmically enhanced.

The expression \eq{DEmedLlllamQCD} for the light parton radiative loss in the BH region was obtained in 
Ref.~\cite{Zakharov:2000iz}. Its origin is the same as 
in QED. Namely, the energy loss arises from radiation angles larger than the {\it typical} scattering 
angle, $\theta^2 \sim 1/(\omega L) \gg \mu^2/E^2$ in QED, and $\theta^2 \sim 1/(\omega L) \gg \mu^2/\omega^2$ 
in QCD. Thus, as was also the case for the QED expression \eq{DEmedLlllam}, the QCD loss \eq{DEmedLlllamQCD}
is specific to a Yukawa scattering potential. 

In the region $\lambda \ll L \ll L^*$ and for small enough frequencies, 
the medium effects come into play and the formation length is given by \eq{LformmedQCD}. 
One can distinguish two (or two and a half, if you will) regions in the spectrum. 

\begin{itemize}
\item[(i)]
If $\ell_f^{\rm med}(\omega) \ll L$, implying $\omega \ll \omega_c$, the spectrum is the same as for the asymptotic particle 
in the LPM regime, see \eq{LPMspectrumQCD} and Fig.~\ref{fig-spec-asym},  
\bea
\left. \omega \frac {dI}{d\omega} \right|_{\rm induced} &\sim& \alpha_s \frac L\lambda \ \ \ \ \ \ \ \ (\omega < \lambda \mu^2) \nn \\ 
\left. \omega \frac {dI}{d\omega} \right|_{\rm induced}  &\sim& \alpha_s \sqrt{\frac {\omega_c}{\omega}} \ \ \ \ 
(\lambda \mu^2 < \omega < \omega_c) \ \ \, .
\label{diffspec1}
\eea
The latter arises from typical emission angles as given in \eq{typangle1sec4},
\be
\label{typangle1}
\theta^2 \ \sim \ \frac{1}{\omega \ell_f^{\rm med}(\omega)} \ \sim \  \sqrt{\frac{\mu^2/\lambda}{\omega^3}} \ \gg \ \frac{1}{\omega L} \, .
\ee
The contribution of the region $\omega < \lambda \mu^2$ to the energy loss is small. The region $\lambda \mu^2 < \omega < \omega_c$
yields
\beq
\label{DEmedintermQCD1}
\Delta E_1(\lambda  \ll L \ll L^*) \ \sim \ \alpha_s \, \omega_c \, . 
\eeq
This contribution is specific to QCD (in QED, $\ell_f^{\rm med}(\omega)$ exceeds $L$ if the latter is smaller than $L^*$)
and was identified in Refs.~\cite{Baier:1996kr,Zakharov:1997uu}.\footnote{In \eq{DEmedintermQCD1} we have not displayed 
the logarithmic factor $\sim \ln (L/\lambda)$ which might be associated with this contribution. This factor arises in 
Refs.~\cite{Baier:1996kr,Zakharov:1997uu} for a Coulomb scattering potential. 
But we cannot say with certainty that its presence is a model-independent statement.} 

\item[(ii)] When $\omega > \omega_c$ (note that  $\omega_c \ll E$ as long as $L \ll L^*$),
 $\ell_f^{\rm med}(\omega) > L$. 
In this case, the radiation spectrum is the same as for a single {\it effective} scattering of typical 
scattering angle ${N} \mu^2/\omega^2$, with ${N} \sim L/\lambda$. The spectrum is then estimated from 
\eq{omdPom2QCD}. The estimate \eq{omdPom2QCDbis} follows, but is now valid when $1/(\omega L) \gg {N} \mu^2/\omega^2$ 
(which exactly coincides with the condition $\omega \gg \omega_c$),  
\be
\label{diffspec2} 
\left. \omega \frac{dI}{d\omega} \right|_{\rm induced} \ \sim \ \alpha_s \frac{\omega_c}{\omega} \hskip 1.2cm 
(\omega_c < \omega < E) \ \ \, .
\ee
This part of the spectrum arises from the typical angles
\be
\label{typangle2}
\theta^2 \ \sim \ \frac{1}{\omega L} \, 
\ee
(meaning that the formation length is of order $L$), 
and contributes to the energy loss as 
\beq
\label{DEmedintermQCD2}
\Delta E_2(\lambda  \ll L \ll L^*) \ \sim \ \alpha_s \, \omega_c \ln \frac{E}{\omega_c} \ \sim \ \alpha_s \, \omega_c \ln \frac{L^*}{L}\ . 
\eeq
This contribution was discussed in Ref.~\cite{Zakharov:2000iz}. It is the QCD analog of the QED expression 
\eq{DEmedLlllam-extended} and depends
on the presence of a long high-momentum tail in the Coulomb scattering potential.
\end{itemize}
\begin{figure}[t]
\begin{center}
\includegraphics[width=8cm]{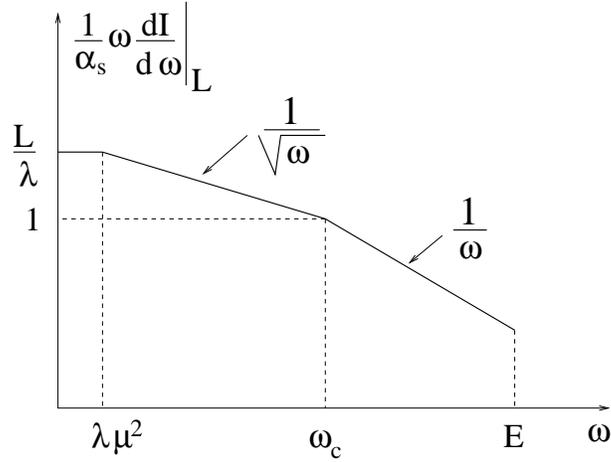}
\end{center}
\caption[*]{Induced gluon radiation spectrum of a light quark produced in a hot QCD medium for $\lambda \ll L \ll L^*$ (double logarithmic plot).}
\label{fig-spec-inmed}
\end{figure}

The full radiation spectrum is represented in 
Fig.~\ref{fig-spec-inmed}. In the region $\omega \ll \omega_c$, it is the same as for an asymptotic quark
(see Fig.~\ref{fig-spec-asym}). In the region $\omega \gg \omega_c$, the spectrum is suppressed compared to the case
of an asymptotic particle due to the constraint $\theta^2 \geq 1/(\omega L)$. 
This region still gives the dominant contribution
to the energy loss.\footnote{Note that other quantities than the average energy loss, involving all moments of the distribution $dI/d\omega$, such as quenching factors \cite{Baier:2001yt}, might be dominated by the soft part \eq{diffspec1} of the spectrum.} 
At large energies, the contribution \eq{DEmedintermQCD2} dominates over 
that of \eq{DEmedintermQCD1}.\footnote{This holds irrespectively of whether \eq{DEmedintermQCD1} involves a factor 
$\sim \ln (L/\lambda)$ or not.}

Finally, when $L \gg L^*$, the parameter $\omega_c$ becomes larger than $E$ and ceases to play a role. The spectrum
is given by \eq{diffspec1} for all $\omega > \lambda \mu^2$, as for an asymptotic particle.  
Integrating over $\omega$ reproduces the estimate \eq{DElargeLQCD} for the asymptotic particle energy loss in a thick medium.

The overall behavior of $\Delta E(L)$ is the same as in the Abelian case,\footnote{As far as $\Delta E(L)$ is concerned, 
the only difference between QED and 
QCD is the smooth logarithmic enhancement contained in \eq{DEmedLlllamQCD} and \eq{DEmedintermQCD2}.} see Fig.~\ref{DEvsLinside}. 
Let us stress again that the quadratic growth at small $L \ll L^*$ 
is a specific feature of the medium-induced energy loss of a newborn particle. 
It displays itself both in Abelian and non-Abelian plasmas.

\subsubsection{Heavy quark}

Finally, let us discuss the radiative losses of a heavy quark created in the plasma. 
As was the case for an asymptotic quark, there are three main mass regions. 
However, the case of a heavy quark produced in the plasma is more complicated. As we will see, the second mass region, 
namely $\alpha_s \sqrt{ET^3} \ll M^2 \ll \alpha_s E^2$, splits into two sub-domains, where the logarithmic dependence
of $\Delta E$ is slightly different. 

\vskip 3mm 
\centerline{{\bf A)} \ \boldmath{$M^2 \ll \alpha_s \sqrt{ET^3}$}} 
\vskip 3mm

In this region, the dependence of $\Delta E(L)$ is the same as for light quarks. 
The reason for that is basically the same as in the Abelian case. The characteristic
gluon radiation angle, which is of the same order as the photon radiation angle, as in \eq{typangle2}, or exceeds it,
as in \eq{typangle1}, is much larger than $\theta_M^2$ in the whole range of $L$ and $\omega$.

\vskip 3mm
\centerline{{\bf B)}\ \boldmath{$\alpha_s \sqrt{ET^3} \ll M^2 \ll \alpha_s E^2$}}
\vskip 3mm

In this range, the relevant in-medium formation length $\tilde{L}^*$ given in \eq{Lstartilde} is larger than $\lambda$, 
but smaller than the light quark in-medium formation length $L^*$. The law $\Delta E(L) \sim L^2$ valid at small $L$ 
is replaced by the linear dependence at the scale $\tilde{L}^*$ rather than $L^*$. 
This is the main effect brought about by the quark mass. 

In addition, there is a more subtle effect: the logarithmic factor multiplying
$\alpha_s \omega_c$ in the estimate of $\Delta E(L)$ at small $L$ might change. Indeed, the logarithmic factors in
\eq{DEmedLlllamQCD} and \eq{DEmedintermQCD2} come from integrating the spectra \eq{omdPom2QCDbis} and \eq{diffspec2} 
over the intervals $L\mu^2 < \omega < E$ and $\omega_c < \omega < E$, respectively. But a (large enough) 
mass brings about an effective cutoff in the spectrum when the characteristic
angle $\sim 1/(\omega L)$ becomes smaller than $\theta_M^2$. This happens when the gluon energy exceeds the scale
\be
\label{ombox}
\omega_\Box\ \equiv \ \frac {E^2}{LM^2} \, . 
\ee
Beyond the scale $\omega_\Box$, the spectrum falls down rapidly as $\sim 1/\omega^2$. This is a characteristic 
behavior of the hard part of energy spectra beyond the mass-induced cutoff, cf. \eq{QCDspectrum} 
and Fig.~\ref{fig-spec-asym-heavy}. 
Evidently, the statement above makes sense only when $\omega_\Box < E$, \ie, when $L > E/M^2$. 
The results for $\Delta E(L)$ are thus slightly different depending on whether $E/M^2 > \lambda$ (\ie, $M^2 < \alpha_s ET$) 
or $E/M^2 < \lambda$ (\ie, $M^2 > \alpha_s ET$). 
Representing the radiative loss as 
\be 
\label{DEviaR}
\Delta E(L \ll \tilde{L}^*) \sim \ \alpha_s \, \omega_c \ln R \ ,
\ee
we quote the estimates for $R$ in the two relevant subregions.\footnote{We do it for completeness, although the modification of the 
logarithm's argument is probably a too subtle effect to be observed in experiment. In addition, there might be logarithmic factors not 
coming from $\int d\omega/\omega$, see the footnote after Eq.~\eq{DEmedintermQCD1}.} 

\vskip 3mm
{\bf B1)} \ {\boldmath{$\alpha_s \sqrt{ET^3} \ll M^2 \ll \alpha_s E T$}}
\vskip 3mm

\begin{figure}[t]
\begin{center}
\includegraphics[width=8cm]{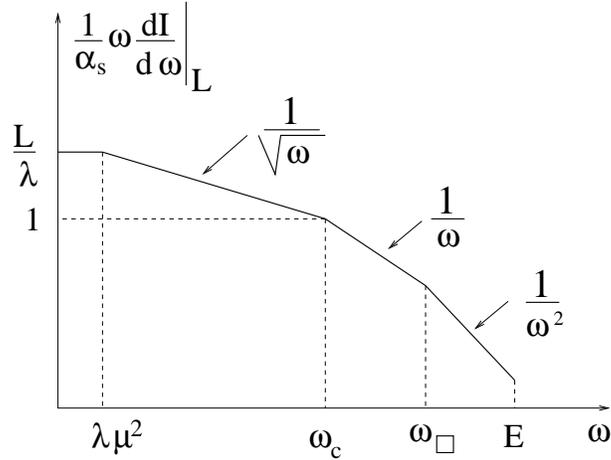}
\end{center}
\caption[*]{Induced gluon radiation spectrum of a heavy quark produced in a hot QGP 
for $\alpha_s \sqrt{ET^3} \ll M^2 \ll \alpha_s E T$ and $E/M^2 \ll L \ll \tilde{L}^*$ (double logarithmic plot).}
\label{fig-spec-inmed-heavy}
\end{figure}

In this subregion we have
\bea
\label{RB1}
R_{\rm B1} \ =\ \left\{ \begin{array}{ccc} 
E/(L\mu^2) & & (L \ll \lambda) \\
E/\omega_c 
& & \left(\lambda \ll L \ll  E/M^2 \right) \\ 
\omega_\Box/\omega_c 
& & (E/M^2 \ll  L \ll \tilde{L}^* ) 
\end{array} \right.\ \ .
\eea
For illustration, we represent in Fig.~\ref{fig-spec-inmed-heavy} the induced radiation spectrum in the last case,
namely $E/M^2 \ll  L  \ll \tilde{L}^*$.

\vskip 3mm
{\bf B2)}\ {\boldmath{$\alpha_s ET \ll M^2 \ll \alpha_s E^2$}}
\vskip 3mm

Here the estimates for $R$ become
\bea
\label{RB2}
R_{\rm B2} \ =\ \left\{ \begin{array}{ccc} 
E/L\mu^2 & &  \left(L \ll E/M^2 \right) \\
\omega_\Box/(L\mu^2) 
& & \left(E/M^2 \ll   L  \ll  \lambda \right) \\ 
\omega_\Box/\omega_c 
& & (\lambda \ll L \ll \tilde{L}^* ) 
\end{array} \right.\ \ .
\eea

The logarithmic enhancement in $\Delta E(L)$ disappears at $L \sim \tilde{L}^*$, where the energy loss is 
\be
\Delta E(L \sim \tilde{L}^*)  \sim \alpha_s \, \omega_c(\tilde{L}^*) \sim \alpha_s E  
\left( \frac {\alpha_s \sqrt{ET^3} }{M^2} \right)^{2/3}\ ,
\ee
the same estimate as \eq{supfactor}, \eq{DELggLstar} for an asymptotic heavy quark. 
As was mentioned, when $L \gg \tilde{L}^*$, the quadratic law \eq{DEviaR} is replaced by the linear one with the
slope given in \eq{DELggLstar}.

\vskip 3mm 
{\bf C)} \ {\boldmath{$M^2 \gg \alpha_s E^2$}}
\vskip 3mm

When the mass is so large, the scale $\tilde{L}^*$ becomes
smaller than $\lambda$. In this case, medium effects do not affect the formation length and the latter is 
given by $E/(\mu M)$ (see \eq{formQCDheavy}) rather than $E/(\mu_{\rm eff} M)$ (see \eq{Lstartilde}). 
At this scale, the quadratic law in $L$  
is replaced by the linear law \eq{BHmassQCD}, the same as for an asymptotic heavy quark. 

At $L \ll E/(\mu M)$, the energy loss is estimated as in \eq{DEviaR}. As in the previous case, the argument $R$ of the logarithm 
depends on whether $L < E/M^2$ (in this case, the light quark estimate  \eq{DEmedLlllamQCD} still holds) or $L >  E/M^2$ 
(in this case, the upper cutoff \eq{ombox} is introduced in the spectrum). To recapitulate,
\bea
\label{RC}
R_{\rm C} \ =\ \left\{ \begin{array}{ccc} 
E/(L\mu^2) & & (L \ll E/M^2 ) \\
\omega_\Box/(L\mu^2) & & (E/M^2 \ll  L \ll E/(\mu M) ) 
\end{array} \right.\ \ .
\eea
 
We observe that the law $\Delta E(L) \sim \alpha_s \, \omega_c \propto L^2$ is universal and is not modified at 
small enough $L$, however large the quark mass is.
On the other hand, the larger the mass, 
the earlier the change of regime between the quadratic law and a linear behavior. 
The slope of the latter decreases as mass grows. For $M^2 \gg \alpha_s E^2$, 
the slope is given by the BH formula \eq{BHmassQCD}. 

\begin{figure}[t]
\begin{center}
\includegraphics[height=7cm]{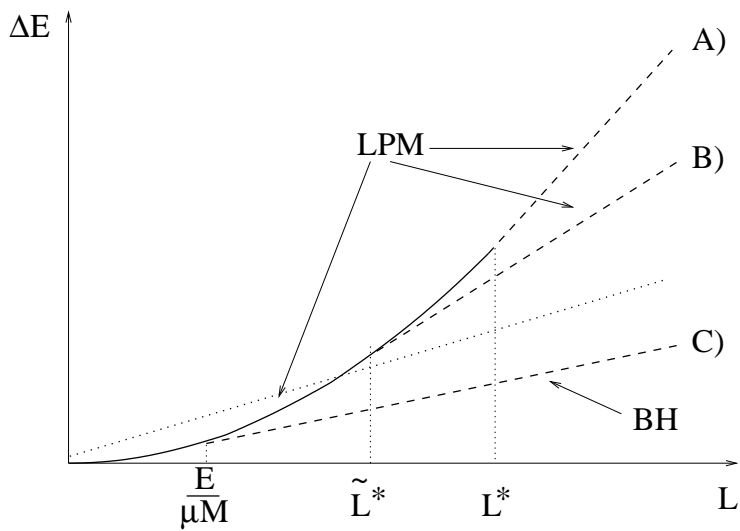}
\end{center}
\caption[*]{Induced radiative energy loss of a heavy quark produced in a QGP. 
{A)} $M^2 \ll \alpha_s \sqrt{ET^3}$; \ 
{B)} $ \alpha_s \sqrt{ET^3} \ll M^2 \ll \alpha_s E^2$; {C)} $M^2 \gg \alpha_s E^2$. ({\it Dotted line:} collisional loss.)}
\label{DEvsLMQCD}
\end{figure}

Our main results for a heavy quark produced in 
the plasma are qualitatively represented in Fig.~\ref{DEvsLMQCD}. The transition between the quadratic and 
linear regimes occurs at a length scale which is $\sim {\rm min}(L^*, \tilde{L}^*, E/(\mu M))$. For 
$M^2 \ll \alpha_s \sqrt{ET^3}$ it is $L^*$, for $ \alpha_s \sqrt{ET^3} \ll M^2 \ll \alpha_s E^2$ it is 
$\tilde{L}^*$, and for $M^2 \gg \alpha_s E^2$ it is $E/(\mu M)$.

\section{Concluding remarks}
\label{sec6}

The primary purpose of this review was, as for any other review, to bring together and 
present in a systematic way the known results scattered in the original papers. Another 
goal was to rederive those results using simple physical arguments rather than coming to grips
with complicated formalisms. 
We tried, for instance, to explain in a semi-heuristic way the origin of the 
quadratic dependence $\Delta E_{\rm rad} \sim L^2$ for thin plasmas or of the law $\Delta E_{\rm rad} \sim L \sqrt{E}$ 
for thick plasmas. However, besides reviewing known results, we have made some new observations. 

First, we found that the $L^2$ law, which was generally believed to be a specific feature of QCD,
is valid also in the Abelian case. The extra suppression at small $L$ compared to 
a linear behavior $\Delta E_{\rm rad}(L) \sim L$ is always present when a particle is created within the medium in a hard
process. It is due to the fact that a newborn particle needs time to grow its radiation field coat and
acquire the capacity to radiate. We stress that, although the Abelian and non-Abelian physical pictures and
results are similar as far as the {\it average} radiative energy loss is concerned, the {\it spectra} of emitted photons 
in a QED plasma and of emitted gluons in a QGP are different (for thick plasmas, the corresponding spectra are given in 
\eq{LPMspectrum} and \eq{LPMspectrumQCD}). The difference is due to different kinematics
of photon and gluon bremsstrahlung, as discussed in sections \ref{sec3} and \ref{sec4}. In QCD the presence of the extra graph 
of Fig.~\ref{3graph} (together with the graphs of Fig.~\ref{QEDemission} with appropriate color factors) broadens the gluon
angular spectrum, see \eq{spectrumlightQCD} and \eq{spectrumlightQED}. 

Another task we tried to accomplish was the systematic analysis of the radiative energy loss 
of a massive particle in the different regions of $M$ and $L$. We have emphasized (see Fig.~\ref{DEvsLMQCD}) 
that the mass effects play no role at small enough length, however large the quark mass is.
For $M^2 \ll \alpha_s \sqrt{ET^3}$, there is no effect also at large $L$. For larger masses   
(one should distinguish the main regions $\alpha_s \sqrt{ET^3} \ll M^2 \ll \alpha_s E^2$ and $M^2 \gg \alpha_s E^2$)
the heavy quark loss $\Delta E_{\rm rad}(L)$ starts to deviate from the light quark loss at some critical length 
which decreases when $M$ increases. 

Let us now discuss the question of the physical meaning of the mean free path $\lambda$. 
The definition \eq{lamlog} relates $\lambda$ to the so-called {\it anomalous damping} $\zeta$ 
of ultrarelativistic collective excitations with quark or gluon quantum numbers \cite{damping}. 
The anomalous damping depends on the total cross section ($\zeta \sim n \sigma^{\rm tot}$), 
rather than on the transport cross section \eq{transport}. It is the latter rather than the former 
which determines the scale of different transport phenomena and a legitimate question is whether
$\zeta$ (or equivalently $\lambda$) is a physical observable quantity. 

This question was studied in Ref.~\cite{CanJPhys}. 
No way to measure $\zeta$ was found there in the ultrarelativistic plasma,
even in a thought experiment, but it was found that, in a nonrelativistic (Boltzmann) plasma and in a certain 
range of parameters, $\zeta$ shows up in the argument of Coulomb logarithms describing transport phenomena. 

Coming back to the energy loss problem, we see that in most formulae $\lambda$ enters not alone, but in the
transport coefficient
\be
\label{qhat}
\hat{q}\ = \ \frac {\mu^2}\lambda \ \sim \ \alpha_{(s)}^2 T^3 \, .
\ee
For example, the estimate \eq{DElargeL}
for the electron radiative loss in a thick plasma is represented as
$\Delta E_{\rm rad}(L \gg L^*) \sim \alpha L \sqrt{\hat{q} E}$. 
The parameter $\hat{q}$ describes how the transverse momentum of the particle grows with distance, 
$\ave{q_\perp^2 (L)} = \hat{q} \, L$. To be more precise, $\hat{q}$ is of the form (compare to \eq{QEDmeanloss}) 
\be
\hat{q} \sim n \, \int_{\mu^2}^{|t|_{\rm max}} \frac{d \sigma}{dt} \, |t| \, d|t| 
\propto \alpha^2 T^3 \ln{\frac{ET}{\mu^2}} \, ,
\label{qhatlog}
\ee
and thus depends logarithmically on the energy of the incoming particle. Recently, the coefficient 
of the logarithm in \eq{qhatlog} was evaluated analytically \cite{Arnold:2008vd}.

Seemingly, the parameter $\lambda$ may show up as a scale that distinguishes a very thin 
plasma $L \ll \lambda$ where the particle undergoes at most one scattering and the intermediate
region $\lambda \ll L \ll L^*$, where the multiple scattering kinematics is effective. 
In our whole discussion, we indeed made a clear distinction between these two regions and treated    
them differently. 

As far as the radiative energy loss of a particle created in a plasma is concerned, nothing essential
happens at $L \sim \lambda$, as can be qualitatively seen on Fig.~\ref{DEvsLinside}. 
However, in the case of a light quark produced in a QGP and at $L \ll \lambda$, the parameter $\lambda$ enters 
the argument of the logarithm in \eq{DEmedLlllamQCD} (use $\mu^2 =  \hat{q} \, \lambda$). 
The region $L \ll \lambda$ is, however, rather marginal -- it corresponds to a quark produced near the edge of the plasma. 
Since it seems unrealistic to ``tag'' such quarks, the observability of $\lambda$ in this case is questionable.
In a more realistic case,  
$\lambda \ll L \ll L^*$, the logarithm in the large $\omega \gg \omega_c$ contribution \eq{DEmedintermQCD2} depends 
only on the combination \eq{qhat}. As was already mentioned, the small $\omega \lsim \omega_c$ contribution 
may provide a logarithmic factor $\sim \ln(L/\lambda)$, and the scale $\sim \lambda$ can in this case be observed 
through weak logarithmic dependence. More studies of this delicate issue are required.

For an asymptotic particle, the situation looks different. As is clear from Fig.~\ref{DEvsLQED}, 
the dependence is essentially modified at $L \sim \lambda$. In addition, the slope of the curve in the BH region 
$L \ll \lambda$ is given by the estimate \eq{BHQCDlight} which explicitly involves $\lambda$. 
On the other hand, it is difficult to imagine how a plasma (in thermal equilibrium) of size $L \sim \lambda$ or less  
could be created, as we already mentioned in the footnote at the beginning of section 3.1. 

Another attempt to pinpoint an explicit dependence on $\lambda$ is associated with the estimate
\eq{BHmassQCD} for the radiative energy loss of a heavy quark. We have seen that, when the mass
is large enough, $M^2 \gg \alpha_s E^2$, this estimate is valid not only for unphysically thin, 
$L \ll \lambda$, but also for thick plasmas, see the dashed curve corresponding to $M^2 \gg \alpha_s E^2$ 
in Fig.~\ref{DEvsLMQCD}. The expression \eq{BHmassQCD} involves the 
combination $\mu/\lambda = \sqrt{\hat{q}/\lambda}$. 
However, \eq{BHmassQCD} describes only the {\it radiative} energy loss. And as we have seen, 
when $M^2 \gg \alpha_s E^2$, the radiative loss is suppressed compared to the {\it collisional} one. For light quarks, 
the radiative and collisional losses have different patterns: the characteristic energy of the radiated gluons is of order
$E$, which is much larger than the characteristic energy transfer in one elastic collision. 
But the radiation spectrum of heavy quarks is soft. The spectrum is cut off beyond the scale $\mu E/M$, 
which for large masses $M^2 \gg \alpha_s E^2$ is smaller than the plasma temperature $T$. 
In other words, it seems impossible to separate for so heavy quarks the radiative component of the
net drag force $dE/dx$ and access $\mu/\lambda$ and thus $\lambda$. Quite curiously, for smaller masses, when radiative losses dominate, 
their value is not sensitive to $\lambda$. For example, a nontrivial estimate \eq{DELggLstar} in the intermediate mass region 
$\alpha_s \sqrt{ET^3} \ll M^2 \ll \alpha_s E^2$ depends only on $\hat{q}$.

Still the parameter $\lambda$ (and not only the combination \eq{qhat}) seems to have an independent physical meaning.
In fact, this parameter appears  under the logarithm in the more refined estimates
\eq{DElargeLCoulomb}, \eq{DElargeLCoulombQCD} for the light particle radiative losses 
(recall that $T \sim \lambda \mu^2 = \hat{q} \lambda^2$). These estimates 
take into account the behavior $\sim ({N} \ln {N}) \, \mu^2$
rather than just ${N} \mu^2$ for the characteristic effective scattering momentum transfer. In the developed LPM regime, 
$N = L^*/\lambda = \sqrt{E/(\lambda \mu^2)}$. Assuming that the estimates \eq{DElargeLCoulomb}, \eq{DElargeLCoulombQCD}
are correct,\footnote{They were obtained in a model where the particle is scattered on a set of static
Coulomb sources separated by the average distance $\lambda$. The presence of the factor $\propto \sqrt{\ln{E}}$ 
in \eq{DElargeLCoulomb} is a robust model-independent feature, the origin of the logarithm being the same as in 
\eq{qhatlog}. On the other hand, in what particular way the argument of the logarithm is formed, whether it is
$E/T$, $ET/\mu^2 \sim E/(\alpha T)$, or some other ratio, is a more delicate and difficult question. Only an exact model-independent 
calculation (possibly using the formalism of Ref.~\cite{Yaffe}) could resolve it.} the situation
is then analogous to that observed in Ref.~\cite{CanJPhys} for Boltzmann plasmas: the parameter 
$\lambda$ affects observable quantities in a weak logarithmic way. 

In other words, the physical status of $\lambda$ (or $\zeta$) in ultrarelativistic
plasmas remains unclear. But it is indisputably very useful as a theoretical instrument allowing one to 
obtain meaningful physical results for radiation spectra and energy losses. 

As a dessert, let us mention the fascinating issue of energy losses in ${\cal N} = 4$ supersymmetric
Yang-Mills (SYM) theory. At weak coupling, there is not much difference with QCD, and we expect the estimates quoted in this
paper to apply also to ${\cal N} = 4$ SYM. The main interest of the ${\cal N} = 4$ SYM theory is that in the 
large $N_c$ limit, many quantities 
can be evaluated also at strong 't Hooft coupling\footnote{Do not confuse it with the mean free path!} 
$\lambda = g^2 N_c \gg 1$, using the duality conjecture \cite{duality}. 
In particular, the drag force $dE/dx$ acting on a heavy quark moving through a thick ${\cal N} = 4$ SYM plasma reads \cite{Yaffe1}
\be
\label{dragheavy}
\frac {dE}{dx} \ =\ - \frac \pi 2 \, \sqrt{\lambda} \, T^2 \, \frac {\sqrt{E^2-M^2}}M \hskip 1cm (\lambda \gg 1) \, . 
\ee
This estimate is valid when $M \gg (\lambda T E^2)^{1/3}$ \cite{Marquet,Fadafan:2008bq}. 
The dependence \eq{dragheavy} resembles the perturbative result \eq{BHmassQCD}. One difference is that 
the latter is valid in a different mass region, namely $M \gg gE$, and that it describes only the radiative energy loss
which happens to be dominated in this region by the collisional loss. 

In Refs.~\cite{stronglight} (see also \cite{Chesler}), 
 the energy losses of light partons in a strongly coupled  ${\cal N} = 4$ plasma were
estimated. The dependence 
\be
\label{draglight}
\frac {dE}{dx} \ \sim\  - \lambda^{1/6} \left( E^2 T^4 \right)^{1/3} \ 
\ee
for the {\it mean} drag force (for light partons, this quantity involves large
fluctuations) was found.
This is different from the perturbative dependence  $dE/dx \propto \sqrt{E}$.
  More studies in this direction
are desirable.

\acknowledgments

We would like to thank Yuri Dokshitzer, Fran\c{c}ois Arleo, Pol-Bernard Gossiaux and Arkady Vainshtein 
for useful 
discussions and comments. S.~P. also thanks Andr\'e Peshier for a fruitful collaboration on collisional 
energy loss, on which most of section \ref{sec2} of the present work is based. 

\eject

\appendix

\section{Typical momentum broadening in Coulomb rescattering}

Here we consider a charged (colored) particle 
with $E \to \infty$ moving in a perturbative QED (QCD) plasma, and undergoing $n$ successive 
Coulomb scatterings. The range $1/\mu$ of the Coulomb potential is assumed to be 
much smaller than the mean free path $\lambda$ between two successive scatterings,
so that the elastic Coulomb rescatterings are independent. We calculate the {typical} transverse momentum $q_{\rm typ}(N)$ of the particle 
after $N$ scatterings. In the case of fixed coupling (QED), the result was found in Ref.~\cite{Baier:1996sk} to be 
$q^2_{\rm typ}(N)\ \sim \mu^2 N \ln N$. We present an alternative derivation of this result and generalize it to running coupling (QCD).

Consider first the Abelian fixed coupling case. Coulomb scattering with transverse momentum exchange $\vec{q}_i$ is associated
to the normalized probability density
\beq
\label{VqQED}
\frac{1}{\sigma_{\rm tot}}  \frac{d\sigma}{d^2 \vec{q}_i} \equiv P(\vec{q}_i) 
= \frac{1}{\pi} \frac{\mu^2}{(\vec{q}_i^{\,\, 2}+\mu^2)^2} \ \ ; \ \ 
\int d^2 \vec{q}_i \, P(\vec{q}_i) = 1 \, .
\eeq
The {\it average} momentum exchange $\ave{\vec{q}^{\,\,2}}$ in a single Coulomb scattering
is logarithmically divergent. (The divergence is cut-off by the kinematical constraint on
the maximal transverse exchange $|\vec{q}|_{\rm max}$, but we focus on the $E\to \infty$ limit where
$|\vec{q}|_{\rm max} \to \infty$.)
On the other hand, the {\it typical} transverse momentum transfer $q_{\rm typ}$, defined as the transfer
such that the probability to have $|\vec{q}| < q_{\rm typ}$ is $1/2$, is well-defined.
Solving the equation
\beq
\int d^2 \vec{q} \, P(\vec{q}) \, \Theta(q_{\rm typ}^2 - \vec{q}^{\, \,2}) = 1/2 \, ,
\eeq
we easily find that $q_{\rm typ}$ in one scattering equals the Debye mass $\mu$. 

We now determine the typical transfer $q_{\rm typ}(N)$ after $N$ scatterings, defined by
\beq
\label{deftyp}
\int d^2 \vec{q} \left(\prod_{i=1}^N d^2 \vec{q}_i \, P(\vec{q}_i) \right) \, \delta^{2}\left( \vec{q}- \sum_{i=1}^N \vec{q}_i \right) \, 
\Theta(q_{\rm typ}^2(N) - \vec{q}^{\,\,2}) = \frac{1}{2} \, .
\eeq
Representing the $\delta$-function as
\beq
\delta^{2}\left(\vec{q}- \sum_{i=1}^N \vec{q}_i\right) = \int \frac{d^2 \vec{r}}{(2\pi)^2} \, \exp{\left[ i \vec{r} \cdot \left( 
\vec{q}- \sum_{i=1}^N \vec{q}_i \right) \right]} \, , 
\eeq
we obtain from \eq{deftyp},
\beq
\label{deftyp2}
\frac{1}{2} = \int \frac{d^2 \vec{r}}{2\pi} \, \left[ {\tilde P}(\vec{r}) \right]^N 
\int \frac{d^2 \vec{q}}{2\pi} \, e^{i \vec{r} \cdot \vec{q}} \, \Theta(q_{\rm typ}^2(N) - \vec{q}^{\,\,2}) \, ,
\eeq
where 
\beq
\label{Vtilde}
{\tilde P}(\vec{r}) = \int d^2 \vec{q} \, P(\vec{q}) e^{- i \vec{r} \cdot \vec{q}} = 
\mu r K_1(\mu r) \, .
\eeq
The $\vec{q}$-integral in \eq{deftyp2} can be done exactly, leading to 
\beq
\label{deftyp3}
\frac{1}{2} = \int_0^\infty dr \, q_{\rm typ}(N) J_1(q_{\rm typ}(N)\, r) \, \left[ r K_1(r) \right]^N 
= - \int_0^\infty dr \left[ r K_1(r) \right]^N 
\frac{\partial}{\partial r} J_0(q_{\rm typ}(N)\,r) \, .
\eeq
In the latter equation and in the following, $r$ is expressed in units of 
$\mu^{-1}$, and $q_{\rm typ}(N)$ in units of $\mu$. 
Integrating by parts and using $(r K_1(r))'= - r K_0(r)$ we arrive at 
\beq
\label{mastereq}
\frac{1}{2} = N \int_0^\infty dr \, r \, J_0(q_{\rm typ}(N) \, r) \left[ r K_1(r) \right]^{N-1} K_0(r) \, .
\eeq

The equation \eq{mastereq} for $q_{\rm typ}(N)$ has been derived from 
\eq{deftyp} without any appro\-xi\-ma\-tion. We now assume a large number of scatterings,
$N \gg 1$, and derive the asymptotic behavior of $q_{\rm typ}(N)$ in this limit.

Clearly, when $N \gg 1$, the integral in \eq{mastereq} is saturated by $r \ll 1$. We can thus approximate
\be
\label{K0K1}
K_0(r) \simeq - \ln{r} \ \ ;  \ \ \ \ \ \  r K_1(r) \simeq 1 - \frac{r^2}{4} \ln{\frac{1}{r^2}} 
\simeq \exp{\left[-\frac{r^2}{4} \ln{\frac{1}{r^2}}\right]}\ .
\ee 
Since $K_0(r)$ is a slowly varying function for $r \ll 1$, we obtain from \eq{mastereq},
\beq
\label{mastereq2}
\frac{1}{2} \simeq N \, \ave{\ln{1/r}} \int_0^1 dr \, r \, J_0(q_{\rm typ}(N)\,r) 
\exp{\left[ -\frac{N r^2}{4} \ln{\frac{1}{r^2}} \right]} \, .
\eeq
The integral is dominated by the region  
\beq
\label{domreg}
N \, r^2 \ln{\frac{1}{r^2}}  \sim 1 \mathop{\ \ \Longleftrightarrow \ \ }_{N \gg 1} r^2 \sim \frac{1}{N \ln{N}} \, .
\eeq
Using this, we can rewrite \eq{mastereq2} as
\beq
1 \simeq N \ln{N}  \int_0^1 dr \, r \, J_0(q_{\rm typ}(N)\,r) \exp{\left[ - (N \ln{N})  \frac{r^2}{4} \right]} \, .
\eeq
Introducing $u= (N \ln{N})\, r^2$, this becomes
\beq
1 \simeq \int_0^\infty \frac{du}{2} \, J_0\left(\frac{q_{\rm typ}(N)}{\sqrt{N \ln{N}}} \sqrt{u}\right) 
\, e^{-u/4} \, .
\eeq
Using
\beq
\int_0^{\infty} du \, J_0( C \sqrt{u}) e^{-u/4} = 4 e^{-C^2} \, ,
\eeq
we finally obtain (reintroducing the dimension of $q_{\rm typ}(N)$)
\beq
\label{nlnn}
q_{\rm typ}^2(N) \simeq (\ln{2}) \cdot (N \ln{N}) \cdot \mu^2 \, .
\eeq

This result immediately follows also from the expression 
\be
\label{probadensity}
f(q_\perp^2,N) \simeq \frac{1}{\pi \mu^2 N \ln{N}} \exp{\left(- \frac{q_\perp^2}{\mu^2 N \ln{N}} \right)} 
\ee 
for the probability distribution of the transverse momentum transfer $q_\perp^2$ after $N$ scatterings, 
derived previously in Ref.~\cite{Baier:1996sk}. Indeed, defining the typical transfer as in \eq{deftyp}, namely 
\be
\int d^2 \vec{q} \,f(q^2,N)\,\Theta(q_{\rm typ}^2(N) - q^{2}) = \frac{1}{2} \, ,
\ee
and using \eq{probadensity}, we recover \eq{nlnn}.

The derivation above was performed for fixed coupling, and the estimate \eq{nlnn} is  thus valid for QED. 
In QCD, the running of the coupling should
be taken into account and this brings about certain modifications. The effective coupling constant depends on the transverse momentum 
transfer $q^2$. The normalized probability density of a single Coulomb scattering is now
\be
\label{VqQCD}
\left. P(\vec{q})\right|_{\rm QCD}= \frac 1\pi  \frac{\mu^2}{(q^2+\mu^2)^2} \, \frac {\alpha_s^2(q^2)}{\alpha_s^2(\mu^2)} 
F\left(\frac \mu {\Lambda_{\rm QCD}}\right) \simeq   \frac 1\pi  \frac{\mu^2}{(q^2+\mu^2)^2} \, \frac 
{\ln^2 \frac {\mu^2} {\Lambda^2_{\rm QCD}}}{\ln^2 \frac {q^2} {\Lambda^2_{\rm QCD}}} 
\ ,
\ee
where $F(x)$ is a smooth function that tends to unity in the limit $\mu \gg \Lambda_{\rm QCD}$ we are interested in, 
and which we have thus been allowed to set to $F = 1$. The analysis is done along the same lines as in QED. 

We first note that the expression \eq{mastereq} can be rewritten for a general scattering probability density ${\tilde P}(\vec{r})$ as 
\beq
\label{mastereq-general}
\frac{1}{2} = - N \int_0^\infty dr \, J_0(q_{\rm typ}(N) \, r) \left[{\tilde P}(\vec{r}) \right]^{N-1} 
\frac{\partial {\tilde P}(\vec{r})}{\partial r} \, .
\eeq
When $N \gg 1$, we have typically $r \ll 1$ ($r$ being expressed in units of $\mu^{-1}$), implying ${\tilde P}(\vec{r}) \simeq 1$
(this can be easily checked a posteriori). 
Thus, \eq{mastereq-general} can be approximated by
\beq
\label{mastereq-general2}
\frac{1}{2} \simeq  N \int_0^\infty dr \, J_0(q_{\rm typ}(N) \, r) \, \exp{\left[- (N-1) \left(1- {\tilde P}(\vec{r}) \right) \right]} 
\, \frac{\partial}{\partial r} \left(1- {\tilde P}(\vec{r}) \right) \, .
\eeq
Using \eq{VqQCD}, we obtain (expressing $q$ in units of $\mu$)
\beqa
\label{1minusVtildeQCD}
1- {\tilde P}(\vec{r}) &=& \int d^2 \vec{q} \, \left. P(\vec{q})\right|_{\rm QCD} \left( 1- e^{- i \vec{r} \cdot \vec{q}} \right) \simeq \int_0^{1/r^2} dq^2 \, \frac{1}{(q^2+1)^2} \, \frac{\ln^2 \frac{\mu^2}{\Lambda^2_{\rm QCD}}}{\ln^2 \frac{q^2 \mu^2}{\Lambda^2_{\rm QCD}}} \, \frac{r^2 q^2}{4}  \nn \\
&\simeq&  \frac{r^2}{4} \ln^2 \frac{\mu^2}{\Lambda^2_{\rm QCD}} \int_{1}^{1/r^2} \frac{dq^2}{q^2 \ln^2 \frac{q^2 \mu^2}{\Lambda^2_{\rm QCD}}} 
\simeq \frac{\alpha_s(\mu^2/r^2)}{\alpha_s(\mu^2)}  \cdot \frac{r^2}{4} \ln{\frac{1}{r^2}} \, .
\eea

Comparing to \eq{K0K1}, we see that the running of $\alpha_s$ displays itself in the factor $\alpha_s(\mu^2/r^2)/\alpha_s(\mu^2)$. 
Using now \eq{domreg}, we infer that the running of the 
coupling modifies the fixed coupling estimate \eq{nlnn} to
\be
\label{nlnnQCD}
\left. q^2_{\rm typ}(N)\right|_{\rm QCD} \ \sim \ \frac{\alpha_s(N\mu^2)}{\alpha_s(\mu^2)} \, (N \ln{N}) \cdot \mu^2  \, .
\ee

It is interesting to mention that the typical momentum transfer at large $N$, in QED and 
QCD (see \eq{nlnn} and \eq{nlnnQCD}), can be heuristically obtained from the following formulae
\bea
q^2_{\rm typ}(N)  \ &\sim& \  \mu^2 N \int_{\mu^2}^{q_{\rm typ}^2(N)} \frac {dq^2}{q^2}    \hskip 2cm  ({\rm QED})  \, ,
\label{qtypnQED} \\
q^2_{\rm typ}(N)  \ &\sim& \  \mu^2 N \int_{\mu^2}^{q_{\rm typ}^2(N)} \frac {dq^2}{q^2} \, \frac{\alpha_{s}^2(q^2)}{\alpha_{s}^2(\mu^2)}  
\ \ \ \ \ ({\rm QCD}) \, .
\label{qtypnQEDQCD}
\eea

Finally, let us remark that \eq{nlnnQCD} can also be represented as
\be
\label{qtypnQCD}
\left. q^2_{\rm typ}(N)\right|_{\rm QCD} 
\ \sim \ \mu^2 N \, \frac{\ln \frac{\mu^2}{\Lambda_{\rm QCD}^2} \cdot \ln N}{\ln \frac {\mu^2}{\Lambda_{\rm QCD}^2}  + \ln N}\ .
\ee  
Thus, for {\it very} large $N$, namely $\ln N  \gg \ln \frac{\mu}{\Lambda_{\rm QCD}}$, we obtain
\be
\label{qtypngg} 
\left. q^2_{\rm typ}(N)\right|_{\rm QCD} \ \sim N \,\mu^2 \ln \frac{\mu}{\Lambda_{\rm QCD}} \sim N \, T^2  \ .
\ee
The scale $T^2$ is nothing but the average momentum transfer $\langle q_\perp^2 \rangle$ associated to the 
QCD probability density \eq{VqQCD}. 
In QED, this quantity is logarithmically divergent, but in QCD the running of $\alpha_s$ makes the 
integral for $\langle q_\perp^2 \rangle$ convergent even when the upper bound is put to infinity, as can be seen from \eq{qtypnQEDQCD}.

\section{LPM effect and Feynman diagrams}

In the main body of the paper, we have operated mostly with heuristic arguments based on formation length estimates 
and single scattering formulae. The same results can be derived by calculating the Feynman diagrams describing photon (gluon) 
radiation in the process of multiple scattering of a fast particle in the plasma. In this Appendix, 
we will not attempt to perform a complete diagrammatic analysis, but will present some illustrative calculations
which might help to understand better the origin of LPM suppression. We will restrict ourselves to 
the Abelian case and mostly follow the discussion of Ref.~\cite{Baier:1996vi}.

\begin{figure}[t]
\begin{center}
\includegraphics[width=5cm]{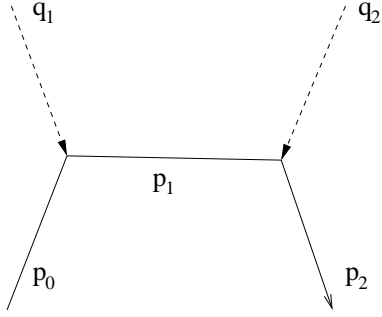}
\end{center}
\caption[*]{Electron elastic scattering on two centers.}
\label{2scat}
\end{figure}

We adopt, as we did in Appendix A, the model where a 
scalar 
massless particle is scattered on static centers with a Yukawa potential,
\be
\label{sumVi}
V(\vec{x}) \ \sim \ \alpha \sum_i \frac {\exp\{-\mu|\vec{x} - \vec{x}_i|\}}{|\vec{x} - \vec{x}_i|} \ ,
  \ee
where $\vec{x}_i$ is the position of the $i$-th center. Consider the case of only two
such centers and assume $\vec{x}_1 = \vec{0}$, $\vec{x}_2 = (\vec{x}_{2 \perp}, z)$. 
Then the elastic scattering amplitude (see Fig.~\ref{2scat}) reads
\be
\label{M2scat}
{\cal M}_{\rm el} \ \propto e^2 \int  \frac {d^3\vec{q}_1 \, d^3\vec{q}_2}{(\vec{q}_1^{\, 2} + \mu^2)(\vec{q}_2^{\, 2} + \mu^2) }
\delta^{(3)}(\vec{q}_1 + \vec{q}_2 - \vec{q}) 
e^{-i \vec{q}_2 \cdot \vec{x}_2} \, \frac 1{p_1^2 + i \epsilon} \ .
\ee
The total momentum transfer $\vec{q}$ and intermediate electron four-momentum $p_1$ are given by 
\be
\label{convpi}
\vec{q} = \vec{p}_2 -  \vec{p}_0 \ \ \ ; \ \ \ p_1 = (E, \vec{q}_{1\perp}, E+q_{1\|}) \, ,
\ee
where we have chosen $p_0 = (E, \vec{0}_\perp, E)$. In the model of static centers, the energy transfer in each
elastic scattering stricly vanishes, $q_1^0 = q_2^0 =0$, implying $p_0^0 = p_1^0 = p_2^0 =E$.

We integrate over $dq_{1\|} dq_{2\|} \delta(q_{1\|} + q_{2\|} - q_\|)$ by closing the contour in the upper $q_{1\|}$-plane 
and picking up the contribution of the pole of $p_1^2$  at the value\footnote{The contributions of the poles of 
$\vec{q}_1^{\, 2} + \mu^2$ and $\vec{q}_2^{\, 2} + \mu^2$ are suppressed 
by $\sim e^{-\mu z} \ll 1$. Indeed, the distance between successive scattering centers is $z \sim \lambda$ where the mean
free path $\lambda$ satisfies $\lambda \sim 1/(e^2 T) \gg \mu^{-1} \sim 1/(e T)$ in a perturbative plasma.} 
\be
\label{pole}
p_{1\|} = E+ q_{1\|} \simeq E - \frac{q_{1 \perp}^2}{2E} \, .
\ee
Using the on-shell condition $p_2^2 = 0$ we get $q_{\|} \simeq - q_{\perp}^2/(2E)$ and we obtain from \eq{M2scat}
\be
\label{M2scatperp}
{\cal M}_{\rm el} \ \propto e^2 \int  \frac {d^2 \vec{q}_{1 \perp} d^2 \vec{q}_{2 \perp}}
{(q_{1 \perp }^2 + \mu^2)(q_{2 \perp }^2 + \mu^2) } 
\delta^{(2)} ( \vec{q}_{1 \perp } + \vec{q}_{2 \perp } - \vec{q}_\perp)
e^{-i \vec{q}_{2 \perp } \cdot \vec{x}_{2 \perp }} e^{i\Phi_{\rm scatt}}
 \ ,
\ee
where 
\be
\label{Phiscat}
\Phi_{\rm scatt} = z (p_{1\|} - p_{2\|}) \approx \frac {z}{2E}\left[ (\vec{q}_{1 \perp } + \vec{q}_{2 \perp })^2 -
\vec{q}_{1 \perp }^{\, 2} \right] \, .
\ee

To evaluate the elastic cross section, we fix the longitudinal distance $z$ between the scattering centers and average 
$|{\cal M}_{\rm el} |^2$ over $\vec{x}_{2 \perp }$. Integrating further over $d^2 \vec{q}_\perp$ yields
\be
\label{sigma2scat}
\sigma_{\rm scatt} \ \propto \ \alpha^2 \int 
\frac{d^2 \vec{q}_{1 \perp } d^2 \vec{q}_{2\perp}}{(q_{1 \perp }^2 + \mu^2)^2(q_{2 \perp }^2 + \mu^2)^2} \ .
\ee

\begin{figure}[t]
\begin{center}
\includegraphics[height=6cm]{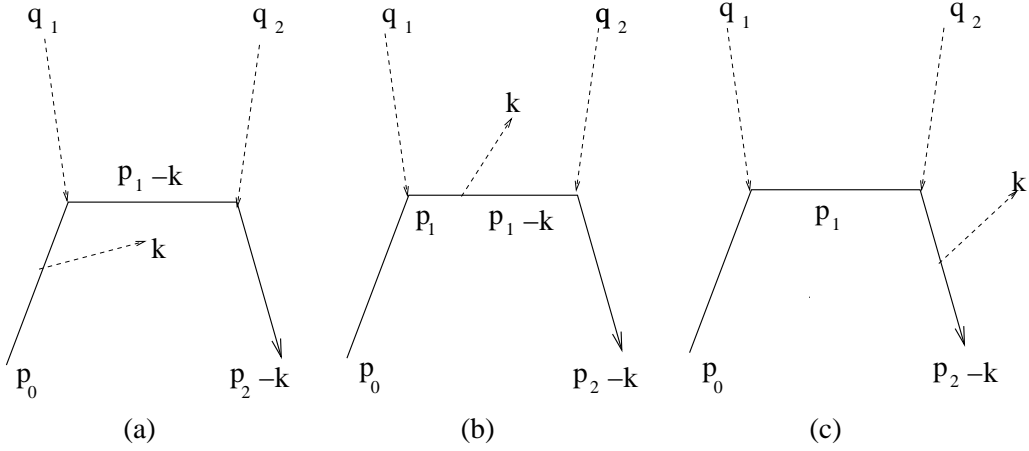}
\end{center}
\caption[*]{Amplitude for photon radiation induced by double elastic scattering.}
\label{2scatrad}
\end{figure}

Consider now the process where the scattered fast electron emits an additional photon. 
There are three graphs depicted in Fig.~\ref{2scatrad}.

It is convenient to define the momenta $p_i$ as in the case of elastic scattering, see for instance \eq{convpi}, 
so that $p_i$ ``do not know'' about the emitted photon. On the other hand, the final momentum is now 
$p_2-k$ rather than $p_2$, and the intermediate momentum $p_1-k$ appears
in the graphs of Fig.~\ref{2scatrad}a and \ref{2scatrad}b. This shift of momenta brings a modification of the phase factors in
the amplitude. For different graphs, this modification is different, and now one cannot suppress the phase factors as we 
did in the elastic case, when going from \eq{M2scatperp} to \eq{sigma2scat}.

Let us see how it works. Consider first the graph of Fig.~\ref{2scatrad}a. The conditions $(p_1 - k)^2 = 0$ and $(p_2 - k)^2 = 0$
imply, respectively, 
\be
\label{p12long}
 p_{1\|} \simeq E - \frac {p_{1 \perp}^2}{2E} - \frac {\omega \theta_1^2}{2} \, , \ \ \ \ \ 
 p_{2\|} \simeq E - \frac {p_{2 \perp}^2}{2E} - \frac {\omega \theta_2^2}{2}\ ,
\ee 
where $\theta_{1,2}$ are the angles between the direction of the emitted photon and $\vec{p}_1$ ($\vec{p}_2$). We assumed that
the angles are small.\footnote{Indeed, the radiation probability is dominated by small angles.} 
Substituting the expressions \eq{p12long} into the phase $\Phi = z(p_{1 \|} - p_{2 \|} )$, we obtain for the graph of 
Fig.~\ref{2scatrad}a, 
\be
\label{Phirada}
\Phi_{\rm rad}^{\rm (a)} \ =\ \Phi_{\rm scatt} + \frac {\omega z}2 (\theta_2^2 - \theta_1^2)\ ,
\ee
where $\Phi_{\rm scatt}$ is given in \eq{Phiscat}. The corresponding contribution to the radiation amplitude is
\be
\label{Mrada}
{\cal M}_{\rm rad}^{\rm (a)} \ =\  - \frac {\vec{\theta}_0 \cdot \vec{\varepsilon}}{\theta_0^2} \, \exp \left \{
\frac {i\omega z}2 (\theta_2^2 - \theta_1^2) \right \} \, {\cal M}_{\rm el} \, ,
\ee
where $\vec{\theta}_0 = \vec{k}_\perp/\omega$ and $\vec{\varepsilon}$ is the photon polarization vector.

For the graph of Fig.~\ref{2scatrad}c describing the emission from the final line, the structure 
$-\vec{\theta}_0 \cdot \vec{\varepsilon}/\theta_0^2$ is transformed into 
$\vec{\theta}_2 \cdot\vec{\varepsilon}/\theta_2^2$. The phase factor is different from that in the graph of Fig.~\ref{2scatrad}a due
to different kinematics. Putting the intermediate momentum on mass shell gives in this case the condition $p_1^2 =0$
rather than $(p_1-k)^2 = 0$, so that the expression for $p_{1 \|}$ is not modified compared to the elastic scattering
case. We have 
\be
\label{Phiradc}
\Phi_{\rm rad}^{\rm (c)} \ =\ \Phi_{\rm scatt} + \frac {\omega z}2 \theta_2^2 
\ee
and 
\be
\label{Mradc}
{\cal M}_{\rm rad}^{\rm (c)} \ =\  \frac {\vec{\theta}_2 \cdot \vec{\varepsilon}}{\theta_2^2} \, \exp \left \{
\frac {i\omega z}2 \theta_2^2 \right \} \, {\cal M}_{\rm el} \, .
\ee

The graph of Fig.~\ref{2scatrad}b provides two different contributions from the poles $p_1^2 = 0$ and $(p_1 - k)^2 = 0$. They 
both involve the structure $\vec{\theta}_1 \cdot \vec{\varepsilon}/\theta_1^2$, but the residues have opposite signs. In addition, 
the phase factors for these two contributions are different. For the pole $p_1^2 = 0$, the phase coincides with \eq{Phiradc}, whereas 
for the pole $(p_1- k)^2 = 0$, it coincides with \eq{Phirada}. The sum of all contributions can be expressed as
${\cal M}_{\rm rad} = e \, {\cal M}_{\rm el} \, \vec{\varepsilon} \cdot \vec{J}$, where 
\bea
\label{sumJ12}
\vec{J} \ &=&\ \vec{J}_1 \exp\left\{i \frac {\omega z}2( \theta_2^2 - \theta_1^2) \right \} 
+ \vec{J}_2 \exp\left\{i \frac {\omega z}2 \theta_2^2 \right \}  \\
&& \vec{J}_1  \ =\ \frac {\vec{\theta}_1}{\theta_1^2} - \frac {\vec{\theta}_0}{\theta_0^2}\ , \ \ \ \ \ \ \ \ \ 
\vec{J}_2  \ =\ \frac {\vec{\theta}_2}{\theta_2^2} - \frac {\vec{\theta}_1}{\theta_1^2} \ \ \, .
\label{J12}
\eea
Each term in the sum \eq{sumJ12} corresponds to the radiation induced by elastic scattering on the associated center. 
The phase difference $\omega z \theta_1^2/2$ between the two terms 
can be interpreted as the phase acquired by the photon of energy $\omega$ in the frame moving with the
fast particle (see \eq{disbalance}). 
 
The result \eq{sumJ12} can be easily generalized to the case of ${N}$ scattering centers. Let us assume 
that their longitudinal positions are $z_n = (n-1)\lambda$, $n = 1,\ldots, {N}$. Then 
\bea
\label{sumJ}
\vec{J}({N}) &=&\ \exp\left \{ i \frac {\omega \lambda ({N} - 1)}2 \theta_{{N}}^2 \right\} 
\sum_{n=1}^{N} \vec{J}_n \, e^{i\Phi_n} \\
\vec{J}_n &=&  \frac {\vec{\theta}_n}{\theta_n^2} - \frac {\vec{\theta}_{n-1}}{\theta_{n-1}^2} \ \ \ ; \ \ \ 
\vec{\theta}_n \equiv \vec{\theta}_0 - \sum_{m=1}^n \frac{\vec{q}_{m \perp}}{E}  \\
\Phi_n &=& -\frac {\omega \lambda}2 \sum_{m=n}^{{N} - 1} \theta_m^2  \ \ (n = 1,\ldots, {N} -1) \ \ \ ; 
\ \ \ \Phi_{N} = 0  \, .
\label{Phin}
\eea
The radiation energy spectrum is 
\be
\label{sumJsquared}
\omega \frac{dI}{d\omega} = \frac{\alpha}{\pi^2} \int d^2 \vec{\theta}_0 
\left| \sum_{n=1}^{N} \vec{J}_n \, e^{i \Phi_n} \right|^2 \ .
\ee
This should be averaged over $\vec{q}_{n \perp}$ with the weight 
\be
\prod_n  \frac{\mu^2 d^2\vec{q}_{n \perp}}{\pi (q_{n \perp}^2 + \mu^2)^2}  \, \ \ \ \ .
\ee

The analysis of the expression \eq{sumJsquared} confirms the physical picture outlined in section \ref{sec3}. In particular,
{(i)} the characteristic total scattering angle is $\theta_{\rm tot}^2 \sim {N} \mu^2/E^2$; 
{(ii)} the contributions of different scattering centers in \eq{sumJsquared} are coherent (so that we are dealing in 
this case with one effective scattering) when $\omega {N}\lambda \theta_{\rm tot}^2 \sim 
\omega L \theta_{\rm tot}^2 \ll 1$, \ie, when $L \ll \sqrt{\lambda E^2 /(\omega \mu^2)}$. The scale
$\sqrt{\lambda E^2 /(\omega \mu^2)}$ coincides with $\ell_f^{\rm med}(\omega)$ defined in \eq{Lformmedofomega}.
At $\omega \sim E$, the latter coincides with the characteristic in-medium formation length $L^*$ (\ie, the coherence length 
whence one photon of energy $\sim E$ is emitted). 

The results \eq{sumJ} and \eq{sumJsquared} have been derived for an asymptotic particle. In  case the particle is created
in the medium, the radiation amplitude can be obtained from \eq{sumJ} by treating the position of the first
scattering center as the creation point, and by suppressing the contribution of the graph analogous to Fig.~\ref{2scatrad}a
describing the emission from the initial line. Suppressing the irrelevant common phase factor in
front of the sum in \eq{sumJ}, changing ${N}\to {N} +1$ and the numeration $1 \to 0$, etc., we derive
\be
\label{sumJ0}
\vec{J}_{{\rm creation} + {N}\, {\rm scatterings}} =  
\frac {\vec{\theta}_0}{\theta_0^2} e^{i\Phi_0}   
+ \sum_{n=1}^{N} \vec{J}_n \, e^{i\Phi_n} \ .
 \ee
For ${N} = 1$, we reproduce the result \eq{JL}. 

The medium-induced radiation spectrum is 
\be
\left. \omega \frac {dI}{d\omega} \right|^{\rm induced}_{N} \sim \int d^2 \vec{\theta}_0 
\left(|\vec{J}|^2 - \frac 1{\theta_0^2} \right)  \, ,
\ee
where the square of the first term in \eq{sumJ0} (corresponding to the vacuum contribution) has been subtracted, as we did in \eq{omdPom}. 
We obtain
\be
\label{omdPomN} 
\left. \omega \frac {dI}{d\omega} \right|^{\rm induced}_{N}  \sim 
  \alpha   \int d^2 \vec{\theta}_0 \left[
2 \frac {\vec{\theta}_0}{\theta_0^2} \cdot \sum_{n=1}^{N} \vec{J}_n \cos(\Phi_0 - \Phi_n)
+ \left|  \sum_{n=1}^{N} \vec{J}_n e^{i\Phi_n} \right|^2 \right] 
\ ,
\ee 
which is convenient to represent as\footnote{We use $\sum_n \vec{J}_n = \vec{\theta}_{N}/\theta_{N}^2 - \vec{\theta}_0/\theta_0^2$ 
and  $\int d^2 \vec{\theta}_0 \, (1/\theta_{N}^2 - 1/\theta_0^2) = 0$. 
}
\bea
\label{omdPomN1} 
\left. \omega \frac {dI}{d\omega} \right|^{\rm induced}_{N} \ &\sim& \ 
\alpha  \int d^2 \vec{\theta}_0 \left\{
2 \frac {\vec{\theta}_0}{\theta_0^2} \cdot  \sum_{n=1}^{N} \vec{J}_n \left[ \cos(\Phi_0 - \Phi_n) -1\right]
\right.  \nonumber \\
&& \hskip 1.5cm  + \left.   \sum_{n \neq m = 1}^{N} \vec{J}_n \cdot \vec{J}_m \left[\cos (\Phi_n - \Phi_m) -1 
\right] \right\} \, .
\eea
It is manifest that the induced spectrum vanishes when $L=0$, since this implies $\Phi_n = 0$ for all $n$. 

Consider first the contribution of the first term of \eq{omdPomN1} and concentrate on a particular term in the sum
$\sum_n$. As can be seen from \eq{Phin}, the phase difference $\Phi_n - \Phi_0$ does not depend on
$\vec{\theta}_n$ and hence on $\vec{q}_{n\perp}$. We can thus average over $\vec{q}_{n\perp}$ before doing
the integral. Averaging over azimuthal directions gives (see \eq{azimint})
\be
\label{Jnaver}
\langle \vec{J}_n \rangle_{\rm azim}  = \left\langle \frac {\vec{\theta}_{n-1} - \vec{q}_{n \perp}/E}{(\vec{\theta}_{n-1} - 
\vec{q}_{n\perp}/E)^2}-  \frac {\vec{\theta}_{n-1}}{\theta_{n-1}^2}  \right\rangle_{\rm azim} 
= - \frac {\vec{\theta}_{n-1}}{\theta_{n-1}^2} \, \, \Theta \left( \frac {q_{n\perp}^2}{E^2} - \theta_{n-1}^2 \right) \, .
\ee
Averaging further over $q_{n\perp}^2$ with the weight \eq{Pthets}, we obtain 
\be
\label{Jnaver1}
\langle \vec{J}_n \rangle \sim - \frac {\mu^2}{E^2}\frac {\vec{\theta}_{n-1}}{\theta_{n-1}^2 (\theta_{n-1}^2 +\mu^2/E^2)} \, \, .
\ee
Thus, we get the contribution
\bea
\label{omdPomN2} 
\left. \omega \frac {dI}{d\omega} \right|^{\rm induced}_{\rm 1st \, term} \sim 
 \alpha  \frac {\mu^2}{E^2} \sum_{n=1}^{N} \int \frac{d^2 \vec{\theta}_0}{\theta_0^2} 
\frac{\vec{\theta}_0 \vec{\theta}_{n-1}}{\theta_{n-1}^2 (\theta_{n-1}^2 + \mu^2/E^2 )}
\left[ 1- \cos \left( \frac {\omega \lambda}2 \sum_{m=0}^{n-1} \theta_m^2 \right) \right] \ , \nn \\
\eea
where, in each term of the sum, the averaging over $\vec{q}_{m\perp}$ with $m \neq n$ should still be performed. 

Suppose $L \ll L^*$  (otherwise the physics is the same as for the asymptotic particle, and there is no point to analyze \eq{omdPomN1} instead of
\eq{sumJsquared}). Then 
$1/(\omega L) >  1/(E L) \gg {N} \mu^2/E^2$. Thus, bearing in mind that $\theta_0^2 \sim 1/(\omega L)$ 
(to be verified {\it a posteriori}) and $|\vec{q}_{(m \neq n)\perp}| \sim \mu$, we derive 
\be
m \neq n \Rightarrow \theta_m^2 \simeq \theta_0^2 \sim \frac{1}{\omega L} \gg {N} \frac {\mu^2}{E^2}\, .
\ee
We arrive at the estimate \eq{omdPom2} of the spectrum, giving (see \eq{omdPom3}):
\be 
\label{spec1stline}
\left. \omega \frac {dI}{d\omega} \right|^{\rm induced}_{\rm 1st \, term} 
\ \sim \ \alpha \, \frac{\omega \, \omega_c}{E^2} \, ,
\ee
which contributes to the average loss as 
\be
\label{firstlineloss}
\left. \Delta E \right|_{\rm 1st\, term} \ \sim \ \alpha \, \omega_c  \, .
\ee
The integral is dominated by the angles of order $1/(\omega L)$, indeed.

Let us show now that, at $L \ll L^*$, the contribution of the second term in \eq{omdPomN1} is suppressed compared to
\eq{spec1stline} and \eq{firstlineloss}. The sum involves $\sim {N}^{\, 2}$ terms. Consider one of these terms, say, the term with 
$m=1, \, n = {N}$. The phase difference 
$$\Phi_{1{N}} = \Phi_{N} - \Phi_1 =  \frac {\omega \lambda }2 \sum_{k=1}^{{N} -1} \theta_k^2$$
does not depend on $\vec{q}_{n\perp}$ and we can  average 
$\vec{J}_n$ over it as before. In addition, choosing $\vec{\theta}_1$ rather than $\vec{\theta}_0$ as an integration
variable, we can observe that $\Phi_{1{N}}$ does not depend on $\vec{q}_{1\perp}$, and we can also average 
$\vec{J}_1$ over the latter. We obtain
\be
\label{termN1} 
\left. \omega \frac {dI}{d\omega} \right|_{1{N}} \sim \alpha \frac {\mu^4}{E^4}  
\int d^2 \vec{\theta}_1 \frac {\vec{\theta}_1 \cdot \vec{\theta}_{{N}-1} (1- \cos \Phi_{1{N}})}
{\theta_1^2  (\theta_1^2 + \mu^2/E^2 ) \theta_{{N}-1}^2 (\theta_{{N}-1}^2 + \mu^2/E^2 )} \, .
\ee
In contrast to the integral \eq{omdPomN2}, this integral is not dominated by large angles with 
$\Phi_{1{N}} \sim 1$. Indeed, assuming $\theta_k \simeq \theta_1$, we are led to the integral
\be
\sim  \int_{{N}\mu^2/E^2}  \frac{d\theta^2}{\theta^6} \left[ 1 - \cos (\omega L \theta^2) \right] \, ,
\ee
which is saturated by
\be
\label{logregion}
{N} \frac {\mu^2}{E^2} \ll \theta^2 \ll \frac 1{\omega L} \, ,
\ee
and exhibits a logarithmic behavior. Expanding $\cos{(\omega L \theta^2)}$ (we are allowed to do it in view of
\eq{logregion}), multiplying the contribution of one term in the sum of the second term of \eq{omdPomN1} by ${N}^2$, 
and neglecting the logarithmic factor which is irrelevant in the present discussion, we obtain the estimate 
\be
\label{secondline} 
\left. \omega \frac {dI}{d\omega} \right|_{\rm 2nd\, term}^{\rm induced} 
\ \sim \ \alpha \, \left( \frac{\omega \, \omega_c}{E^2} \right)^2  \, .
\ee 
Integrating this over $\omega$ gives the average loss 
\be
\label{secondlineloss}
\left. \Delta E \right|_{\rm 2nd\, term} \ \sim \ \alpha \, \frac{\omega_c^2}{E} \ 
\sim \ \alpha E \left( \frac{L}{L^*} \right)^4 \, .
\ee
This is indeed suppressed by $\sim \omega_c/E \sim (L/L^*)^2 \ll 1$ compared to the contribution \eq{firstlineloss} 
of the first term of \eq{omdPomN1}.


\end{document}